\documentclass{jfm}
\usepackage{graphicx}
\usepackage{epstopdf, epsfig}
\usepackage{hyperref}
\usepackage{color}
\usepackage{subcaption}
\usepackage{lipsum}
\usepackage[ruled,vlined]{algorithm2e}
\usepackage{lipsum}
\usepackage{tikz}
\usepackage{scalerel}
\newcommand\smallsquare{\scaleobj{0.7}{\Box}}

\newcommand\mapsfrom{\mathrel{\reflectbox{\ensuremath{\mapsto}}}}

\shorttitle{Joint reconstruction and segmentation of noisy velocity images}
\shortauthor{A. Kontogiannis, S. V. Elgersma, A. J. Sederman and M. P. Juniper}

\title{Joint reconstruction and segmentation of noisy velocity images as an inverse Navier--Stokes problem}

\author{Alexandros Kontogiannis\aff{1}
  \corresp{\email{ak2239@cam.ac.uk}},
  Scott V. Elgersma\aff{2}
  Andrew J. Sederman\aff{2}
  \and Matthew P. Juniper \aff{1}}

\affiliation{\aff{1}Department of Engineering, University of Cambridge, Trumpington Street, Cambridge CB2 1PZ, UK
\aff{2}Department of Chemical Engineering \& Biotechnology, University of Cambridge, Philippa Fawcett Drive, Cambridge CB3 0AS, UK}

\usepackage{mathrsfs}
\usepackage{mathtools}

\let\bm\boldsymbol

\DeclarePairedDelimiter{\abs}{\lvert}{\rvert}
\DeclarePairedDelimiter{\norm}{\big\lVert}{\big\rVert}

\DeclarePairedDelimiter{\binner}{\big\langle}{\big\rangle}
\DeclarePairedDelimiter{\Binner}{\Big\langle}{\Big\rangle}

\newcommand{\cu}{{\mathcal{C}_{\bm{u}}}}
\newcommand{\invcu}{{\mathcal{C}^{-1}_{\bm{u}}}}
\newcommand{\cgi}{{\mathcal{C}_{\bm{g}_i}}}
\newcommand{\invcgi}{{\mathcal{C}^{-1}_{\bm{g}_i}}}
\newcommand{\cgo}{{\mathcal{C}_{\bm{g}_o}}}
\newcommand{\invcgo}{{\mathcal{C}^{-1}_{\bm{g}_o}}}
\newcommand{\cnu}{{\Sigma_{\nu}}}
\newcommand{\invcnu}{{\Sigma^{-1}_{\nu}}}

\newcommand{\cov}{\mathcal{C}}
\newcommand{\R}{\mathbb{R}}

\newcommand{\eps}{\epsilon}

\newcommand{\spd}{\mathscr{V}}
\newcommand{\spdext}{{\mathring{\spd}}}
\newcommand{\nuext}{{\mathring{\bm{\nu}}}}
\newcommand{\zetaext}{{\mathring{\zeta}}}

\newcommand{\sdist}{{\phi_\pm}}

\newcommand*\mean[1]{\bar{#1}}

\definecolor{navyblue}{rgb}{0.0, 0.0, 0.5}
\definecolor{indigo}{rgb}{0.29, 0.0, 0.51}
\definecolor{linkpurple}{rgb}{0.58,0.2,0.82}

\hypersetup{
  colorlinks=true,
  urlcolor=indigo,
  citecolor=indigo,
  linkcolor=navyblue
}

\newcommand{\revvv}[1]{\textcolor{black}{#1}}
\newcommand{\rev}[1]{\textcolor{black}{#1}}
\newcommand{\revv}[1]{\textcolor{black}{#1}}

\begin{document}

\maketitle

\begin{abstract}
We formulate and solve a generalized inverse Navier--Stokes problem \revv{for the joint velocity field reconstruction and boundary segmentation} of noisy flow velocity images. To regularize the problem we use a Bayesian framework with Gaussian random fields. This allows us to estimate the uncertainties of the unknowns by approximating their posterior covariance with a quasi-Newton method. We first test the method for synthetic noisy images of 2D flows and observe that the method successfully reconstructs and segments the noisy synthetic images with a signal-to-noise ratio (SNR) of {3}. Then we conduct a magnetic resonance velocimetry (MRV) experiment to acquire images of an axisymmetric flow for low ($\simeq 6$) and high ($>30$) SNRs. We show that the method is capable of reconstructing and segmenting the low SNR images, producing noiseless velocity fields and a smooth segmentation, with negligible errors compared with the high SNR images. This amounts to a reduction of the total scanning time by a factor of 27. At the same time, the method provides additional knowledge about the physics of the flow (e.g. pressure), and addresses the shortcomings of MRV (low spatial resolution and partial volume effects) \revv{that otherwise hinder the accurate estimation of wall shear stresses}. \revvv{Although the implementation of the method is restricted to 2D steady planar and axisymmetric flows, the formulation applies immediately to 3D steady flows and naturally extends to 3D periodic and unsteady flows}.
\end{abstract}


\section{Introduction}
Experimental measurements of fluid flows inside or around an object often produce velocity images that contain noise. These images may be post-processed in order to either reveal obscured flow patterns or to extract a quantity of interest (e.g. pressure or wall shear stress). For example, magnetic resonance velocimetry (MRV) \citep{Fukushima1999,Mantle2003,Elkins2007,Markl2012,Demirkiran2021} can measure all three components of a time varying velocity field but the measurements become increasingly noisy as the spatial resolution is increased. To achieve an image of acceptable signal-to-noise ratio (SNR), repeated scans are often averaged, leading to long signal acquisition times. To address that problem, fast acquisition protocols (pulse sequences) can be used, but these may be difficult to implement and can lead to artefacts depending on the magnetic relaxation properties and the magnetic field homogeneity of the system studied. Another way to accelerate signal acquisition is by using sparse sampling techniques in conjunction with a reconstruction algorithm. The latter approach is an active field of research, commonly referred to as \textit{compressed sensing} \citep{Donoho2006,Lustig2007,Benning2014,Peper2019,Corona2021}. Compressed sensing (CS) algorithms exploit \textit{a priori} knowledge about the structure of the data, which is encoded in a regularization norm (e.g. total variation, wavelet bases), but without considering the physics of the problem. \revv{Even though the present study concerns the reconstruction of fully-sampled, noisy MRV images, the method that we present here can be applied to sparsely-sampled MRV data}.

For images depicting fluid flow, \textit{a priori} knowledge can come in the form of a \mbox{Navier--Stokes} problem. The problem of reconstructing and segmenting a flow image then can be expressed as a generalized inverse Navier--Stokes problem whose flow domain, boundary conditions, and model parameters have to be inferred in order for the modeled velocity to approximate the measured velocity in an appropriate metric space. This approach not only produces a reconstruction that is an accurate fluid flow inside or around the object (a solution to a Navier--Stokes problem), but also provides additional physical knowledge (e.g. pressure), which is otherwise difficult to measure. Inverse Navier--Stokes problems have been intensively studied during the last decade, mainly enabled by the increase of available computing power. Recent applications in fluid mechanics range from the forcing inference problem \citep{Hoang2014}, to the reconstruction of scalar image velocimetry (SIV) \citep{Gillissen2018,Sharma2019} and particle image velocimetry (PIV) \citep{Gillissen2019} signals, and the identification of optimal sensor arrangements \citep{Mons2017,Verma2019}. Regularization methods that can be used for model parameters are reviewed by \cite{Stuart2010} from a Bayesian perspective and by \cite{Benning2018} from a variational perspective. The well-posedness of Bayesian inverse Navier--Stokes problems is addressed by \cite{Cotter2009}.

Recently, \cite{Koltukluoglu2018} treat the reduced inverse Navier--Stokes problem of finding only the Dirichlet boundary condition for the inlet velocity that matches the modeled velocity field to MRV data for a steady 3D flow in a glass replica of the human aorta. They measure the model-data discrepancy using the $L^2$-norm and introduce additional variational regularization terms for the Dirichlet boundary condition. The same formulation is extended to periodic flows by \cite{Koltukluoglu2019,Koltukluoglu2019b}, using the harmonic balance method for the temporal discretization of the Navier--Stokes problem. \cite{Funke2019} address the problem of inferring both the inlet velocity (Dirichlet) boundary condition and the initial condition, for unsteady blood flows and 4D MRV data, with applications to cerebral aneurysms. We note that the above studies consider rigid boundaries and require \textit{a priori} an accurate, and time-averaged, geometric representation of the blood vessel.

To find the shape of the flow domain, e.g. the blood vessel boundaries, computed tomography (CT) or magnetic resonance angiography (MRA) is often used. The acquired image is then reconstructed, segmented, and smoothed. This process not only requires substantial effort and the design of an additional experiment (e.g. CT, MRA), but it also introduces geometric uncertainties \citep{Morris2016,Sankaran2016}, which, in turn, affect the predictive confidence of arterial wall shear stress distributions and their mappings \citep{Katritsis2007,Sotelo2016}. For example, \cite{Funke2019} report discrepancies between the modeled and the measured velocity fields near the flow boundaries, and they suspect they are caused by geometric errors that were introduced during the segmentation process. In general, the assumption of rigid boundaries either implies that a time-averaged geometry has to be used, or that an additional experiment (e.g. CT, MRA) has to be conducted to register the moving boundaries to the flow measurements.

A more consistent approach to this problem is to treat the blood vessel geometry as an unknown when solving the generalized inverse Navier--Stokes problem. In this way, the inverse Navier--Stokes problem simultaneously reconstructs and segments the velocity fields and can better adapt to the MRV experiment by correcting the geometric errors and improving the reconstruction.

In this study, we address the problem of \revv{simultaneous velocity field reconstruction and boundary segmentation} by formulating a generalized inverse Navier--Stokes problem, whose flow domain, boundary conditions, and model parameters are all considered unknown. To regularize the problem, we use a Bayesian framework and Gaussian measures in Hilbert spaces. This further allows us to estimate the posterior Gaussian distributions of the unknowns using a quasi-Newton method, which has not yet been addressed for this type of problem. We provide an algorithm for the solution of this generalized inverse Navier--Stokes problem, and demonstrate it on synthetic images of 2D steady flows and real MRV images of a steady axisymmetric flow.

This paper consists of two parts. In section \ref{sec:inverse_ns_problem}, we formulate the generalized inverse Navier--Stokes problem and an algorithm that solves it. In section \ref{sec:rec_and_segm_of_images} we test the method using both synthetic and real MRV velocity images and describe the setup of the MRV experiment.

\section{An inverse Navier--Stokes problem for noisy flow images}
\label{sec:inverse_ns_problem}
In this section, we formulate the generalized inverse Navier--Stokes problem and provide an algorithm for its solution. In what follows, $L^2(\Omega)$ denotes the space of square-integrable functions in $\Omega$, with inner product $\binner{\cdot,\cdot}$ and norm $\norm{\cdot}_{L^2(\Omega)}$, and $H^k(\Omega)$ the space of square-integrable functions with $k$ square-integrable derivatives in $\Omega$. For a given covariance operator, $\mathcal{C}$, we also define the covariance-weighted $L^2$ spaces, endowed with the inner product ${\binner{\cdot,\cdot}_\mathcal{C} := \binner{\cdot,\mathcal{C}^{-1}\cdot}}$, which generates the norm $\norm{\cdot}_{\mathcal{C}}$. The Euclidean norm in the space of real numbers $\R^n$ is denoted by $\abs{\cdot}_{\R^n}$. We use the superscript $(\cdot)^\star$ to denote a measurement, $(\cdot)^\circ$ to denote a reconstruction, and $(\cdot)^\bullet$ to denote the ground truth.

\subsection{The inverse Navier--Stokes problem}
\label{sec:rec_error}

\begin{figure}
  \centerline{\includegraphics[width=\textwidth]{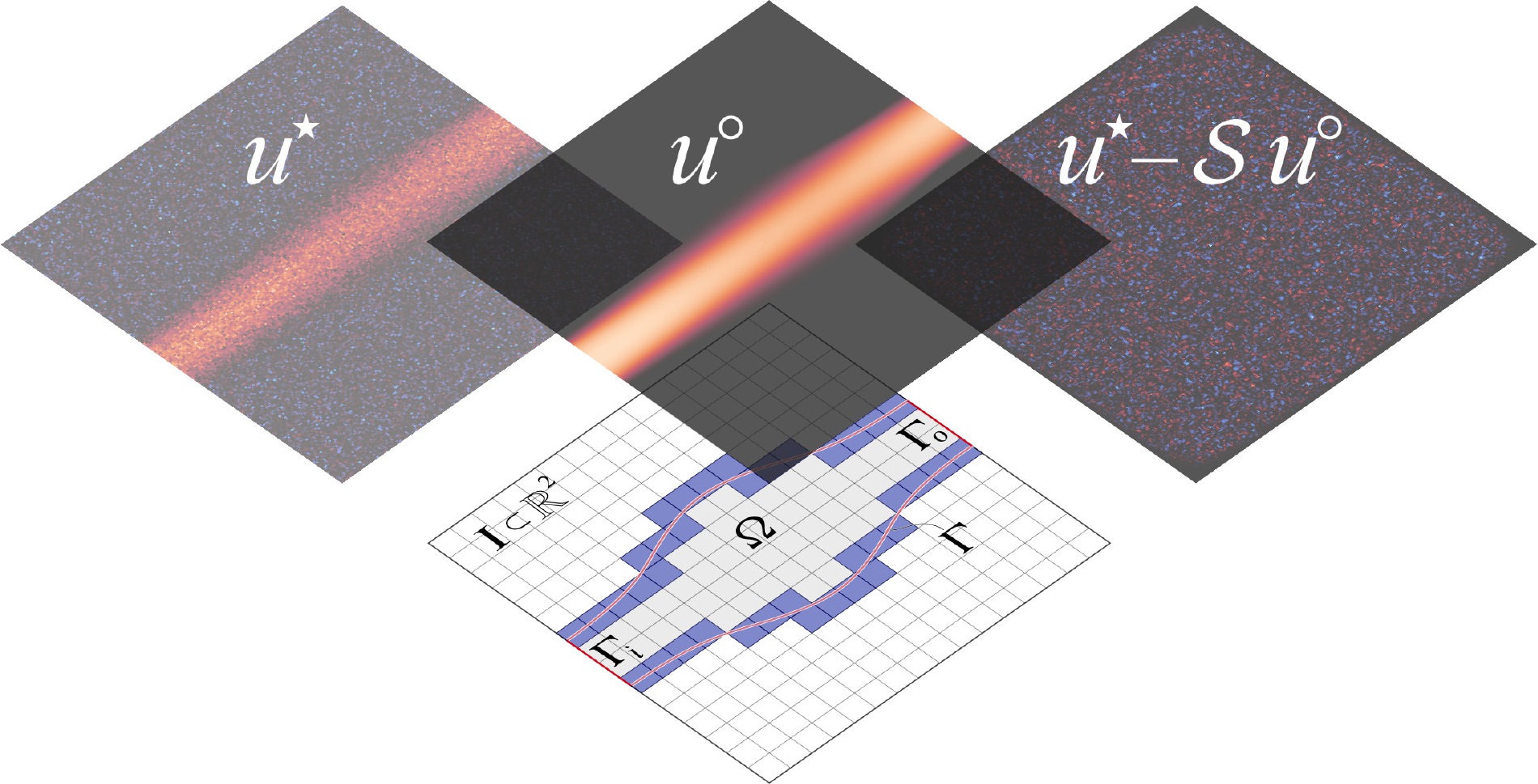}}
  \caption{Given the images of a measured velocity field $\bm{u}^\star$, we solve an inverse {Navier--Stokes} problem to infer the boundary $\Gamma$ (or $\partial\Omega$), the kinematic viscosity, and the inlet velocity profile on $\Gamma_i$. The solution to this inverse problem is a reconstructed velocity field $\bm{u}^\circ$, from which the noise and the artefacts $(\bm{u}^\star-\mathcal{S}\bm{u}^\circ)$ have been filtered out.}
\label{fig:theoretical_twin}
\end{figure}

A $n$-dimensional velocimetry experiment usually provides noisy flow velocity images on a domain $I \subset \R^n$, depicting the measured flow velocity $\bm{u}^\star$ inside an object $\Omega \subset I$ with boundary $\partial\Omega = \Gamma \cup \Gamma_i \cup \Gamma_o$ (figure \ref{fig:theoretical_twin}). An appropriate model is the Navier--Stokes problem 
\begin{equation}
\left\{\begin{alignedat}{2}
    \bm{u}\bm{\cdot}\nabla\bm{u}-\nu{\rmDelta} \bm{u}  + \nabla p &= \bm{0} \quad &&\textrm{in}\quad \Omega \\
    \nabla \bm{\cdot} \bm{u} &= 0 \quad &&\textrm{in}\quad \Omega\\
    \bm{u} &= \bm{0} \quad &&\textrm{on}\quad \Gamma \\
   \bm{u} &= \bm{g}_i \quad &&\textrm{on}\quad \Gamma_i\\
   -\nu\partial_{\bm{\nu}}\bm{u}+p\bm{\nu} &= \bm{g}_o \quad &&\textrm{on}\quad \Gamma_o
  \end{alignedat}\right.\quad,
  \label{eq:navierstokes_bvp}
\end{equation}
where $\bm{u}$ is the velocity, $p\mapsfrom p/\rho$ is the reduced pressure, $\rho$ is the density, $\nu$ is the kinematic viscosity, $\bm{g}_i$ is the Dirichlet boundary condition at the inlet $\Gamma_i$, $\bm{g}_o$ is the natural boundary condition at the outlet $\Gamma_o$, $\bm{\nu}$ is the unit normal vector on $\partial\Omega$, and ${\partial_{\bm{\nu}}\equiv\bm{\nu}\bm{\cdot}\nabla}$ is the normal derivative. 

We denote the data space by $\bm{D}$ and the model space by $\bm{M}$, and assume that both spaces are subspaces of $\bm{L}^2$. In the 2D case, $\bm{u}^\star = (u^\star_x,u^\star_y)$, and we introduce the covariance operator
\begin{equation}
{\cu = \mathrm{diag}\Big(\sigma^2_{u_x} \mathrm{I},~\sigma^2_{u_y} \mathrm{I}\Big)}\quad,
\end{equation}
 where $\sigma^2_{u_x},\sigma^2_{u_y}$ are the Gaussian noise variances of $u^\star_x,u^\star_y$, respectively, and $\mathrm{I}$ is the identity operator. 
 The discrepancy between the measured velocity field ${\bm{u}^\star} \in \bm{D}$ and the modeled velocity field $\bm{u} \in \bm{M}$ is measured on the data space $\bm{D}$ using the reconstruction error functional
\begin{equation}
\mathscr{E}(\bm{u}) \equiv \frac{1}{2}\norm{\bm{u}^\star-\mathcal{S}\bm{u}}^2_\cu := \frac{1}{2}\int_I \big(\bm{u}^\star-\mathcal{S}\bm{u}\big)\mathcal{C}_{\bm{u}}^{-1}\big(\bm{u}^\star-\mathcal{S}\bm{u}\big)\quad,
\label{eq:rec_error}
\end{equation}
where $\mathcal{S}: \bm{M}\to \bm{D}$ is the $L^2$-projection\footnote{\revv{Since the discretized space consists of bilinear quadrilateral finite elements (see section 
\ref{sec:numerics}), this projection is a linear interpolation.}} from the model space $\bm{M}$ to the data space $\bm{D}$.

Our goal is to infer the unknown parameters of the Navier--Stokes problem \eqref{eq:navierstokes_bvp} such that the model velocity $\bm{u}$ approximates the noisy measured velocity $\bm{u}^\star$ in the covariance-weighted $L^2$-metric defined by $\mathscr{E}$. In the general case, the unknown model parameters of \eqref{eq:navierstokes_bvp} are the shape of $\Omega$, the kinematic viscosity $\nu$, and the boundary conditions $\bm{g}_i,\bm{g}_o$. This inverse Navier--Stokes problem leads to the nonlinearly constrained optimization problem
\begin{equation}
\text{find} \quad \bm{u}^\circ \equiv \underset{\Omega,\bm{x}}{\mathrm{argmin}}~\mathscr{E}\big(\bm{u}(\Omega;\bm{x})\big), \quad \text{such that $\bm{u}$ satisfies}\ \eqref{eq:navierstokes_bvp}\quad,
\label{eq:inv_prob1}
\end{equation}
where $\bm{u}^\circ$ is the reconstructed velocity field, and $\bm{x}=(\bm{g}_i,\bm{g}_o,\nu)$. Like most inverse problems, \eqref{eq:inv_prob1} is ill-posed and hard to solve. To alleviate the ill-posedness of the problem we need to restrict our search of the unknowns $(\Omega,\bm{x})$ to function spaces of sufficient regularity.  

\subsection{Regularization}
\label{sec:regularization}
If $x(t) \in L^2(\R)$ is an unknown parameter, one way to regularize the inverse problem \eqref{eq:inv_prob1} is to search for minimizers of the augmented functional $\mathscr{J} \equiv \mathscr{E} + \mathscr{R}$, where
\begin{equation}
\mathscr{R}(x) = \sum_{j=0}^k \int_{\R} \alpha_j \big|\partial_x^j(x-\mean{x})\big|^2
\label{eq:arbitrary_reg}
\end{equation} 
is a regularization norm for a given (and fixed) prior assumption $\mean{x}(t) \in H^k(\R)$, weights ${\alpha_j\in \mathbb{R}}$, and positive integer $k$. This simple idea can be quite effective because by minimizing $\mathscr{R}$ we force $x$ to lie in a subspace of $L^2$ having higher regularity, namely $H^k$, and as close to the prior value $\mean{x}$ as $\alpha_j$ allow\footnote{The regularization term, given by \eqref{eq:arbitrary_reg}, can be further extended to fractional Hilbert spaces by defining the norm $\norm{x}_{H^s(\R)} := \norm{(1+\abs{t}^s)\mathcal{F}x}_{L^2(\R)}$ for noninteger $s$, with $0<s<\infty$, and where $\mathcal{F}$ denotes the Fourier transform. Interestingly, under certain conditions, which are dictated by Sobolev's embedding theorem \citep[Chapter~5]{Evans2010}, these Hilbert spaces can be embedded in the more familiar spaces of continuous functions.}. However, as \cite{Stuart2010} points out, in this setting, the choice of $\alpha_j$, and even the form of $\mathscr{R}$, is arbitrary. 

There is a more intuitive approach that recovers the form of the regularization norm $\mathscr{R}$ from a probabilistic viewpoint. In the setting of the Hilbert space $L^2$, the Gaussian measure $\gamma \sim \mathcal{N}(m,\mathcal{C})$ has the property that its finite-dimensional projections are multivariate Gaussian distributions, and it is uniquely defined by its mean $m \in L^2$, and its covariance operator $\mathcal{C}:L^2\to L^2$ (appendix \ref{app:gaussian_meas}). It can be shown that  there is a natural Hilbert space $H_\gamma$ that corresponds to $\gamma$, and that \citep{Bogachev1998,Hairer2009}
$$H_\gamma = \sqrt{\cov}\big(L^2\big) \quad.$$
In other words, if $x$ is a random function distributed according to $\gamma$, any realization of $x$ lies in $H_\gamma$, which is the image of $\sqrt{\cov}$. Furthermore, the corresponding inner product
\begin{gather}
\binner{x,x'}_\cov = \binner{\cov^{-1/2}x,\ \cov^{-1/2}x'}
\label{eq:cm_inner_prod}
\end{gather}
is the covariance between $x$ and $x'$, and the norm $\norm{x}^2_\cov = \binner{x,x}_\cov$ is the variance of $x$. Therefore, if $x$ is an unknown parameter for which \textit{a priori} statistical information is available, and if the Gaussian assumption can be justified, we can choose
\begin{equation}
\mathscr{R}(x) = \frac{1}{2}\norm{x-\mean{x}}^2_\cov \quad.
\end{equation}
In this way, $\mathscr{J}\equiv \mathscr{E}+\mathscr{R}$ increases as the variance of $x$ increases. Consequently, minimizing $\mathscr{J}$ penalizes improbable realizations.

As mentioned in section \ref{sec:rec_error}, the unknown model parameters of the Navier--Stokes problem \eqref{eq:navierstokes_bvp} are the kinematic viscosity $\nu$, the boundary conditions $\bm{g}_i,\bm{g}_o$, and the shape of $\Omega$. Since we consider the kinematic viscosity $\nu$ to be constant, the regularizing norm is simply
\begin{gather}
\frac{1}{2} \big|\nu - \mean{\nu}\big|^2_{\Sigma_\nu} = \frac{1}{2\sigma^2_{\nu}} \big|\nu - \mean{\nu}\big|^2_\R\quad,
\end{gather}
where $\mean{\nu}\in\R$ is a prior guess for $\nu$, and $\sigma^2_{\nu}\in\R$ is the variance. For the Dirichlet boundary condition, $\bm{g}_i \in \bm{L}^2(\Gamma_i)$, we choose the exponential covariance function
\begin{gather}
C(x,x') = \frac{\sigma_{\bm{g}_i}^2}{2\ell} \exp\bigg(-\frac{\abs{x-x'}}{\ell}\bigg)
\label{eq:cov_exp}
\end{gather}
with variance $\sigma_{\bm{g}_i}^2 \in \R$ and characteristic length $\ell \in \R$. For zero-Dirichlet (no-slip) or zero-Neumann boundary conditions on $\partial\Gamma_i$, \eqref{eq:cov_exp} leads to the norm \cite[Chapter~7.21]{Tarantola2005}
\begin{gather}
\norm{\bm{g}_i}_{\cov_{\bm{g}_i}}^2 \simeq \frac{1}{\sigma_{\bm{g}_i}^2}\int_{\Gamma_i}\bm{g}^2_i + \ell^2~\big(\nabla\bm{g}_i\big)^2 \quad.
\end{gather}
Using integration by parts we find that the covariance operator is
\begin{gather}
\cov_{\bm{g}_i} = \sigma_{\bm{g}_i}^2\Big(\mathrm{I} - \ell^2\widetilde{\rmDelta}\Big)^{-1}\quad,
\end{gather}
where $\widetilde{\rmDelta}$ is the $L^2$-extension of the Laplacian $\rmDelta$ that incorporates the boundary condition $\bm{g}_i = \bm{0}$ on ${\partial\Gamma_i}$. For the natural boundary condition, $\bm{g}_o \in \bm{L}^2(\Gamma_o)$, we can use the same covariance operator, but equip $\widetilde{\rmDelta}$ with zero-Neumann boundary conditions, i.e. $\partial_{\bm{\nu}}\bm{g}_o = 0$ on $\partial\Gamma_o$. Lastly, for the shape of $\Omega$, which we implicitly represent with a signed distance function $\sdist$ (defined in section \ref{sec:geom_flow}), we choose the norm
\begin{gather}
\frac{1}{2}\norm{\mean{\phi}_\pm-\sdist}^2_{\mathcal{C}_\sdist}=\frac{1}{2\sigma^{2}_\sdist}\norm{\mean{\phi}_\pm-\sdist}^2_{L^2(I)},
\label{eq:zeta_reg}
\end{gather}
where $\sigma_\sdist \in \R$ and $\mean{\phi}_\pm \in L^2(I)$. Additional regularization for the boundary of $\Omega$ (i.e. the zero level-set of $\sdist$) is needed and it is described in section \ref{sec:geom_flow}. 
Based on the above results, the regularization norm for the unknown model parameters is
\begin{align}
\mathscr{R}(\bm{x},\sdist) = ~&\frac{1}{2}\big|\nu - \mean{\nu}\big|^2_{\Sigma_\nu} + \frac{1}{2}\norm{\bm{g}_i-\mean{\bm{g}}_i}^2_{\cov_{\bm{g}_i}} \nonumber\\ +&\frac{1}{2}\norm{\bm{g}_o-\mean{\bm{g}}_o}^2_{\cov_{\bm{g}_o}}+\frac{1}{2}\norm{\mean{\phi}_\pm-\sdist}^2_{\mathcal{C}_\sdist}\quad.
\end{align}

\subsection{Euler--Lagrange equations for the inverse Navier--Stokes problem}
\label{sec:eul_lag_system}
Testing the Navier--Stokes problem \eqref{eq:navierstokes_bvp} with functions ${(\bm{v},q) \in \bm{H}^1(\Omega) \times L^2(\Omega)}$, and after integrating by parts, we obtain the weak form
\begin{align}
\mathscr{M}(\Omega)(\bm{u},p,\bm{v},q;\bm{x}) \equiv\int_\Omega\Big(\bm{v}\bm{\cdot}\big( \bm{u}\bm{\cdot}\nabla\bm{u}\big) &+ \nu\nabla\bm{v}\bm{:}\nabla \bm{u} - (\nabla\bm{\cdot}\bm{v}) p - q(\nabla\bm{\cdot} \bm{u})\Big) +\int_{\Gamma_o}\bm{v}\bm{\cdot}\bm{g}_o  \nonumber\\  +\int_{\Gamma\cup\Gamma_i} \bm{v}\bm{\cdot}(-\nu\partial_{\bm{\nu}}\bm{u}+p\bm{\nu}) +& \mathscr{N}_{\Gamma_i}(\bm{v},q,\bm{u};\bm{g}_i)+\mathscr{N}_{\Gamma}(\bm{v},q,\bm{u};\bm{0})={0}\quad,
\label{eq:navier_stokes_weak}
\end{align}
where $\mathscr{N}$ is the \cite{Nitsche1971} penalty term
\begin{gather}
\mathscr{N}_{T}(\bm{v},q,\bm{u};\bm{z}) \equiv \int_{T} (-\nu\partial_{\bm{\nu}}\bm{v}+q\bm{\nu}+ \eta\bm{v})\bm{\cdot}(\bm{u}-\bm{z})\quad,
\end{gather}
which weakly imposes the Dirichlet boundary condition $\bm{z} \in \bm{L}^2(T)$ on a boundary $T$, given a penalization constant $\eta$\footnote{\revv{The penalization $\eta$ is a numerical parameter with no physical significance (see section \ref{sec:numerics}).}}.
We define the augmented reconstruction error functional
\begin{align}
\mathscr{J}(\Omega)(\bm{u},p,\bm{v},q;\bm{x}) \equiv\ &  \mathscr{E}(\bm{u})+ \mathscr{R}(\bm{x},\sdist) + \mathscr{M}(\Omega)(\bm{u},p,\bm{v},q;\bm{x})\quad,
\label{eq:aug_rec_func}
\end{align}
which contains the regularization terms $\mathscr{R}$ and the model constraint $\mathscr{M}$, such that $\bm{u}$ weakly satisfies \eqref{eq:navierstokes_bvp}. To reconstruct the measured velocity field $\bm{u}^\star$ and find the unknowns $(\Omega,\bm{x})$, we minimize $\mathscr{J}$ by solving its associated Euler--Lagrange system. 
\subsubsection{Adjoint Navier--Stokes problem}
In order to derive the Euler--Lagrange equations for $\mathscr{J}$, we first define 
\begin{gather}
\bm{\mathcal{U}}' = \Big\{\bm{u}' \in \bm{H}^1(\Omega): \bm{u}'\big\vert_{\Gamma\cup\Gamma_i} \equiv \bm{0} \Big\}
\end{gather}
to be the space of admissible velocity perturbations $\bm{u}'$, and $\mathcal{P}'\subset L^2(\Omega)$ to be the space of admissible pressure perturbations $p'$, such that $(-\partial_{\bm{\nu}}\bm{u}'+p'\bm{\nu})\big\vert_{\Gamma_o}\equiv \bm{0}$. We start with
\begin{align}
 \delta_{\bm{u}}\mathscr{E}\equiv\frac{d}{d\tau}\mathscr{E}(\bm{u}+\tau\bm{u}')\Big\vert_{\tau=0}&=\int_\Omega -\invcu\big(\bm{u}^\star-\mathcal{S}\bm{u}\big)\bm{\cdot}\mathcal{S}\bm{u}'\nonumber\\
&= \int_\Omega -\mathcal{S}^\dagger\invcu\big(\bm{u}^\star-\mathcal{S}\bm{u}\big)\bm{\cdot}\bm{u}'\equiv\Binner{D_{\bm{u}}\mathscr{E},\bm{u}'}_\Omega\quad .
\label{eq:dEdu}
\end{align}
Adding together the first variations of $\mathscr{M}$ with respect to $(\bm{u},p)$,
\begin{gather*}
\delta_{\bm{u}}\mathscr{M}\equiv\frac{d}{d\tau}\mathscr{M}(\cdot)(\bm{u}+\tau\bm{u}',\dots)\Big\vert_{\tau=0} \quad, \quad \delta_{{p}}\mathscr{M}\equiv\frac{d}{d\tau}\mathscr{M}(\cdot)(\dots,p+\tau p',\dots)\Big\vert_{\tau=0},
\end{gather*}
and after integrating by parts, we find
\begin{align}
\delta_{\bm{u}}\mathscr{M}+\delta_{p}\mathscr{M}=&\int_\Omega \Big(-\bm{u}\bm{\cdot}\big(\nabla\bm{v} + (\nabla\bm{v})^\dagger\big) -\nu\rmDelta \bm{v} + \nabla q \Big)\bm{\cdot}\bm{u}' + \int_\Omega (\nabla\bm{\cdot}\bm{v})p' \nonumber\\
+&\int_{\partial\Omega} \big({(\bm{u}\bm{\cdot}\bm{\nu})\bm{v}}+{(\bm{u}\bm{\cdot}\bm{v})\bm{\nu}}+\nu\partial_{\bm{\nu}}\bm{v}-q\bm{\nu}\big)\bm{\cdot} \bm{u}' \nonumber\\ +&\int_{\Gamma\cup\Gamma_i} \bm{v}\bm{\cdot}(-\nu\partial_{\bm{\nu}}\bm{u}'+p'\bm{\nu})+ \mathscr{N}_{\Gamma_i\cup\Gamma}(\bm{v},q,\bm{u}';\bm{0}) \quad.
\label{eq:dMdu}
\end{align}
Since $\mathscr{R}$ does not depend on $(\bm{u},p)$, we can use \eqref{eq:dEdu} and \eqref{eq:dMdu} to assemble the optimality conditions of $\mathscr{J}$ for $(\bm{u},p)$  
\begin{gather}
\Binner{D_{\bm{u}}\mathscr{J},\bm{u}'}_\Omega = {0} \quad,\quad \Binner{D_{{p}}\mathscr{J},{p}'}_\Omega = 0\quad .
\label{eq:dJdu_dJdp}
\end{gather}
For \eqref{eq:dJdu_dJdp} to hold true for all perturbations ${(\bm{u}',p')\in \bm{\mathcal{U}}' \times \mathcal{P}'}$, we deduce that $(\bm{v},q)$ must satisfy the following adjoint Navier--Stokes problem
\begin{gather}
\left\{\begin{alignedat}{2}
- \bm{u}\bm{\cdot}\big(\nabla\bm{v} + (\nabla\bm{v})^\dagger\big) -\nu\rmDelta \bm{v} + \nabla q &= -D_{\bm{u}}\mathscr{E} \quad &&\textrm{in}\quad \Omega \\ 
\nabla \bm{\cdot} \bm{v} &= 0 \quad &&\textrm{in}\quad \Omega\\
\bm{v} &= \bm{0} \quad &&\textrm{on}\quad \Gamma\cup\Gamma_i\\
{(\bm{u}\bm{\cdot}\bm{\nu})\bm{v}}+{(\bm{u}\bm{\cdot}\bm{v})\bm{\nu}}+\nu\partial_{\bm{\nu}}\bm{v}-q\bm{\nu} &= \bm{0} \quad &&\textrm{on}\quad \Gamma_o
\end{alignedat}\right. \quad.
\label{eq:navier-stokes_adjoint_problem}
\end{gather}
In this context, $\bm{v}$ is the adjoint velocity and $q$ is the adjoint pressure, which both vanish when $ \bm{u}^\star \equiv \mathcal{S}\bm{u}$. Note also that we choose boundary conditions for the adjoint problem \eqref{eq:navier-stokes_adjoint_problem} that make the boundary terms of \eqref{eq:dMdu} vanish, and that these boundary conditions are subject to the choice of $\bm{\mathcal{U}}'$, which, in turn, depends on the boundary conditions of the (primal) Navier--Stokes problem.  

\subsubsection{Shape derivatives for the Navier--Stokes problem}
To find the shape derivative of an integral defined in $\Omega$, when the boundary $\partial\Omega$ deforms with speed $\spd$, we use Reynold's transport theorem. For the bulk integral of $f:\Omega \to \R$, we find
\begin{gather}
\frac{d}{d\tau}\bigg(\int_{\Omega(\tau)} f\bigg)\bigg\vert_{\tau=0} = \int_\Omega f' + \int_{\partial\Omega}f~(\spd\bm{\cdot}\bm{\nu})\quad,
\label{eq:reynolds_bulk}
\end{gather} 
while for the boundary integral of $f$ we find \citep[Chapter~5.6]{Walker2015}
\begin{gather}
\frac{d}{d\tau}\bigg(\int_{\partial\Omega(\tau)} f\bigg)\bigg\vert_{\tau=0} = \int_{\partial\Omega} f' + (\partial_{\bm{\nu}}+\kappa)f ~(\spd\bm{\cdot}\bm{\nu})\quad,
\label{eq:reynolds_bound}
\end{gather}
where $f'$ is the shape derivative of $f$ (due to $\spd$), $\kappa$ is the summed curvature of $\partial\Omega$, and $\spd \equiv \zeta\bm{\nu}$, with $\zeta \in L^2(\partial\Omega)$, is the Hadamard parameterization of the speed field. Any boundary that is a subset of $\partial I$, i.e. the edge of the image $I$, is non-deforming and therefore the second term of the above integrals vanishes. The only boundary that deforms is $\Gamma \subset \partial\Omega$. For brevity, let $\delta_\spd I$ denote the shape perturbation of an integral $I$. Using \eqref{eq:reynolds_bulk} on $\mathscr{E}$, we compute
\begin{gather}
\delta_\spd\mathscr{E} = \Binner{D_{\bm{u}}\mathscr{E},\bm{u}'}_\Omega\quad,
\end{gather}
where $D_{\bm{u}}\mathscr{E}$ is given by \eqref{eq:dEdu}.
Using \eqref{eq:reynolds_bulk} and \eqref{eq:reynolds_bound} on $\mathscr{M}$, we obtain the shape derivatives problem for $(\bm{u}',p')$
\begin{equation}
\left\{\begin{alignedat}{2}
    \bm{u}'\bm{\cdot}\nabla\bm{u}+\bm{u}\bm{\cdot}\nabla\bm{u}'-\nu{\rmDelta} \bm{u}'  + \nabla p' &= \bm{0} \quad &&\textrm{in}\quad \Omega \\
    \nabla \bm{\cdot} \bm{u}' &= 0 \quad &&\textrm{in}\quad \Omega\\
    \bm{u}' &= -\partial_{\bm{\nu}}\bm{u}(\spd\bm{\cdot}\bm{\nu}) \quad &&\textrm{on}\quad \Gamma \\
    \bm{u}' &= \bm{0} \quad &&\textrm{on}\quad \Gamma_i \\
   -\nu\partial_{\bm{\nu}}\bm{u}'+p'\bm{\nu} &= \bm{0} \quad &&\textrm{on}\quad \Gamma_o
  \end{alignedat}\right.\quad,
  \label{eq:shape_deriv_ns_bvp}
\end{equation}
which can be used directly to compute the velocity and pressure perturbations for a given speed field $\spd$. We observe that $(\bm{u}',p')\equiv \bm{0}$ when $\zeta\equiv\spd\bm{\cdot}\bm{\nu}\equiv{0}$.
Testing the shape derivatives problem \eqref{eq:shape_deriv_ns_bvp} with $(\bm{v},q)$, and adding the appropriate Nitsche terms for the weakly enforced Dirichlet boundary conditions, we obtain
\begin{align}
\delta_\spd\mathscr{M} =&\int_\Omega\Big(\bm{v}\bm{\cdot}\big( \bm{u}'\bm{\cdot}\nabla\bm{u}+\bm{u}\bm{\cdot}\nabla\bm{u}'\big) + \nu\nabla\bm{v}\bm{:}\nabla \bm{u}' - (\nabla\bm{\cdot}\bm{v}) p' - q(\nabla\bm{\cdot} \bm{u}')\Big)  \nonumber\\  +\int_{\Gamma\cup\Gamma_i}& \bm{v}\bm{\cdot}(-\nu\partial_{\bm{\nu}}\bm{u}'+p'\bm{\nu}) + \mathscr{N}_{\Gamma_i}(\bm{v},q,\bm{u}';\bm{0})+\mathscr{N}_{\Gamma}(\bm{v},q,\bm{u}';-\zeta\partial_{\bm{\nu}}\bm{u})={0}\quad.
\label{eq:shape_deriv_ns_weak1}
\end{align}
If we define $I_i$ to be the first four integrals in \eqref{eq:dMdu}, integrating \eqref{eq:shape_deriv_ns_weak1} by parts yields
\begin{align}
\delta_\spd\mathscr{M}=& \sum_{i=1}^4 I_i + \mathscr{N}_{\Gamma_i}(\bm{v},q,\bm{u}';\bm{0})+\mathscr{N}_{\Gamma}(\bm{v},q,\bm{u}';-\zeta\partial_{\bm{\nu}}\bm{u})={0}\quad,
\end{align}
and, due to the adjoint problem \eqref{eq:navier-stokes_adjoint_problem}, we find
\begin{align}
\delta_\spd\mathscr{M}=&-\delta_\spd\mathscr{E} + \int_\Gamma (-\nu\partial_{\bm{\nu}}\bm{v}+q\bm{\nu})\bm{\cdot}\zeta\partial_{\bm{\nu}}\bm{u}={0}\quad,
\label{eq:dMdV_formula}
\end{align}
since $\bm{u}'\big\vert_{\Gamma_i} \equiv \bm{0}$, and $\bm{u}\big\vert_{\Gamma} \equiv \bm{0}$. Therefore, the shape perturbation of $\mathscr{J}$ is
\begin{gather}
\delta_\spd \mathscr{J} \equiv \Binner{D_{\spd}\mathscr{J},\spd\bm{\cdot}\bm{\nu}}_\Gamma \equiv \Binner{D_{\zeta}\mathscr{J},\zeta}_\Gamma = \delta_\spd \mathscr{E} + \delta_\spd \mathscr{M} + \delta_\spd \mathscr{R} = 0\quad,
\end{gather}
which, due to \eqref{eq:dMdV_formula} and $\delta_\spd \mathscr{R} \equiv 0$, takes the form
\begin{equation}
\Binner{D_{\zeta}\mathscr{J},\zeta}_\Gamma = \Binner{\partial_{\bm{\nu}}\bm{u}\bm{\cdot}\big(-\nu\partial_{\bm{\nu}}\bm{v}+q\bm{\nu}\big),\zeta}_\Gamma \label{eq:grad_shape}\quad,
\end{equation}
where $D_{\zeta}\mathscr{J}$ is the shape gradient. Note that the shape gradient depends on the normal gradient of the (primal) velocity field and the pseudotraction, $\big(-\nu\nabla\bm{v} + q\mathrm{I}\big)\bm{\cdot}\bm{\nu}$, that the adjoint flow exerts on $\Gamma$.

\subsubsection{Generalized gradients for the unknown model parameters $\bm{x}$}
The unknown model parameters $\bm{x}$ have an explicit effect on $\mathscr{M}$ and $\mathscr{R}$, and can therefore be obtained by taking their first variations. For the Dirichlet-type boundary condition at the inlet we find
\begin{align}
\Binner{D_{\bm{g}_i}\mathscr{J},~\bm{g}_i'}_{\Gamma_i} &= \Binner{\nu\partial_{\bm{\nu}}\bm{v}-q\bm{\nu}- \eta\bm{v} + \invcgi\big(\bm{g}_i-{\mean{\bm{g}}_i} \big),~\bm{g}_i'}_{\Gamma_i} \label{eq:grad_gi}\\ &=\Binner{\cgi\big(\nu\partial_{\bm{\nu}}\bm{v}-q\bm{\nu}- \eta\bm{v}\big) + \bm{g}_i-\mean{\bm{g}}_i,~\bm{g}_i'}_{\cgi}=\Binner{\widehat{D}_{\bm{g}_i}\mathscr{J},~\bm{g}_i'}_{\cgi},\nonumber
\end{align}
where $-\widehat{D}_{\bm{g}_i}\mathscr{J}$ is the steepest descent direction that corresponds to the covariance-weighted norm. For the natural boundary condition at the outlet we find
\begin{align}
\Binner{D_{\bm{g}_o}\mathscr{J},~\bm{g}_o'}_{\Gamma_o} &= \Binner{\bm{v} + \invcgo\big(\bm{g}_o-\mean{\bm{g}}_o\big),~\bm{g}_o'}_{\Gamma_o} \label{eq:grad_go}\\&=\Binner{\cgo\bm{v} + \bm{g}_o-\mean{\bm{g}}_o,~\bm{g}_o'}_{\cgo} = \Binner{\widehat{D}_{\bm{g}_o}\mathscr{J},~\bm{g}_o'}_{\cgo}. \nonumber
\end{align}
Lastly, since the kinematic viscosity is considered to be constant within $\Omega$ its generalized gradient is
\begin{align}
\Binner{D_{\nu}\mathscr{J},~\nu'}_\mathbb{R} &= \Binner{\int_\Omega \nabla\bm{v}\bm{:}\nabla\bm{u} + \invcnu \big(\nu-\mean{\nu}\big),~\nu'}_\mathbb{R} \label{eq:grad_nu} \\
& = \Binner{\cnu \int_\Omega \nabla\bm{v}\bm{:}\nabla\bm{u} + \nu-\mean{\nu},~\nu'}_{\cnu} = \Binner{\widehat{D}_{\nu}\mathscr{J},~\nu'}_\cnu .\nonumber
\end{align}
For a given step size $\mathbb{R} \ni\tau > 0$, the steepest descent directions \eqref{eq:grad_gi}--\eqref{eq:grad_nu} can be used either to update an unknown parameter $x$ through
\begin{gather}
x_{k+1} = x_k + \tau s_k,
\label{eq:simple_update}
\end{gather}
with $s_k = -\widehat{D}_{x}\mathscr{J}$, or to reconstruct an approximation $\widetilde{H}$ of the inverse Hessian matrix, in the context of a quasi-Newton method, and thereby to compute $s_k = -\widetilde{H}\widehat{D}_{x}\mathscr{J}$. We adopt the latter approach, which is discussed in section \ref{sec:quasi_newton}.

\subsection{Geometric flow}
\label{sec:geom_flow}

To deform the boundary $\partial\Omega$ using the simple update formula \eqref{eq:simple_update} we need a parametric surface representation. Here we choose to implicitly represent $\partial\Omega$ using signed distance functions $\sdist$. The object $\Omega$ and its boundary $\partial\Omega$ are then identified with a particular function $\sdist$ so that
\begin{gather*}
\Omega =\big\{x \in \Omega:\ \sdist(x) < 0 \big\} \quad,\quad \partial\Omega = \big\{x \in \Omega:\ \sdist(x) = 0 \big\}.
\label{eq:sdf_of_domain}
\end{gather*}

\subsubsection{Implicit representation of $\Omega$ using signed distance functions}
\begin{figure}
\begin{subfigure}{.33\textwidth}
  \includegraphics[height=.95\linewidth]{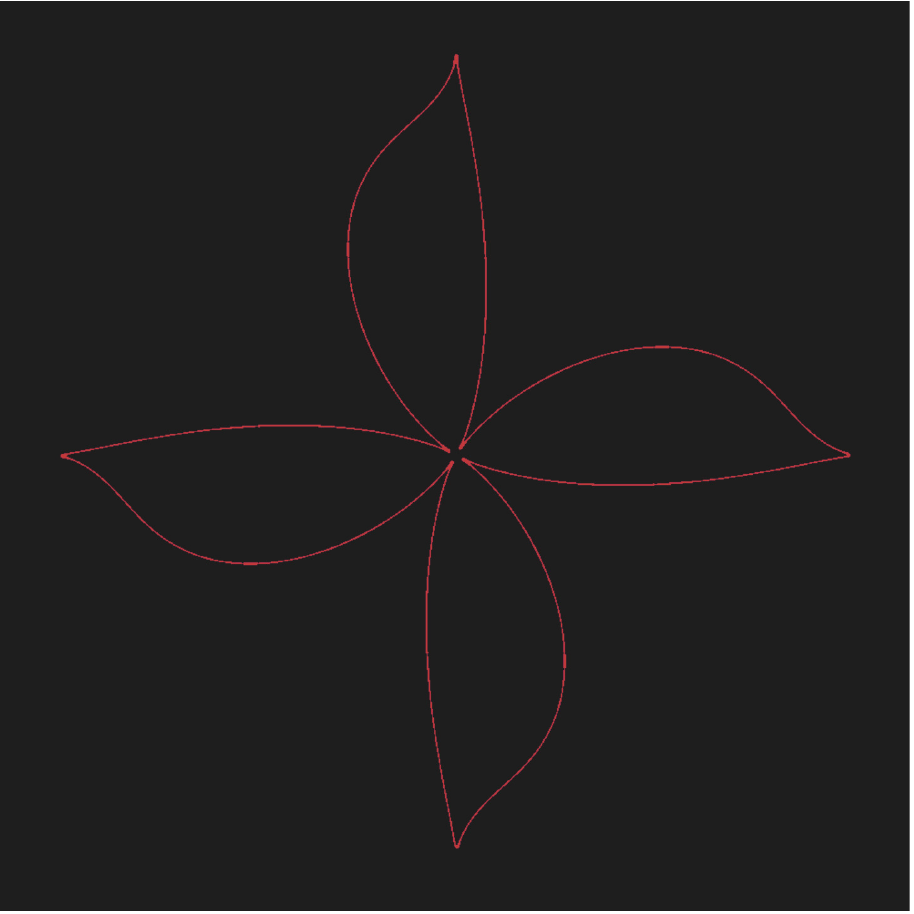}
  \caption{Shape $\partial\Omega$}
  \label{fig:geom_flow_geom}
\end{subfigure}%
\hfill
\begin{subfigure}{.33\textwidth}
  \includegraphics[height=.95\linewidth]{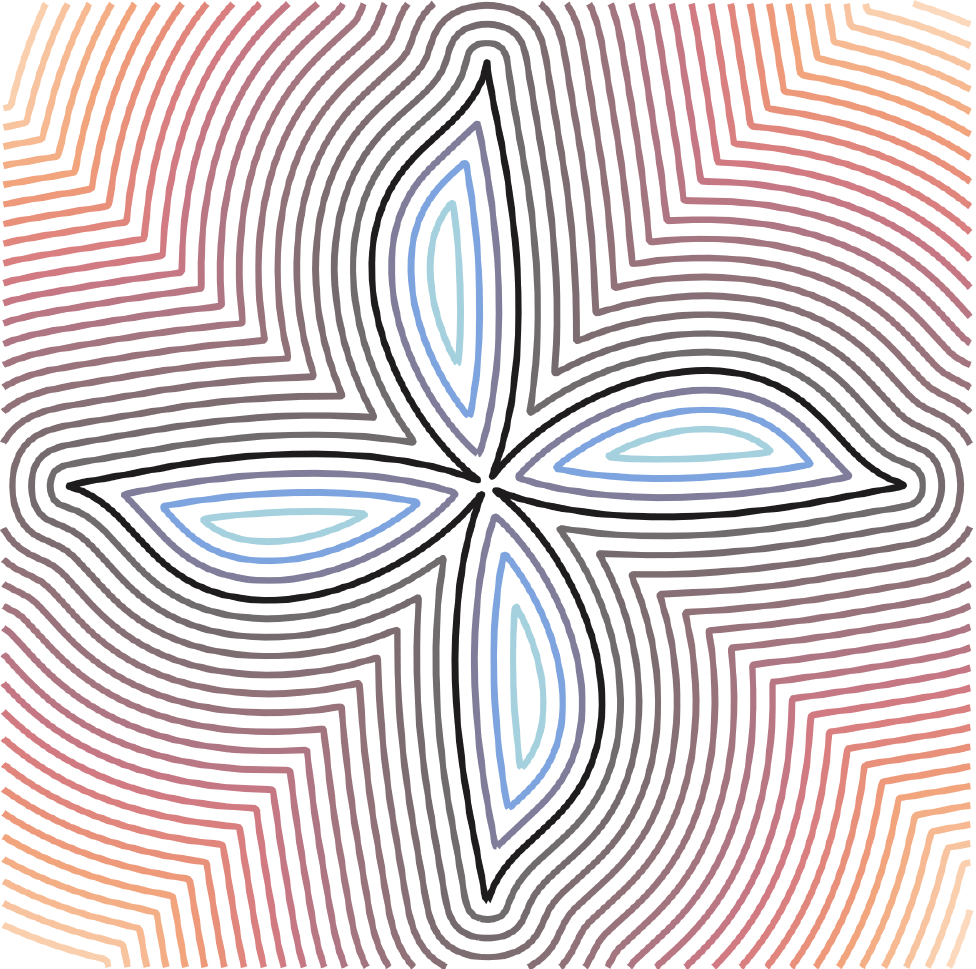}
  \caption{Level-sets of $\sdist$}
  \label{fig:geom_flow_sdist}
\end{subfigure}%
\hfill
\begin{subfigure}{.33\textwidth}
  \includegraphics[height=0.95\linewidth]{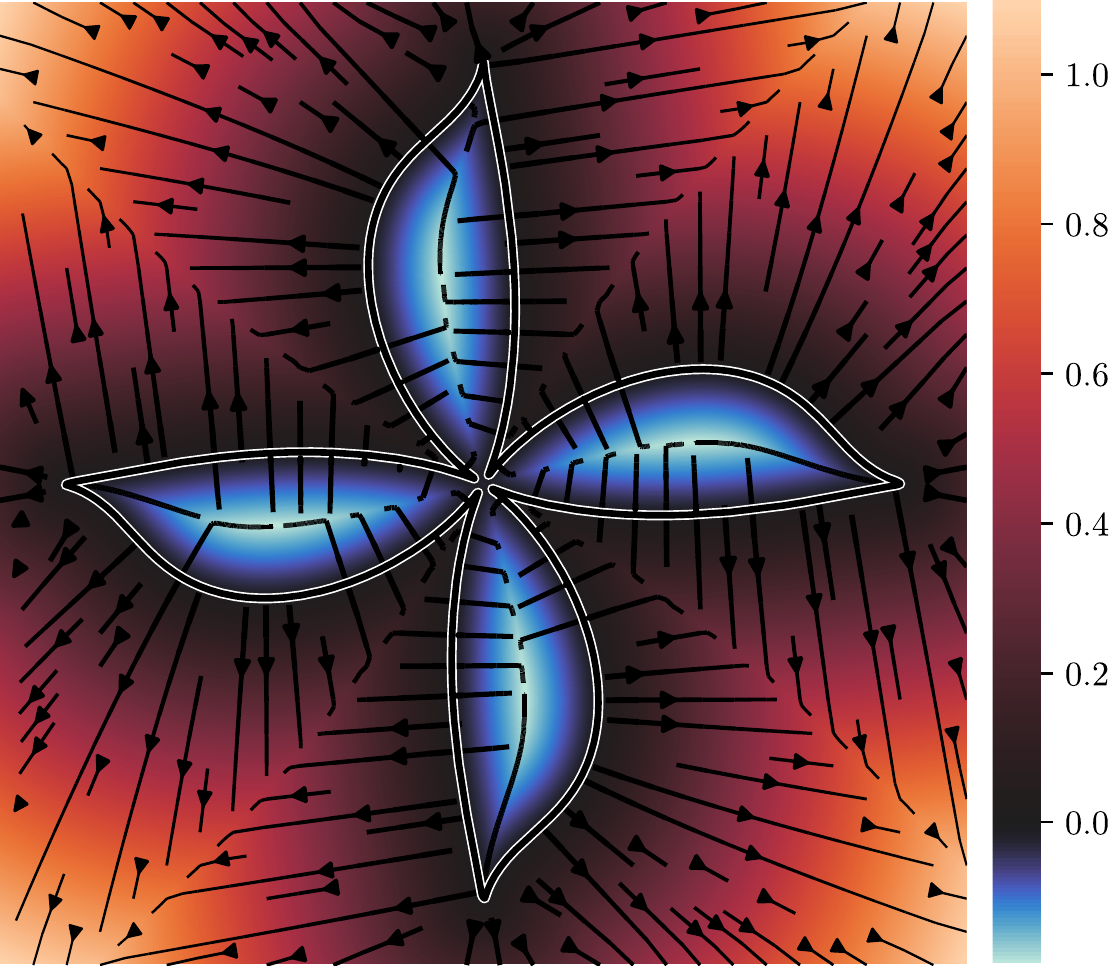}
  \caption{Magnitude of $\sdist$ and $\nuext$}
  \label{fig:geom_flow_nuext}
\end{subfigure}%
\hfill
\begin{subfigure}{.33\textwidth}
  \includegraphics[height=.95\linewidth]{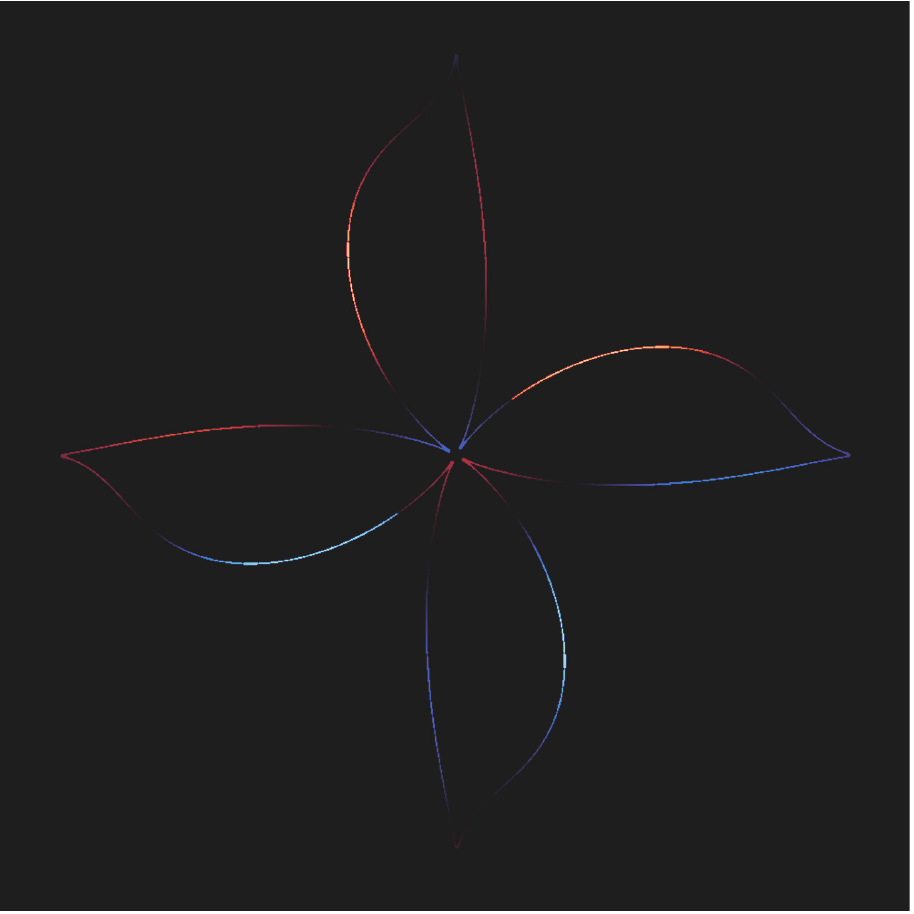}
  \caption{Shape gradient $\zeta$ on $\partial\Omega$}
  \label{fig:geom_flow_zeta}
\end{subfigure}%
\hfill
\begin{subfigure}{.33\textwidth}
  \includegraphics[height=.95\linewidth]{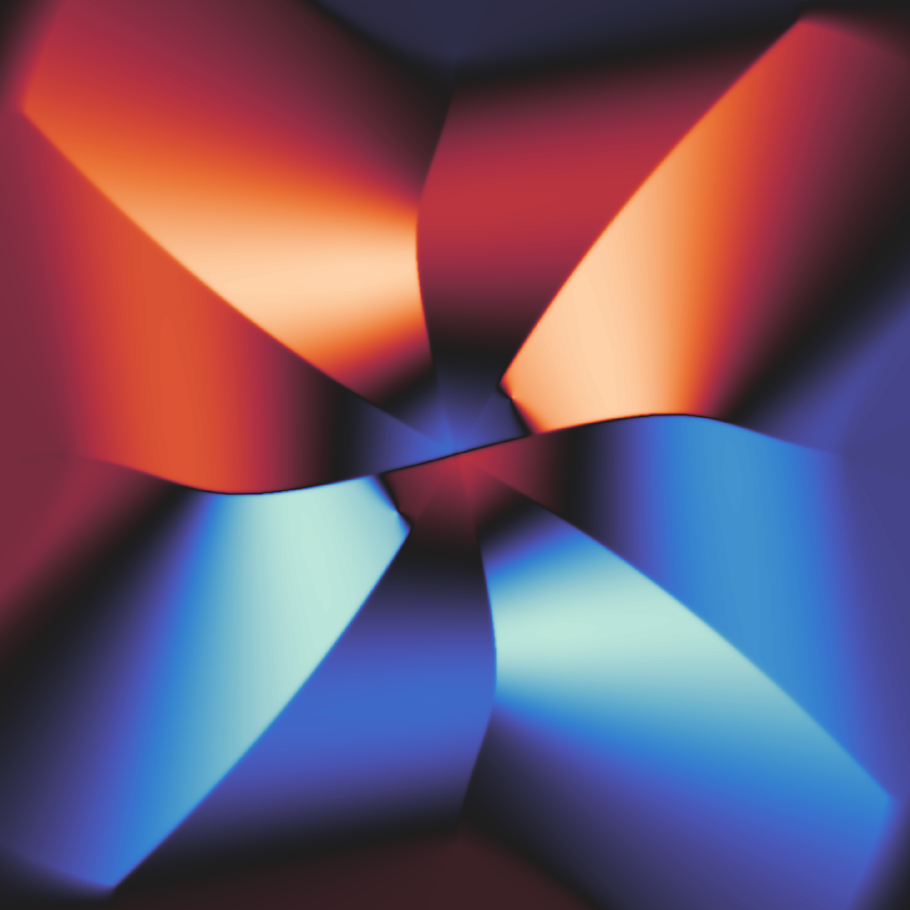}
  \caption{$\zetaext$ in $I$ ($\Rey_\zeta=1$)}
  \label{fig:geom_flow_zetaext_higheps}
\end{subfigure}%
\hfill
\begin{subfigure}{.33\textwidth}
  \includegraphics[height=0.95\linewidth]{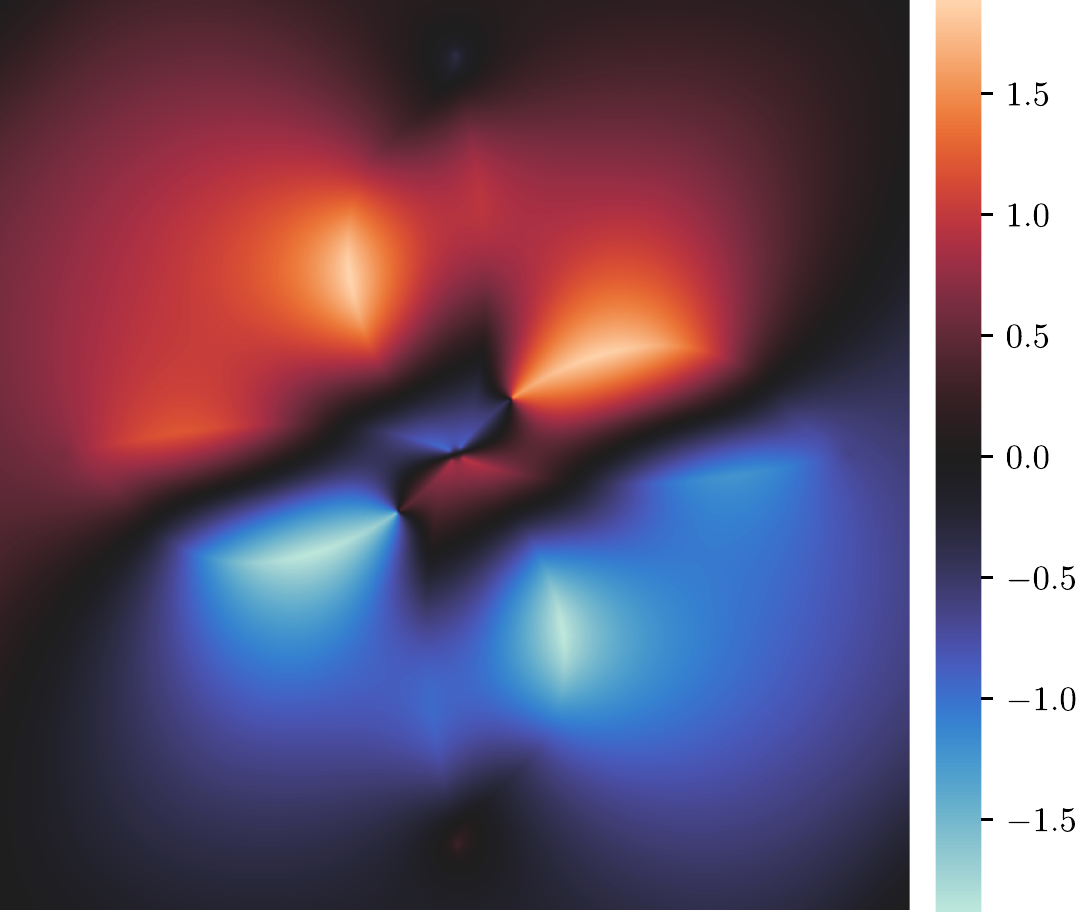}
  \caption{$\zetaext$ in $I$ ($\Rey_\zeta=0.01$)}
  \label{fig:geom_flow_zetaext_loweps}
\end{subfigure}%
\label{fig:geom_flow}
\caption{The geometric flow of $\partial\Omega$ (figure \ref{fig:geom_flow_geom}) relies on the computation of its signed distance field $\sdist$ (figure \ref{fig:geom_flow_sdist}) and its normal vector extension $\nuext$ (figure \ref{fig:geom_flow_nuext}). The shape gradient $\zeta$ (figure \ref{fig:geom_flow_zeta}), which is initially defined on $\partial\Omega$, is extended to the whole image $I$ ($\zetaext$ in figures \ref{fig:geom_flow_zetaext_higheps}, \ref{fig:geom_flow_zetaext_loweps}). Shape regularization is achieved by increasing the diffusion coefficient $\eps_\zeta$ in order to mitigate small scale perturbations when assimilating noisy velocity fields $\bm{u}^\star$. Figure \ref{fig:geom_flow_zetaext_loweps} shows results at a lower value of $\Rey_\zeta$ than figure \ref{fig:geom_flow_zetaext_higheps}.}
\end{figure}

A signed distance function $\sdist$ for $\Omega$ can be obtained by solving the Eikonal equation
\begin{gather}
\abs{\nabla \sdist(x)} = 1 \quad \text{subject to} \quad \sdist\big\vert_{\partial\Omega} = 0 \quad,\quad x \in I.
\label{eq:eikonal}
\end{gather} 
One way to solve this problem is with level-set methods \citep{Osher1988,Sethian1996,Burger2001,Burger2003,Burger2005,Yu2019}. There is, however, a different approach, which relies on the heat equation \citep{Varadhan1967_analytic,Varadhan1967_stochastic,Crane2017}. The main result that we draw from \cite{Varadhan1967_analytic}, in order to justify the use of the heat equation for the approximation of $\sdist$, states that
\begin{gather}
d(x,\partial\Omega) = \lim_{\tau_1 \to 0} \big(-\frac{\sqrt{\tau_1}}{2}\log u(x,\tau_1) \big) \quad,\quad x \in I\quad,
\end{gather}
where $d(x,\partial\Omega)$ is the Euclidean distance between any point $x \in I$ and $\partial\Omega$, and $u$ is the solution of heat propagation away from $\partial\Omega$ 
\begin{gather}
\left\{\begin{alignedat}{2}
\big(I-\tau_1\rmDelta\big) u  &= 0 \quad &&\textrm{in}\ I \\ 
u &= 1 \quad &&\textrm{on}\ \partial \Omega
\end{alignedat}\right.\quad.
\label{eq:varadhans_3rd_thm}
\end{gather}
\cite{Crane2017} used the above result to implement a smoothed distance function computation method which they called the `\textit{heat method}'. Here, we slightly adapt this method to compute  \textit{signed} distance functions $\sdist$ in truncated domains (figure \ref{fig:geom_flow_sdist}). To compute $\sdist$ we therefore solve \eqref{eq:varadhans_3rd_thm} for $\tau_1 \ll 1$, and then obtain $\sdist$ by solving
\begin{gather}
\left\{\begin{alignedat}{2}
\nabla\bm{\cdot}\nabla \sdist  &= \nabla \bm{\cdot} X \quad &&\textrm{in}\ I \\ 
\partial_{\bm{\nu}} \sdist &= X \bm{\cdot} \bm{\nu} \quad &&\textrm{on}\ \partial I\\
 \sdist &= 0 \quad &&\textrm{on}\ \partial \Omega
\end{alignedat}\right.\quad, \quad X = -\text{sgn}(\psi) \frac{\nabla u}{\abs{\nabla u}}\quad,
\label{eq:heat_method}
\end{gather}
with $X$ being the normalized heat flux and $\psi$ being a signed function such that $\psi(x)$ is negative for points $x$ in $\Omega$ and positive for points $x$ outside ${\Omega}$. This intermediate step (the solution of two Poisson problems \eqref{eq:varadhans_3rd_thm}-\eqref{eq:heat_method} instead of one) is taken to ensure that $\abs{\nabla\sdist} = 1$. 

\subsubsection{Propagating the boundary of $\Omega$}
\label{sec:prop_boundary_geomflow}
To deform the boundary $\partial\Omega$ we transport $\sdist$ under the speed field ${\spd \equiv \zeta\bm{\nu}}$. The convection-diffusion problem for $\sdist(x,t)$ reads
\begin{gather}
\left\{\begin{alignedat}{2}
\partial_t\sdist + \spdext\bm{\cdot}\nabla\sdist -\eps_\sdist\rmDelta\sdist  &= 0 \quad &&\textrm{in}\ I\times (0,\tau] \\
\sdist          &= {\sdist}_0 \quad &&\textrm{in}\ I\times \{t=0\}
\end{alignedat}\right.
\quad,\quad \epsilon_\sdist = \frac{\abs{\spd}_\infty \iota}{\Rey_\sdist} \quad,
\label{eq:adv_diff_sdist}
\end{gather}
where ${\sdist}_0$ denotes the signed distance function of the current domain $\Omega$, $\eps_\sdist$ is the diffusion coefficient, $\iota$ is a length scale, $\Rey_\sdist$ is a Reynolds number, and ${\spdext:I \to \R\times\R}$ is an extension of $\spd: \partial\Omega \to \R\times\R$. If we solve \eqref{eq:adv_diff_sdist} for $\sdist(x,\tau)$ we obtain the implicit representation of the perturbed domain $\Omega_\tau$, at time $t=\tau$ (the step size), but to do so we first need to extend $\spd$ to the whole space of the image $I$.

To extend $\spd$ to $I$ we extend the normal vector $\bm{\nu}$ and the scalar function $\zeta$, which are both initially defined on $\partial\Omega$. The normal vector extension (figure \ref{fig:geom_flow_nuext}) is easily obtained by
\begin{gather}
\nuext(x) = \frac{\nabla \sdist}{\abs{\nabla\sdist}} = \nabla \sdist \quad ,\quad x \in I\quad,
\end{gather}
since $\abs{\nabla\sdist} = 1$, and an outward-facing extension is given by 
\begin{gather}
\nuext_o = \text{sgn}(\sdist)~\nuext.
\label{eq:nuext_formula}
\end{gather}
We then use the extended normal vector $\nuext_o$ to extend $\zeta \in L^2(\partial\Omega)$ to $\zetaext \in L^2(I)$, using the convection-diffusion problem
\begin{gather}
\left\{\begin{alignedat}{2}
\partial_t \zetaext + {\nuext_o}\bm{\cdot}\nabla\zetaext - \epsilon_\zeta \rmDelta \zetaext &= 0 \quad &&\textrm{in}\quad I\times (0,\tau_\zeta] \\ 
\zetaext &= \zeta \quad &&\textrm{on}\quad \partial \Omega\times (0,\tau_\zeta] \\
\zetaext &\equiv 0 \quad &&\textrm{in}\quad I\times\{t=0\}
\end{alignedat}\right.
\label{eq:zeta_ext}
\quad,\quad \epsilon_\zeta = \frac{\abs{\nuext_o}_\infty \iota}{\Rey_\zeta}\quad.
\end{gather}
In other words, we convect $\zeta$ along the predefined $\nuext_o$-streamlines and add isotropic diffusion for regularization (figures \ref{fig:geom_flow_zetaext_higheps}, \ref{fig:geom_flow_zetaext_loweps}). The choice of $\epsilon_\sdist$ in \eqref{eq:adv_diff_sdist} and $\epsilon_\zeta$ in \eqref{eq:zeta_ext} has been made in order for the shape regularization to depend only on the length scale $\iota$ and the Reynolds numbers $\Rey_\sdist,\Rey_\zeta$. More precisely, the shape regularization depends only on $\Rey_\sdist$ and $\Rey_\zeta$ because we fix the length scale $\iota$ to equal the smallest possible length scale of the modelled flow, which is the numerical grid spacing $h$ for a uniform cartesian grid. For illustration, if we consider $\zeta$ to be the concentration of a dye on $\partial\Omega$ (figure \ref{fig:geom_flow_zeta}), using a simplified scaling argument similar to the growth of a boundary layer on a flat plate, we observe that the diffusing dye at distance $d$ from $\partial\Omega$ will extend over a width $\delta$ such that
\begin{gather}
\delta \sim \sqrt{\frac{\epsilon_\zeta d}{\abs{\nuext_o}_\infty}}=\sqrt{\frac{\iota d}{\Rey_\zeta}}\quad,\quad \text{or}\quad \frac{\delta}{\iota} \sim \sqrt{\frac{\alpha}{\Rey_\zeta}}, \quad \text{when}\quad d = \alpha \iota\quad.
\end{gather}
The above scaling approximation describes the dissipation rate of small-scale features such as roughness away from $\partial\Omega$. This is therefore how $\Rey_\sdist$ and $\Rey_\zeta$ control the regularity of the boundary $\partial\Omega_\tau$ at time $t = \tau$, which is given by \eqref{eq:adv_diff_sdist}. We take $\tau_\zeta$ to be large enough to find a steady-state for \eqref{eq:zeta_ext}. We recast the linear initial value problems \eqref{eq:adv_diff_sdist} and \eqref{eq:zeta_ext} into their corresponding boundary value problems using backward-Euler temporal discretization because the time dependent solution does not interest us here.

The extended shape gradient \eqref{eq:grad_shape}, after taking into account the regularizing term for $\sdist$, is therefore given by
\begin{align}
\Binner{D_\zetaext\mathscr{J},~\zetaext'}_I &= \Binner{\zetaext + \mathcal{C}^{-1}_\sdist\big(\mean{\phi}_\pm-\sdist \big),~\zetaext'}_I \label{eq:grad_ext_shape}\\
&= \Binner{\mathcal{C}_\sdist\zetaext + \mean{\phi}_\pm-\sdist,~\zetaext'}_{\mathcal{C}_\sdist} = \Binner{\widehat{D}_\zetaext\mathscr{J},~\zetaext'}_{\mathcal{C}_\sdist}\nonumber\quad, 
\end{align}
where $\zetaext$ is the extension of the shape gradient $\zeta(x) = \partial_{\bm{\nu}}\bm{u}\bm{\cdot}\big(-\nu\partial_{\bm{\nu}}\bm{v}+q\bm{\nu}\big)$, for $x$ on $\Gamma$.

\subsection{Segregated approach for the Euler--Lagrange system}
\label{sec:quasi_newton}
The inverse Navier--Stokes problem for the reconstruction and the segmentation of noisy velocity images $\bm{u}^\star$ can be written as the saddle point problem \citep{Benzi2005}
\begin{equation}
\text{find} \quad \bm{u}^\circ \equiv \mathrm{arg}\ \underset{\Omega,\bm{x}}{\mathrm{min}}\ \underset{\bm{v},q}{\mathrm{max}}~\mathscr{J}(\Omega)(\bm{u},p,\bm{v},q;\bm{x}),
\label{eq:inv_prob_aug}
\end{equation}
where $\mathscr{J}$ is given by \eqref{eq:aug_rec_func}. The above optimization problem leads to an Euler--Lagrange system whose optimality conditions were formulated in section \ref{sec:eul_lag_system}. We briefly describe our segregated approach to solve this Euler--Lagrange system in algorithm \ref{algo:reconstruction}.

\begin{algorithm}[h]
\textbf{Input:} $\bm{u}^\star$, initial guesses for the unknowns $(\Omega_0, x_0)$, regularization parameters.\\\
\Begin{
$k \leftarrow 0$\\
$\makebox[0pt][l]{$(\sdist)_k$}\phantom{banana} \leftarrow$ {signed distance field} (eq. \eqref{eq:varadhans_3rd_thm}-\eqref{eq:heat_method})\\
$\makebox[0pt][l]{$(\bm{u},p)_k$}\phantom{banana} \leftarrow$ {Navier--Stokes problem for $(\sdist, \bm{x})_k$} (eq. \eqref{eq:navierstokes_bvp})\\
\While{\texttt{convergence criterion is not met}}{
$\makebox[0pt][l]{($\bm{v},q$)}\phantom{avocadoss} \leftarrow$ {adjoint Navier--Stokes problem with $\bm{u}_k$} (eq. \eqref{eq:navier-stokes_adjoint_problem})\\
$\makebox[0pt][l]{$\widehat{D}_{(\cdot)}\mathscr{J}$}\phantom{avocadoss} \leftarrow $ steepest ascent directions (eq. \eqref{eq:grad_gi}--\eqref{eq:grad_nu} and \eqref{eq:grad_ext_shape})\\
$\makebox[0pt][l]{$\bm{s},\tau$}\phantom{avocadoss} \leftarrow$ {search directions and step-size} (eq. \eqref{eq:search_directions})\\
$\makebox[0pt][l]{$(\sdist,\bm{x})_{k+1}$}\phantom{avocadoss} \leftarrow$ {perturb $\sdist$ (eq. \eqref{eq:adv_diff_sdist}) and model parameters $\bm{x}$} (eq. \eqref{eq:simple_update})\\
$\makebox[0pt][l]{$(\bm{u},p)_{k+1}$}\phantom{avocadoss} \leftarrow$ {linearized Navier--Stokes problem for $(\sdist,\bm{x})_{k+1}$} (eq. \eqref{eq:oseen_problem})\\
$k\leftarrow k+1$\\
}
\makebox[0pt][l]{$(\bm{u}^\circ,p^\circ)$}\phantom{avocados}$\leftarrow (\bm{u},p)_k$\\
\makebox[0pt][l]{$(\Omega^\circ,\bm{x}^\circ)$}\phantom{avocados}$\leftarrow (\sdist,\bm{x})_k$
}
\textbf{Output:} reconstruction $(\bm{u}^\circ,p^\circ)$ and inferred model parameters $(\Omega^\circ,\bm{x}^\circ)$.\\
{\color{darkgray}
\textbf{Optional output:} wall shear rate $\gamma_w^\circ$ from $\bm{u}^\circ$ and $\partial\Omega^\circ$.}
\caption{Reconstruction and segmentation of noisy flow velocity images.}
\label{algo:reconstruction}
\end{algorithm}
To precondition the steepest descent directions \eqref{eq:grad_gi}--\eqref{eq:grad_nu} and \eqref{eq:grad_ext_shape}, we reconstruct the approximated inverse Hessian $\widetilde{H}$ of each unknown using the BFGS quasi-Newton method \citep{Fletcher2000} with damping \citep{Nocedal2006}. Due to the large scale of the problem, it is only possible to work with the matrix-vector product representation of $\widetilde{H}$. Consequently, the search directions are given by
\begin{gather}
\bm{s} = -
\begin{pmatrix}
\widetilde{H}_\zetaext     & \cdot & \cdot & \cdot \\
\cdot &     \widetilde{H}_{\bm{g}_i} & \cdot & \cdot \\
\cdot & \cdot & \widetilde{H}_{\bm{g}_o}     & \cdot \\
\cdot & \cdot & \cdot & \widetilde{H}_\nu
\end{pmatrix}
\begin{pmatrix}
\widehat{D}_\zetaext\mathscr{J}\\
\widehat{D}_{\bm{g}_i}\mathscr{J}\\
\widehat{D}_{\bm{g}_o}\mathscr{J}\\
\widehat{D}_\nu\mathscr{J}
\end{pmatrix}\quad,
\label{eq:search_directions}
\end{gather}
and the unknown variables $\bm{x}$ are updated according to \eqref{eq:simple_update}. The signed distance function $\sdist$ is perturbed according to \eqref{eq:adv_diff_sdist}, with $\spdext \equiv -\big(\widetilde{H}_\zetaext\widehat{D}_\zetaext\mathscr{J}\big)\nuext$. 
We start every line search with a global step size $\tau = 1$, and halve the step size until ${\mathscr{J}((\sdist,\bm{x})_{k+1}) < \mathscr{J}((\sdist,\bm{x})_{k})}$. To update the flowfield $\bm{u}_k$ to $\bm{u}_{k+1}$ we solve the Oseen problem for the updated parameters $(\sdist,\bm{x})_{k+1}$ 
\begin{gather}
 \bm{u}_k\bm{\cdot}\nabla\bm{u}_{k+1}-\nu{\rmDelta} \bm{u}_{k+1}  + \nabla p_{k+1} = \bm{0},\quad \nabla\bm{\cdot}\bm{u}_{k+1} = 0\quad,
 \label{eq:oseen_problem}
\end{gather}
with the boundary conditions given by \eqref{eq:navierstokes_bvp}. Algorithm \ref{algo:reconstruction} terminates if either the covariance-weighted norm for the perturbations of the model parameters is below the user-specified tolerance, or the line search fails to reduce $\mathscr{J}$.

\subsection{Uncertainty estimation}
\label{sec:uncertainty}
We now briefly describe how the reconstructed inverse Hessian $\widetilde{H}$ can provide estimates for the uncertainties of the model parameters. To simplify the description, let $x$ denote an unknown parameter distributed according to $\mathcal{N}(x_k,\mathcal{C}_x)$. The linear approximation to the data $\bm{u}^\star$ is given by
\begin{gather}
\bm{u}^\star = \mathcal{Z}x + \varepsilon\quad,\quad \varepsilon \sim \mathcal{N}(0,\cu)\quad,
\end{gather}
where $\bm{u} = \mathcal{Z}x$, where $\mathcal{Z}$ is the operator that encodes the linearized Navier--Stokes problem around the solution $\bm{u}_k$. To solve \eqref{eq:inv_prob_aug}, we update $x$ as
\begin{gather}
x_{k+1} = x_k + \mathcal{C}\mathcal{Z}^\dagger\invcu\big(\bm{u}^\star-\mathcal{Z}x_k\big)\quad,\quad \text{with}\quad \mathcal{C} = \big(\mathcal{Z}^\dagger\invcu\mathcal{Z}+\mathcal{C}^{-1}_x\big)^{-1}\quad,
\end{gather}
where $\mathcal{Z}^\dagger$ is the operator that encodes the adjoint Navier--Stokes problem, and $\mathcal{C}$ is the \textit{posterior} covariance operator. It can be shown that \citep[Chapter~6.22.8]{Tarantola2005}
\begin{gather}
\mathcal{C} = \big(\mathcal{Z}^\dagger\invcu\mathcal{Z}+\mathcal{C}^{-1}_x\big)^{-1} = \big(\mathcal{C}_x\mathcal{Z}^\dagger\invcu\mathcal{Z}+\mathrm{I}\big)^{-1}\mathcal{C}_x \simeq \widetilde{H}_x\mathcal{C}_x\quad,
\end{gather}
where $\widetilde{H}_x$ is the reconstructed inverse Hessian for $x$. Note that $\widetilde{H}$ by itself approximates $\big(\mathcal{C}_x\mathcal{Z}^\dagger\invcu\mathcal{Z}+\mathrm{I}\big)^{-1}$, and not $\mathcal{C}$, because we use the steepest ascent directions $\widehat{D}_{(\cdot)}\mathscr{J}$ (prior-preconditioned gradients), instead of the gradients ${D}_{(\cdot)}\mathscr{J}$, in the BFGS formula. Therefore, if $\widetilde{\mathcal{C}}\equiv\widetilde{H}_x\mathcal{C}_x$ is the approximated covariance matrix, then samples $x^s_{k+1}$ from the posterior distribution can be drawn using the Karhunen--Lo\`eve expansion
\begin{gather}
x^s_{k+1} = x_k + \sum_k \eta_k~\sqrt{\lambda_k}\varphi_k\quad, \quad \text{with}\quad \eta_k \sim \mathcal{N}(0,1)\quad,
\end{gather}
where $(\lambda,\varphi)_k$ is the eigenvalue/eigenvector pair of $\widetilde{\mathcal{C}}$. The variance of $x_{k+1}$ can then be directly computed from the samples.

\subsection{Numerics}
\label{sec:numerics}
To solve the above boundary value problems numerically, we use an immersed boundary finite element method. In particular, we implement the fictitious domain cut-cell finite element method (FEM), introduced by \cite{Burman2010,Burman2012} for the Poisson problem, and later on extended to the Stokes and the Oseen problems \citep{Schott2014,Massing2014,Burman2015,Massing2018}. We define $\mathcal{T}_h$ to be a tessellation of $I$ produced by square cells (pixels) $K \in \mathcal{T}_h$, having sides of length $h$. We also define the set of cut-cells $\mathcal{T}_h^\triangleright$ consisting of the cells that are cut by the boundary $\partial\Omega$, and $\mathcal{T}_h^{\smallsquare}$ the set of cells that are found inside $\Omega$ and which remain intact (not cut) (see figure \ref{fig:theoretical_twin}). We assume that the boundary $\partial\Omega$ is well-resolved, i.e. $\ell_{\partial\Omega}/h \gg 1$ where $\ell_{\partial\Omega}$ is the smallest length scale of $\partial\Omega$. For the detailed assumptions on $\partial\Omega$ we cite \cite{Burman2012}. The discretized space is generated by assigning a bilinear quadrilateral finite element $\mathcal{Q}_1$ to every cell $K$. To compute the integrals we use standard Gaussian quadrature for cells $K \in \mathcal{T}_h^{\smallsquare}$, while for cut-cells $K \in \mathcal{T}_h^\triangleright$, where integration must be considered only for the intersection $K\cap\Omega$, we use the approach of \cite{Mirtich1996}, which relies on the divergence theorem and simply replaces the integral over $K \cap \Omega$ with an integral over $\partial\big(K\cap\Omega\big)$. The boundary integral on $\partial\big(K\cap\Omega\big)$ is then easily computed using one-dimensional Gaussian quadrature \citep{Massing2013}. Since we use an inf-sup unstable finite element pair ($\mathcal{Q}_1$-$\mathcal{Q}_1$) \citep{BrennerSusanneScott2008} we use a pressure-stabilizing Petrov-Galerkin formulation \citep{Tezduyar1991,Codina2002} and $\nabla$-div stabilization for preconditioning \citep{Benzi2006,Heister2013}. \revv{Typical values and formulas for numerical parameters, e.g. Nitsche's penalization $\eta$, are given by \cite{Massing2014,Massing2018}. Here, we take $\eta = \gamma\nu/h$ \citep{Massing2018}, with $\gamma = 100$.} To solve the Navier--Stokes problem we use fixed-point iteration (Oseen linearization), and at each iteration we solve the coupled system using the Schur complement; with an iterative solver (LGMRES) for the outer loops, and a direct sparse solver (UMFPACK) for the inner loops. The immersed FEM solver, and all the necessary numerical operations of algorithm \ref{algo:reconstruction}, are implemented in Python, using its standard libraries for scientific computing, namely SciPy \citep{2020SciPy-NMeth} and NumPy \citep{harris2020array}. Computationally intensive functions are accelerated using Numba \citep{lam2015numba} and CuPy \citep{cupy_learningsys2017}.

\section{Reconstruction and segmentation of flow images}
\label{sec:rec_and_segm_of_images}
In this section we reconstruct and segment noisy flow images by solving the inverse Navier--Stokes problem \eqref{eq:inv_prob_aug} using algorithm \ref{algo:reconstruction}. We then use the reconstructed velocity field to estimate the wall shear rate on the reconstructed boundary. First, we apply this to three test cases with known solutions by generating synthetic 2D Navier--Stokes data. Next, we perform a magnetic resonance velocimetry experiment in order to acquire images of a 3D axisymmetric Navier--Stokes flow, and apply algorithm \ref{algo:reconstruction} to these images.

We define the signal-to-noise ratio (SNR) of the $u^\star_x$ image as
\begin{gather}
\text{SNR}_x = \frac{\mu_x}{\sigma_{u_x}} \quad,\quad \mu_x \equiv \frac{1}{\abs{\Omega^\bullet}}\int_{\Omega^\bullet} \abs{u_x^\bullet}\quad,
\label{eq:snr_def}
\end{gather}
where $\sigma_{u_x}$ is the standard deviation, $\Omega^\bullet$ is the ground truth domain, $\abs{\Omega^\bullet}$ is the volume of this domain, and $\abs{u_x^\bullet}$ is the magnitude of the ground truth $x$-velocity component in $\Omega^\bullet$. We also define the componentwise averaged, noise relative reconstruction error $\mathscr{E}^\bullet_x$, and the total relative reconstruction error $\mathcal{E}^\bullet$ by
\begin{gather}
\mathscr{E}^\bullet_x \equiv \log \bigg( \frac{1}{\abs{\Omega}}\int_\Omega \frac{\abs{u^\bullet_x-\mathcal{S}u^\circ_x}}{\sigma_{u_x}}\bigg)\quad\text{and} \quad \mathcal{E}^\bullet \equiv  \frac{\norm{\bm{u}^\bullet-\mathcal{S}\bm{u}^\circ}_{L^1(I)}}{\norm{\bm{u}^\bullet}_{L^1(I)}}\quad.
\end{gather}
respectively. Similar measures also apply for the $u^\star_y$ image.

We define the volumetric flow rate $Q$, the cross-section area at the inlet $A$, and the diameter at the inlet $D$. The Reynolds number is based on the reference velocity ${U\equiv Q/A}$, and the reference length $D$.

\subsection{Synthetic data for 2D flow in a converging channel}
\label{sec:converging_channel}

We start by testing algorithm \ref{algo:reconstruction} on a flow through a symmetric converging channel having a taper ratio of $0.67$. To generate synthetic 2D Navier--Stokes data we solve the Navier--Stokes problem \eqref{eq:navierstokes_bvp} for a parabolic inlet velocity profile ($\bm{g}_i$), zero-pseudotraction boundary conditions at the outlet ($\bm{g}_o \equiv \bm{0}$), and $\Rey \simeq 534$, in order to obtain the ground truth velocity $\bm{u}^\bullet$. We then generate the synthetic data $\bm{u}^\star$ by corrupting the components of $\bm{u}^\bullet$ with white Gaussian noise such that $\text{SNR}_x=\text{SNR}_y=3$. For this test case, we are only trying to infer $\Omega$ and $\bm{g}_i$. Note that, in our method, the initial guess $x_0$ of an unknown $x$ equals the mean of its prior distribution $\bar{x}$, i.e. $x_0\equiv\bar{x}$. We start the algorithm using bad initial guesses (high uncertainty in priors) for both the unknown parameters (see table \ref{tab:input_params_1}). The initial guess for $\Omega$, labelled $\Omega_0$, is a rectangular domain with height equal to $0.7D$, centered in the image domain. For ${\bm{g}_i}_0$ we take a parabolic velocity profile with a peak velocity of approximately $2U$ that fits the inlet of $\Omega_0$. For comparison, $\bm{g}_i^\bullet$ has a peak velocity of $1.5U$, while it is also defined on a different domain, namely $\Omega^\bullet$. 

\begin{table}
  \begin{center}
\def~{\hphantom{0}}
  \begin{tabular}{llcccccc}
        & & image dimension   &   model dimension & \multicolumn{2}{c}{$\sigma_{u_x}/U$} & \multicolumn{2}{c}{$\sigma_{u_y}/U$}\\[3pt]
       converging channel  & (2D)  & $192^2$ & $200^2$  & \multicolumn{2}{c}{$3.97\times10^{-1}$} & \multicolumn{2}{c}{$8.97\times10^{-3}$}\\

        &   &  &   &  &  & &\\
         \multicolumn{2}{c}{\textit{Regularization}} & $\sigma_\sdist/D$ & $\sigma_{\bm{g}_i}/U$ & $\sigma_\nu/UD$ & $\Rey_\sdist$ & $\Rey_\zeta$ & $\ell/h$\\[3pt]
         converging channel  & (2D) & $1.0$ & $2.0$ & $\cdot$ & 0.025 & 0.025 & 3\\
  \end{tabular}
  \caption{Input parameters for the inverse 2D Navier--Stokes problem.}
  \label{tab:input_params_1}
  \end{center}
\end{table}
\begin{figure}
\begin{subfigure}{.32\textwidth}
  \includegraphics[height=0.68\linewidth]{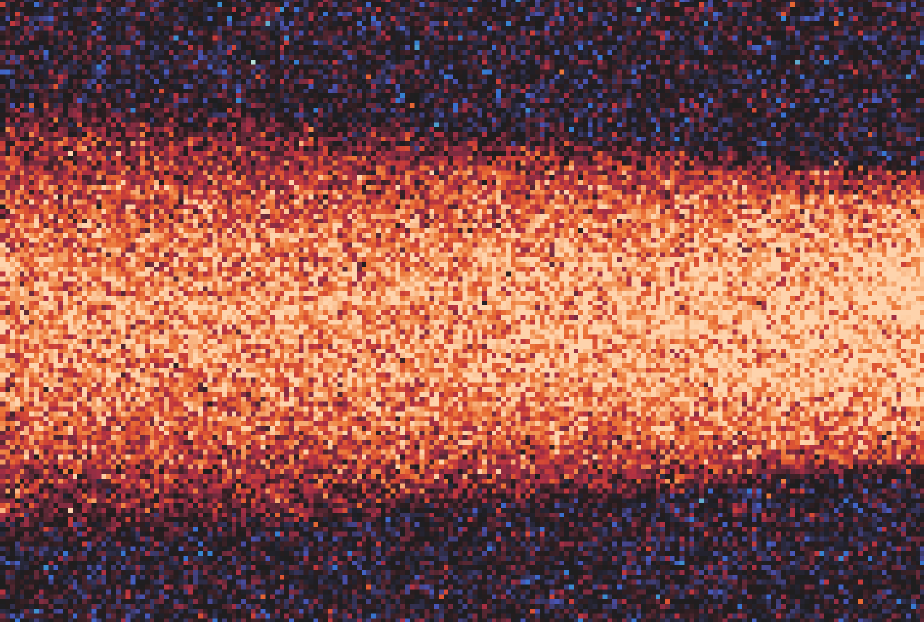}
  \caption{Synthetic image $u^\star_x$}
  \label{fig:tap_ux_sig}
\end{subfigure}%
\hfill
\begin{subfigure}{.32\textwidth}
  \includegraphics[height=0.68\linewidth]{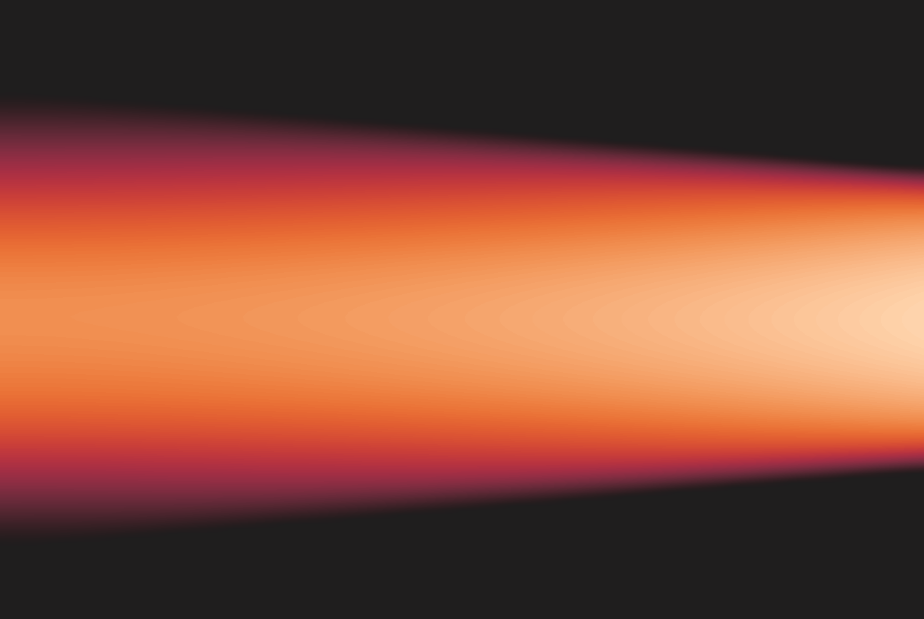}
  \caption{Our reconstruction $u_x^\circ$}
  \label{fig:tap_ux_rec}
\end{subfigure}%
\hfill
\begin{subfigure}{.32\textwidth}
  \includegraphics[height=0.68\linewidth]{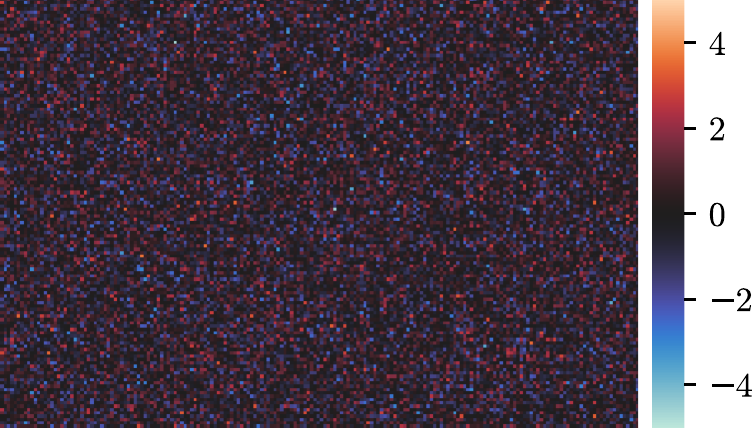}
  \caption{Discrepancy $\sigma_{u_x}^{-1}\big(u^\star_x - \mathcal{S}u_x^\circ\big)$}
  \label{fig:tap_ux_rec_err}
\end{subfigure}
\vfill
\begin{subfigure}{.32\textwidth}
  \includegraphics[height=0.68\linewidth]{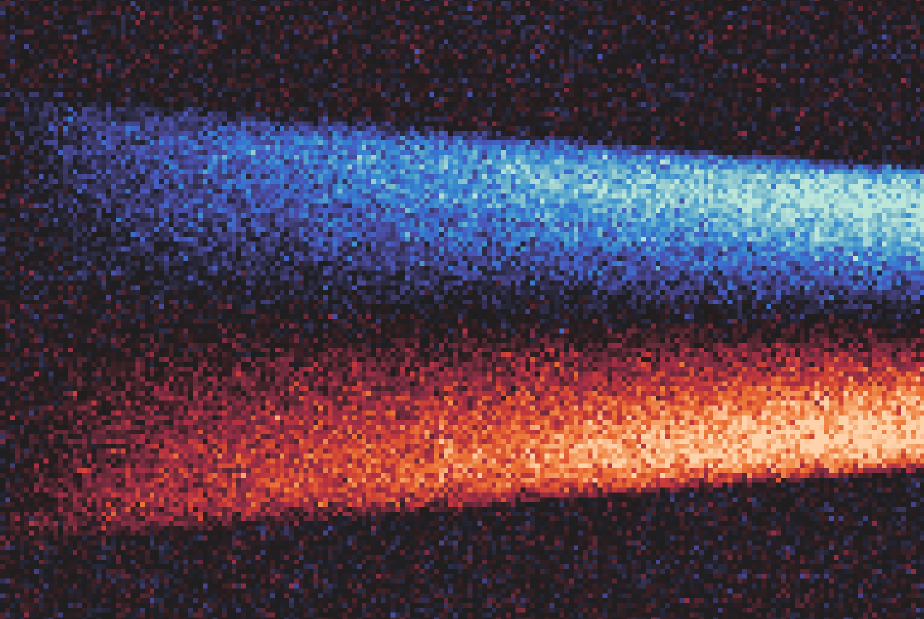}
  \caption{Synthetic image $u^\star_y$}
  \label{fig:tap_uy_sig}
\end{subfigure}%
\hfill
\begin{subfigure}{.32\textwidth}
  \includegraphics[height=0.68\linewidth]{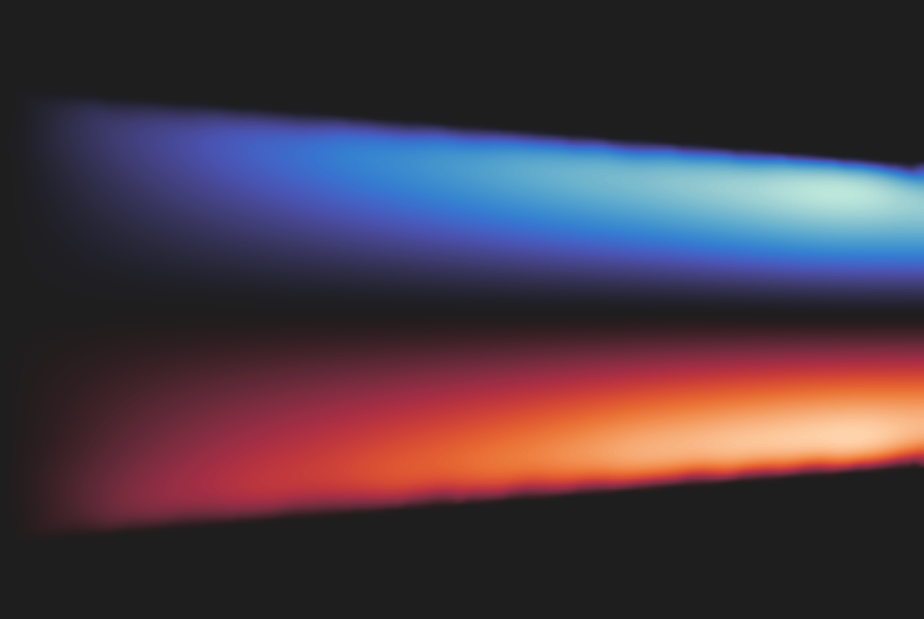}
  \caption{Our reconstruction $u_y^\circ$}
  \label{fig:tap_uy_rec}
\end{subfigure}%
\hfill
\begin{subfigure}{.32\textwidth}
  \includegraphics[height=0.68\linewidth]{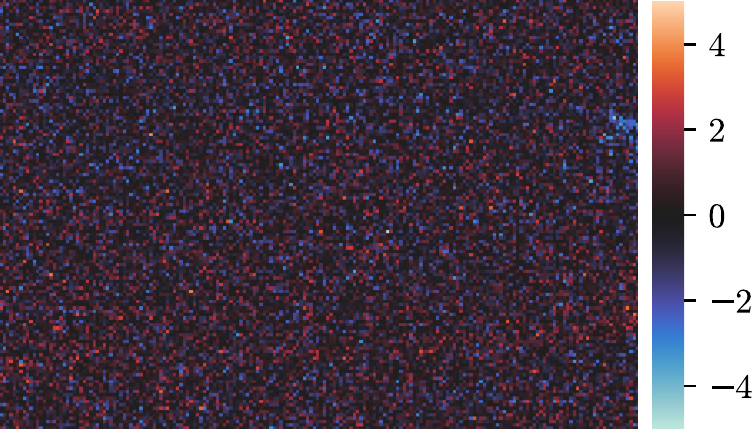}
  \caption{Discrepancy $\sigma_{u_y}^{-1}\big(u^\star_y - \mathcal{S}u_y^\circ\big)$}
  \label{fig:tap_uy_rec_err}
\end{subfigure}%
\caption{Reconstruction (algorithm \ref{algo:reconstruction}) of synthetic noisy velocity images depicting the flow (from left to right) in a converging channel. Figures \ref{fig:tap_ux_sig}-\ref{fig:tap_ux_rec} and \ref{fig:tap_uy_sig}-\ref{fig:tap_uy_rec} show the horizontal, $u_x$, and vertical, $u_y$, velocities and share the same colormap (colorbar not shown). Figures \ref{fig:tap_ux_rec_err} and \ref{fig:tap_uy_rec_err} show the discrepancy between the noisy velocity images and the reconstruction (colorbars apply only to figures \ref{fig:tap_ux_rec_err} and \ref{fig:tap_uy_rec_err}).}
\label{fig:tap_rec_results1}
\end{figure}

\begin{figure}
\begin{subfigure}{0.55\textwidth}
\centering
  \includegraphics[height=0.525\linewidth]{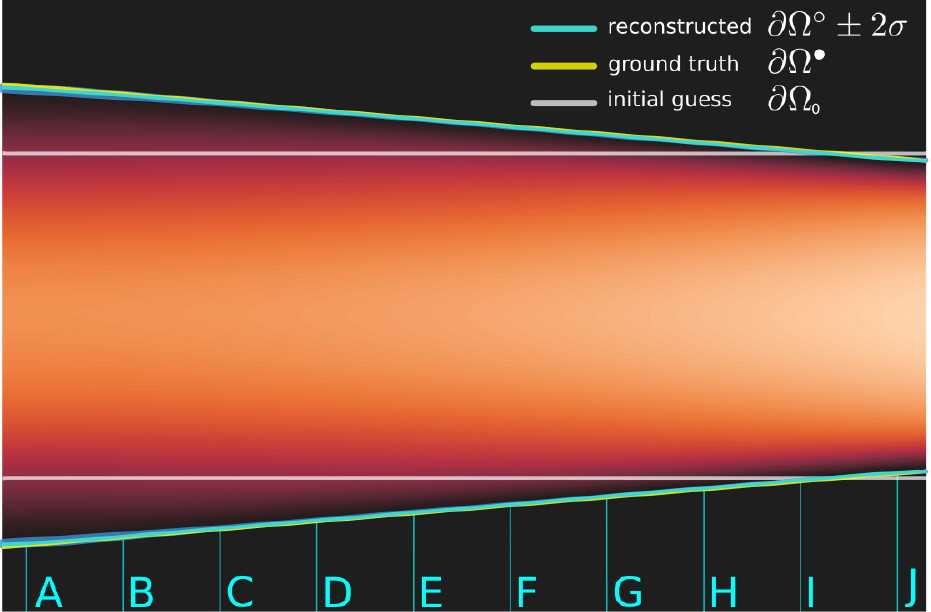}
  \caption{Velocity magnitude $\abs{\bm{u}^\circ}$ and shape $\partial\Omega^\circ$}
  \label{fig:tap_shape_rec}
\end{subfigure}%
\hfill
\begin{subfigure}{0.45\textwidth}
\centering
  \includegraphics[height=0.6416\linewidth,trim=0 0 0 -5,clip]{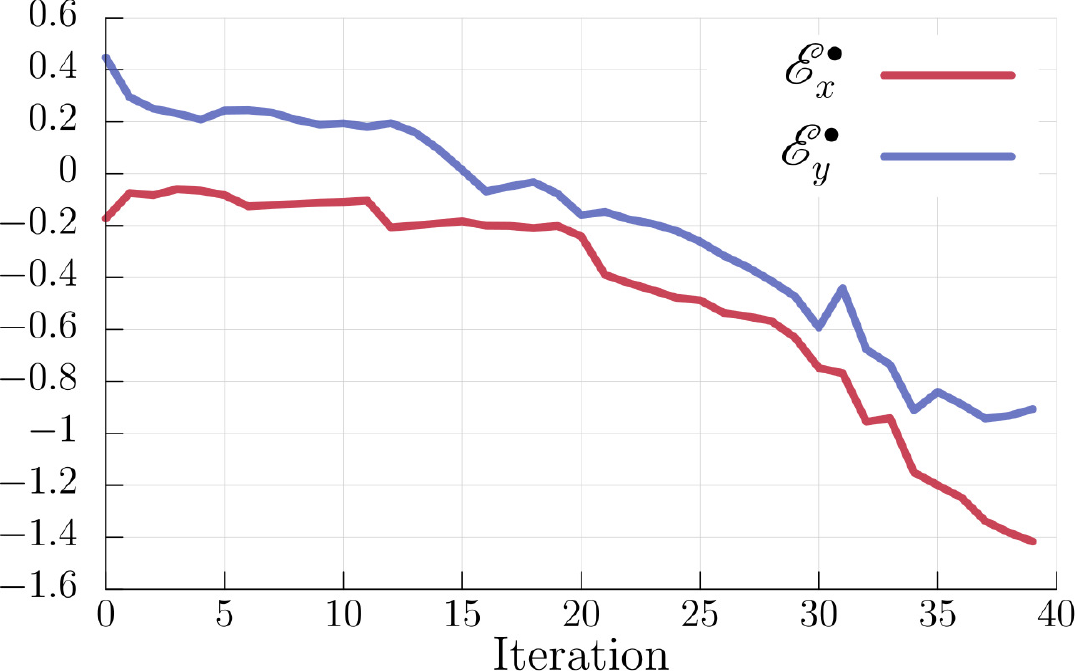}
  \caption{Reconstruction error history}
  \label{fig:tap_conv}
\end{subfigure}%
\hfill
\begin{subfigure}{0.49\textwidth}
\centering
\includegraphics[height=0.82\linewidth]{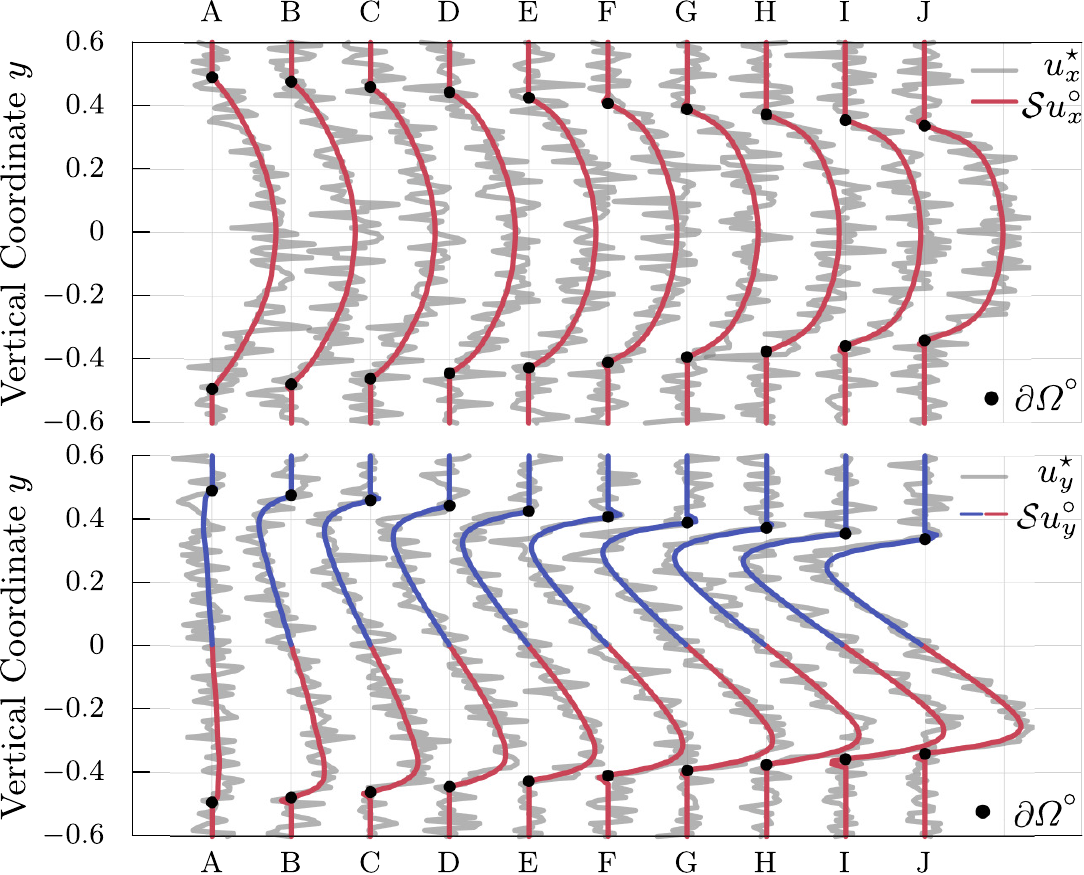}
\caption{Noisy data (grey) and reconstruction}
\label{fig:tap_slices_rec}
\end{subfigure}%
\hfill
\begin{subfigure}{0.49\textwidth}
\centering
\includegraphics[height=0.82\linewidth]{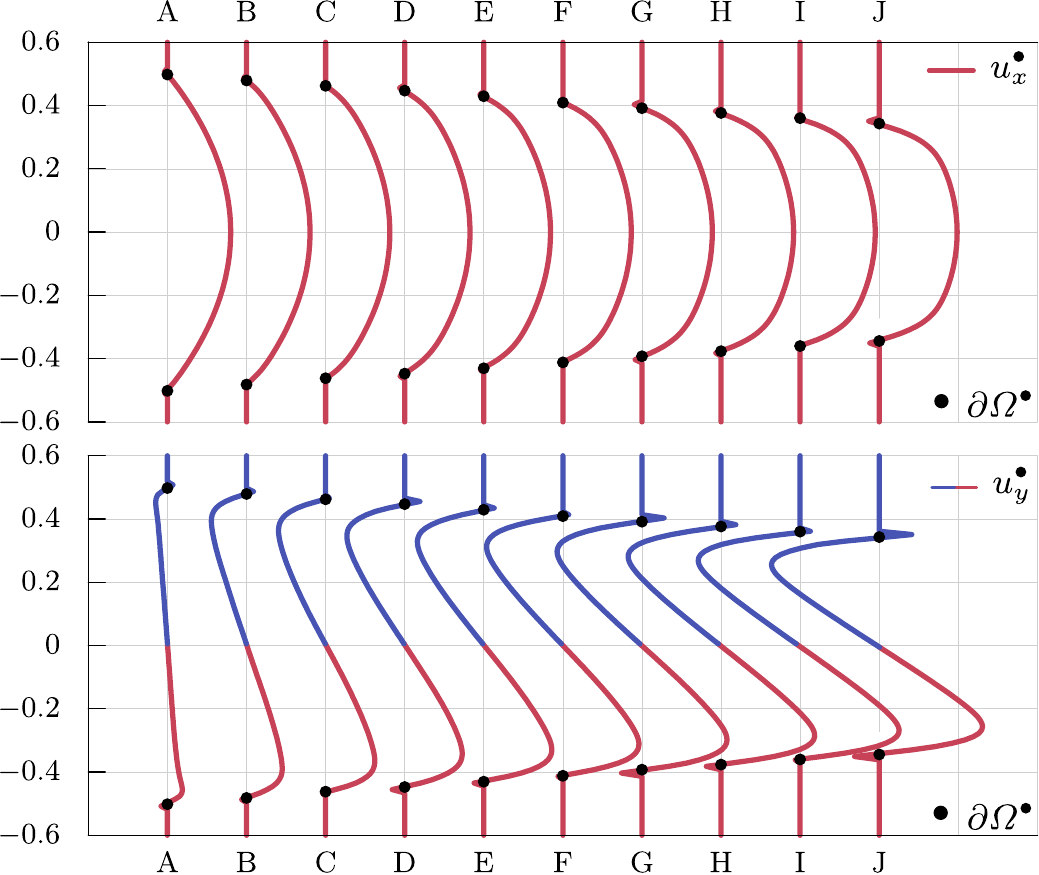}
\caption{Ground truth velocity distributions}
\label{fig:tap_slices_gt}
\end{subfigure}%
\caption{Reconstruction (algorithm \ref{algo:reconstruction}) of synthetic images depicting the flow (from left to right) in a converging channel. Figure \ref{fig:tap_shape_rec} depicts the reconstructed boundary $\partial\Omega^\circ$ (cyan line), the $2\sigma$ confidence region computed from the approximated posterior covariance $\widetilde{\mathcal{C}}_\zetaext \equiv \widetilde{H}_\zetaext\mathcal{C}_\sdist$ (blue region), the ground truth boundary $\partial\Omega^\bullet$ (yellow line), and the initial guess $\partial\Omega_0$ (white line). Figure \ref{fig:tap_conv} shows the reconstruction error as a function of iteration number. Velocity slices are drawn for 10 equidistant cross-sections (labelled with the letters A to J) for both the reconstructed images (figure \ref{fig:tap_slices_rec}) and the ground truth (figure \ref{fig:tap_slices_gt}), colored red for positive values and blue for negative.}
\label{fig:tap_rec_results2}
\end{figure}

\begin{figure}
\centering
\begin{subfigure}{0.49\textwidth}
\centering
\includegraphics[height=0.525\textwidth,trim=-20 0 0 0, clip]{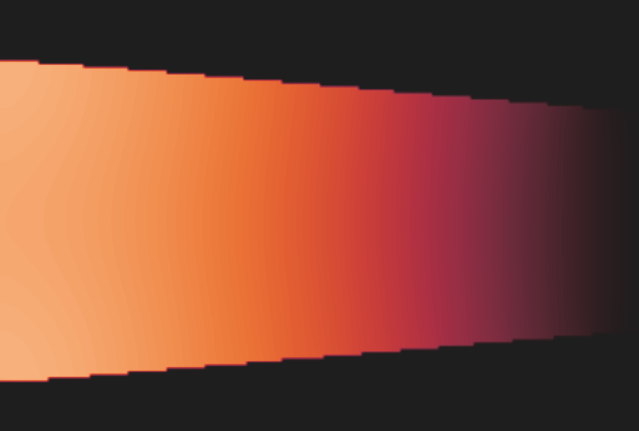}
\caption{Our reconstruction $p^\circ$}
\end{subfigure}%
\hfill
\begin{subfigure}{0.49\textwidth}
\centering
\includegraphics[height=0.525\textwidth,trim=-20 0 0 0, clip]{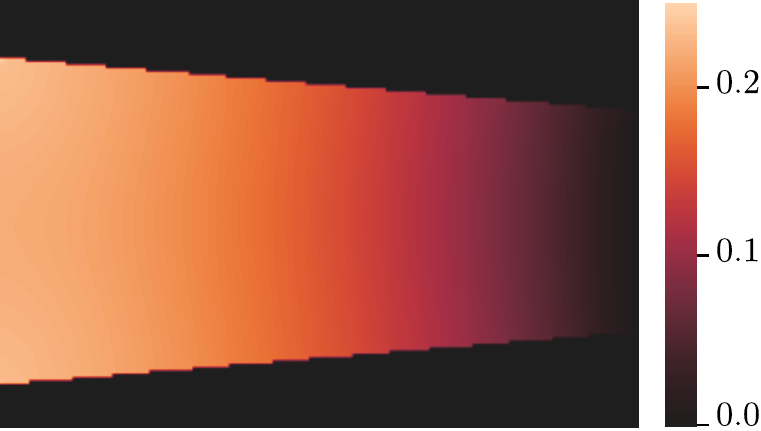}
\caption{Ground truth $p^\bullet$}
\end{subfigure}%
\caption{(a) Reconstructed and (b) ground truth reduced hydrodynamic pressure ($p$) for the flow (from left to right) in the converging channel in figure \ref{fig:tap_rec_results2}.}
\label{fig:pres_toy_problem_1}
\end{figure}

The algorithm manages to reconstruct and segment the noisy flow images in 39 iterations, with total reconstruction error $\mathcal{E}^\bullet \simeq 1.44\%$. The results are presented in figures \ref{fig:tap_rec_results1} and \ref{fig:tap_rec_results2}. We observe that the inverse Navier--Stokes problem performs very well in filtering the noise $(\bm{u}^\star-\mathcal{S}\bm{u}^\circ)$ (figures \ref{fig:tap_ux_rec_err}, \ref{fig:tap_uy_rec_err}), providing noiseless images for each component of the velocity (figures \ref{fig:tap_ux_rec}, \ref{fig:tap_uy_rec}). As we expect, the discrepancies $\invcu(\bm{u}^\star-\mathcal{S}\bm{u}^\circ)$ (figures \ref{fig:tap_ux_rec_err}, \ref{fig:tap_uy_rec_err}) consist mainly of Gaussian white noise, except at the corners of the outlet (figure \ref{fig:tap_uy_rec_err}), where there is a weak correlation. For a more detailed presentation of the denoising effect, we plot slices of the reconstructed velocity (figure \ref{fig:tap_slices_rec}) and the ground truth velocity (figure \ref{fig:tap_slices_gt}). \revv{The reconstructed pressure $p^\circ$, which is consistent with the reconstructed velocity $\bm{u}^\circ$ to machine precision accuracy, is, in effect, indistinguishable from the ground truth $p^\bullet$ (figure \ref{fig:pres_toy_problem_1}).}

\begin{figure}
\centering
\begin{subfigure}{0.49\textwidth}
\includegraphics[height=0.45\textwidth]{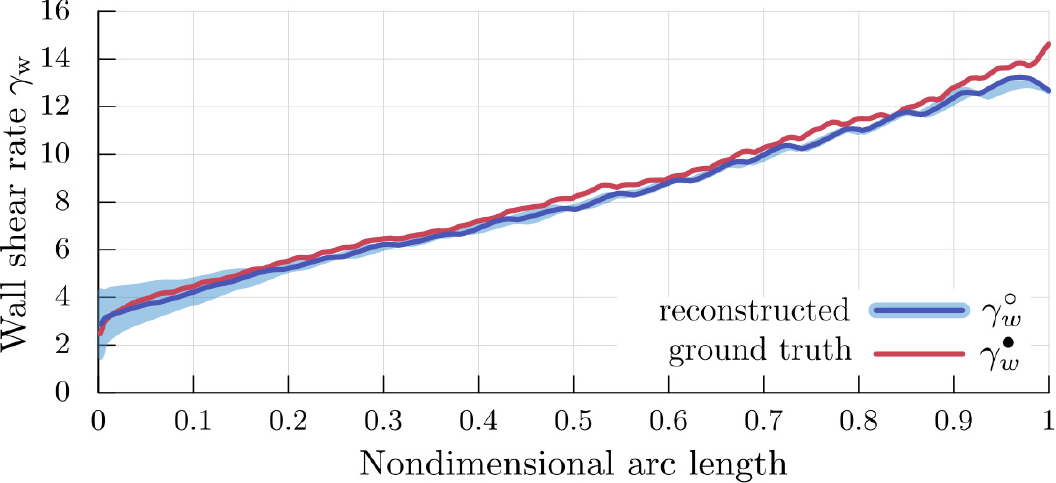}
\caption{Lower boundary}
\label{fig:wss_tap_lower}
\end{subfigure}%
\hfill
\begin{subfigure}{0.49\textwidth}
\includegraphics[height=0.45\textwidth]{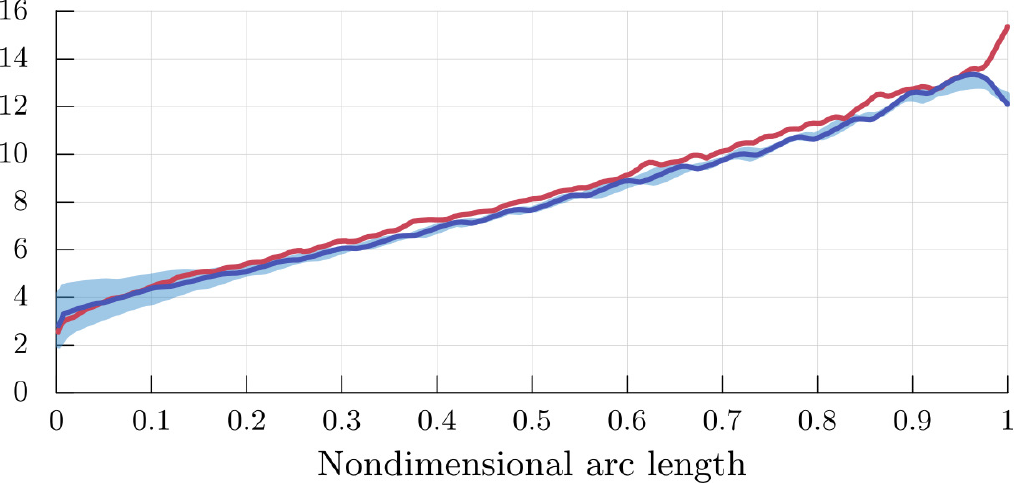}
\caption{Upper boundary}
\label{fig:wss_tap_upper}
\end{subfigure}%
\caption{Wall shear rate $\gamma_w \equiv \bm{\tau}\bm{\cdot}\partial_{\bm{\nu}}\bm{u}$, where $\bm{\tau}$ is the unit tangent vector of $\partial\Omega$, for the converging channel flow in figure \ref{fig:tap_rec_results2}. The wall shear stress is found by multiplying this by the viscosity. The reconstructed wall shear rate ($\gamma_w^\circ$) is calculated on $\partial\Omega^\circ$ and for $\bm{u}^\circ$, while the ground truth ($\gamma_w^\bullet$) is calculated on $\partial\Omega^\bullet$ and for $\bm{u}^\bullet$. The blue region is bounded by the two wall shear rate distributions for $\bm{u}^\circ$, calculated on the upper ($\partial\Omega^\circ_+$) and lower ($\partial\Omega^\circ_-$) limits of the $2\sigma$ confidence region of $\partial\Omega^\circ$. Note that the reconstructed solution can sometimes be found outside the blue region because the reconstructed shape $\partial\Omega^\circ$ may be less regular than $\partial\Omega^\circ_+$ or $\partial\Omega^\circ_-$.}
\label{fig:wss_toy_problem_1}
\end{figure}

Having obtained the reconstructed velocity $\bm{u}^\circ$, we can compute the wall shear rate $\gamma_w^\circ$ on the reconstructed boundary $\partial\Omega^\circ$, which we compare with the ground truth $\gamma_w^\bullet$ in figure \ref{fig:wss_toy_problem_1}. Using the upper ($\partial\Omega^\circ_+$) and lower ($\partial\Omega^\circ_-$) limits of the $2\sigma$ confidence region for $\partial\Omega^\circ$ (figure \ref{fig:tap_shape_rec}) we estimate a confidence region for $\gamma_w^\circ$; although this has to be interpreted carefully. Note that, for example, $\partial\Omega^\circ_+$ and $\partial\Omega^\circ_-$ can be smoother than the mean $\partial\Omega^\circ$, and, therefore, $\partial\Omega^\circ$ may be found outside this confidence region. A better estimate of the confidence region could be obtained by sampling the posterior distribution of $\partial\Omega^\circ$ in order to solve a Navier--Stokes problem for each sample $\partial\Omega_k$ and find the distribution of $\gamma^\circ_w$. Since the latter approach would be computationally intensive, we only provide our estimate, which requires the solution of only two Navier--Stokes problems.

\subsection{Synthetic data for 2D flow in a simulated abdominal aortic aneurysm}
\label{sec:blood_vessel_dummy}

Next, we test algorithm \ref{algo:reconstruction} in a channel that resembles the cross-section of a small abdominal aortic aneurysm, with $D_\text{max}/D\simeq 1.5$, where $D_\text{max}$ is the maximum diameter at the midsection. We generate synthetic images for $\bm{u}^\star$ as in section \ref{sec:converging_channel}, again for ${\text{SNR}_x=\text{SNR}_y=3}$, but now for $\Rey = 153$. The ground truth domain $\Omega^\bullet$ has horizontal symmetry but the inlet velocity profile deliberately breaks this symmetry. The inverse problem is the same as that in section \ref{sec:converging_channel} but with different input parameters (see table \ref{tab:input_params_2}). The initial guess $\Omega_0$ is a rectangular domain with height equal to $0.85D$, centered in the image domain. For ${\bm{g}_i}_0$ we take a skewed parabolic velocity profile with a peak velocity of approximately $2U$ that fits the inlet of $\Omega_0$.

\begin{table}
  \begin{center}
\def~{\hphantom{0}}
  \begin{tabular}{llcccccc}
        & & image dim.   &   model dim. & \multicolumn{2}{c}{$\sigma_{u_x}/U$} & \multicolumn{2}{c}{$\sigma_{u_y}/U$}\\[3pt]
       simul. abd. aortic aneurysm  & (2D)  & $192^2$ & $200^2$  & \multicolumn{2}{c}{$2.80\times10^{-1}$} & \multicolumn{2}{c}{$5.26\times10^{-3}$}\\

        &   &  &   &  &  & &\\
         \multicolumn{2}{c}{\textit{Regularization}} & $\sigma_\sdist/D$ & $\sigma_{\bm{g}_i}/U$ & $\sigma_\nu/UD$ & $\Rey_\sdist$ & $\Rey_\zeta$ & $\ell/h$\\[3pt]
         simul. abd. aortic aneurysm  & (2D) & $1.0$ & $2.0$ & $\cdot$ & 0.1 & 0.1 & 3\\
  \end{tabular}
  \caption{Input parameters for the inverse 2D Navier--Stokes problem.}
  \label{tab:input_params_2}
  \end{center}
\end{table}
\begin{figure}
\begin{subfigure}{.32\textwidth}
  \includegraphics[height=0.68\linewidth]{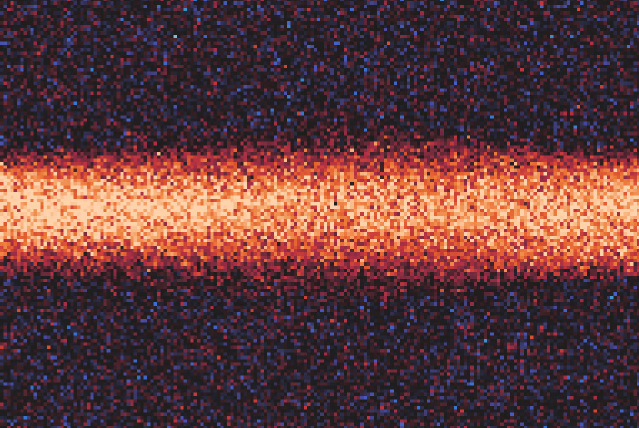}
  \caption{Synthetic image $u^\star_x$}
  \label{fig:exp_ux_sig}
\end{subfigure}%
\hfill
\begin{subfigure}{.32\textwidth}
  \includegraphics[height=0.68\linewidth]{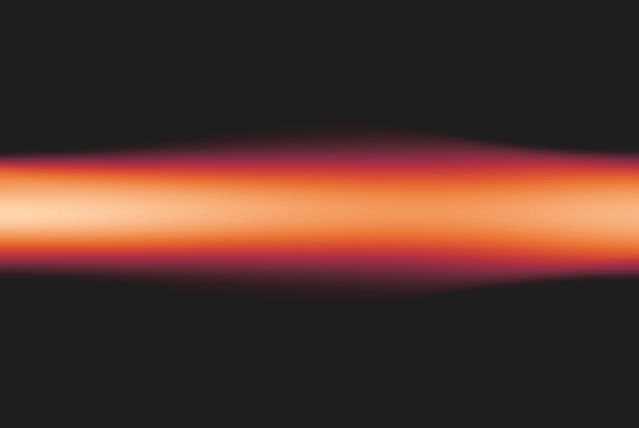}
  \caption{Our reconstruction $u_x^\circ$}
  \label{fig:exp_ux_rec}
\end{subfigure}%
\hfill
\begin{subfigure}{.32\textwidth}
  \includegraphics[height=0.68\linewidth]{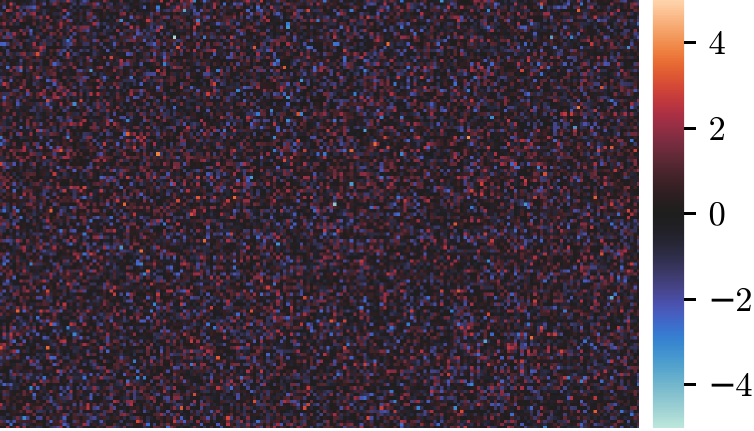}
  \caption{Discrepancy $\sigma_{u_x}^{-1}\big(u^\star_x - \mathcal{S}u_x^\circ\big)$}
  \label{fig:exp_ux_rec_err}
\end{subfigure}
\label{fig:test}
\begin{subfigure}{.32\textwidth}
  \includegraphics[height=0.68\linewidth]{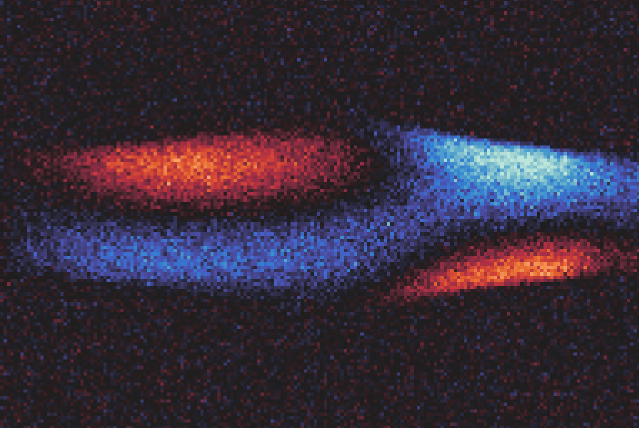}
  \caption{Synthetic image $u^\star_y$}
  \label{fig:exp_uy_sig}
\end{subfigure}%
\hfill
\begin{subfigure}{.32\textwidth}
  \includegraphics[height=0.68\linewidth]{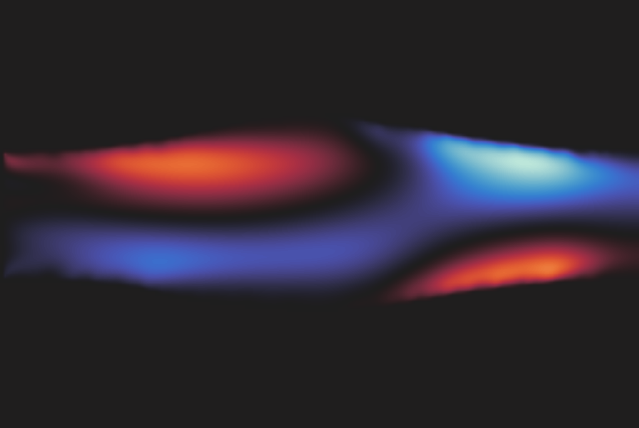}
  \caption{Our reconstruction $u_y^\circ$}
  \label{fig:exp_uy_rec}
\end{subfigure}%
\hfill
\begin{subfigure}{.32\textwidth}
  \includegraphics[height=0.68\linewidth]{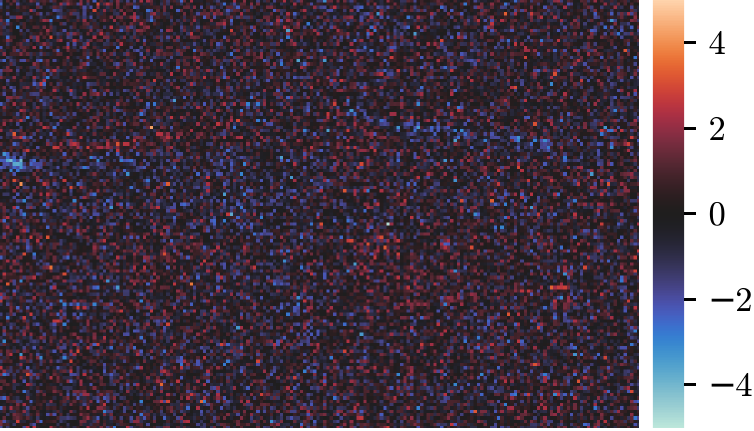}
  \caption{Discrepancy $\sigma_{u_y}^{-1}\big(u^\star_y - \mathcal{S}u_y^\circ\big)$}
  \label{fig:exp_uy_rec_err}
\end{subfigure}%
\caption{As for figure \ref{fig:tap_rec_results1}, but for the synthetic images depicting the flow (from left to right) in the simulated 2D model of an abdominal aortic aneurysm.}
\label{fig:exp_rec_results1}
\end{figure}

\begin{figure}
\begin{subfigure}{0.55\textwidth}
\centering
  \includegraphics[height=0.525\linewidth]{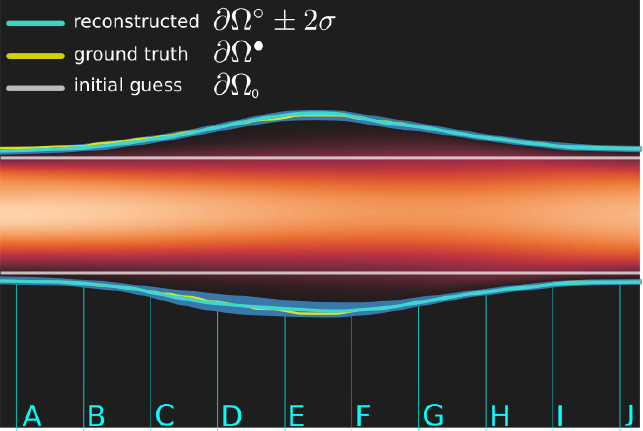}
  \caption{Velocity magnitude $\abs{\bm{u}^\circ}$ and shape $\partial\Omega^\circ$}
  \label{fig:exp_shape_rec}
\end{subfigure}%
\hfill
\begin{subfigure}{0.45\textwidth}
\centering
  \includegraphics[height=0.6416\linewidth,trim=0 0 0 -5,clip]{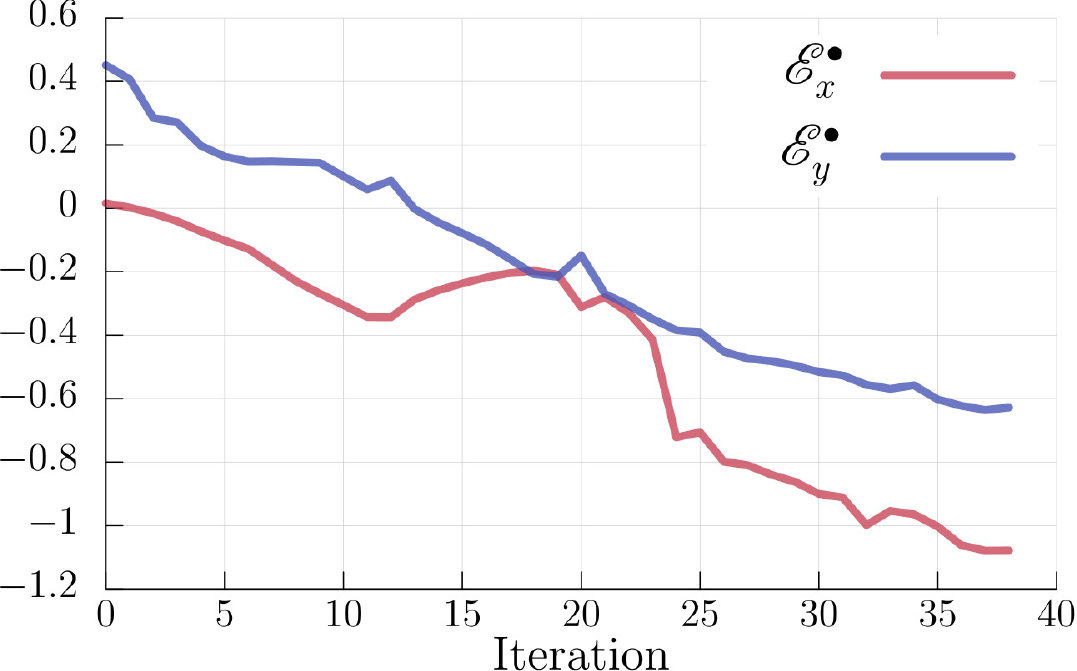}
  \caption{Reconstruction error history}
  \label{fig:exp_conv}
\end{subfigure}%
\hfill
\begin{subfigure}{0.49\textwidth}
\centering
\includegraphics[height=0.82\linewidth]{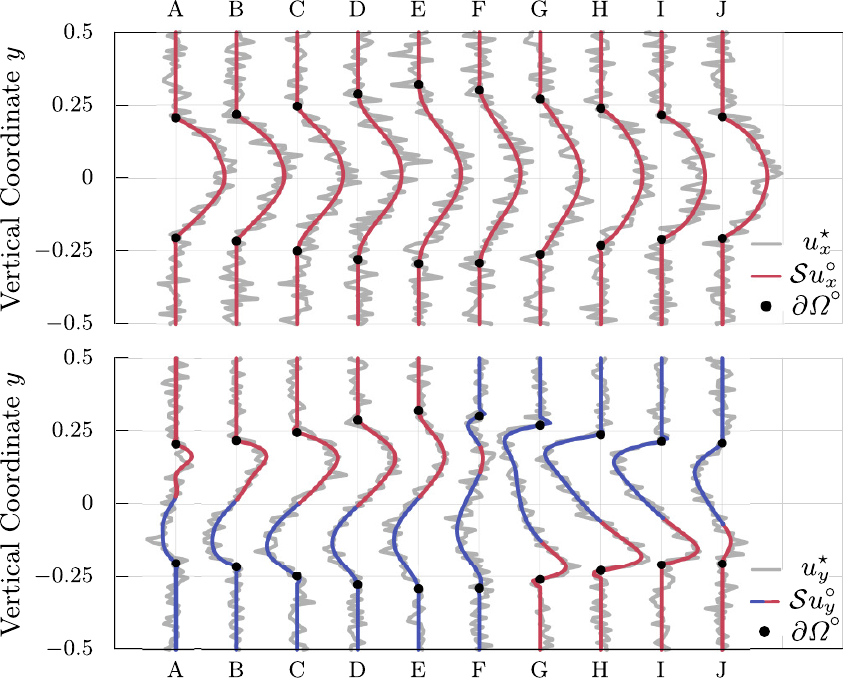}
\caption{Noisy data (grey) and reconstruction}
\label{fig:exp_slices_rec}
\end{subfigure}%
\hfill
\begin{subfigure}{0.49\textwidth}
\centering
\includegraphics[height=0.82\linewidth]{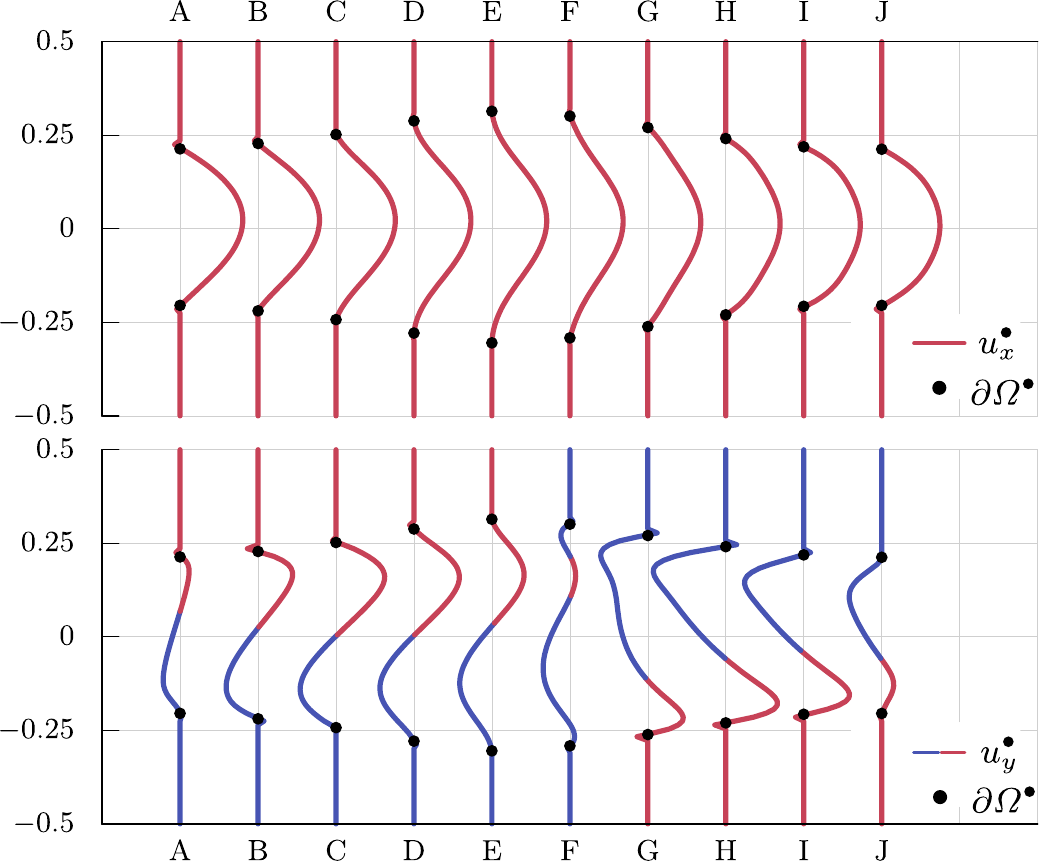}
\caption{Ground truth velocity distributions}
\label{fig:exp_slices_gt}
\end{subfigure}%
\caption{As for figure \ref{fig:tap_rec_results2}, but for the synthetic images depicting the flow (from left to right) in the simulated 2D model of an abdominal aortic aneurysm.}
\label{fig:exp_rec_results2}
\end{figure}

\begin{figure}
\centering
\begin{subfigure}{0.49\textwidth}
\centering
\includegraphics[height=0.525\textwidth,trim=-20 0 0 0, clip]{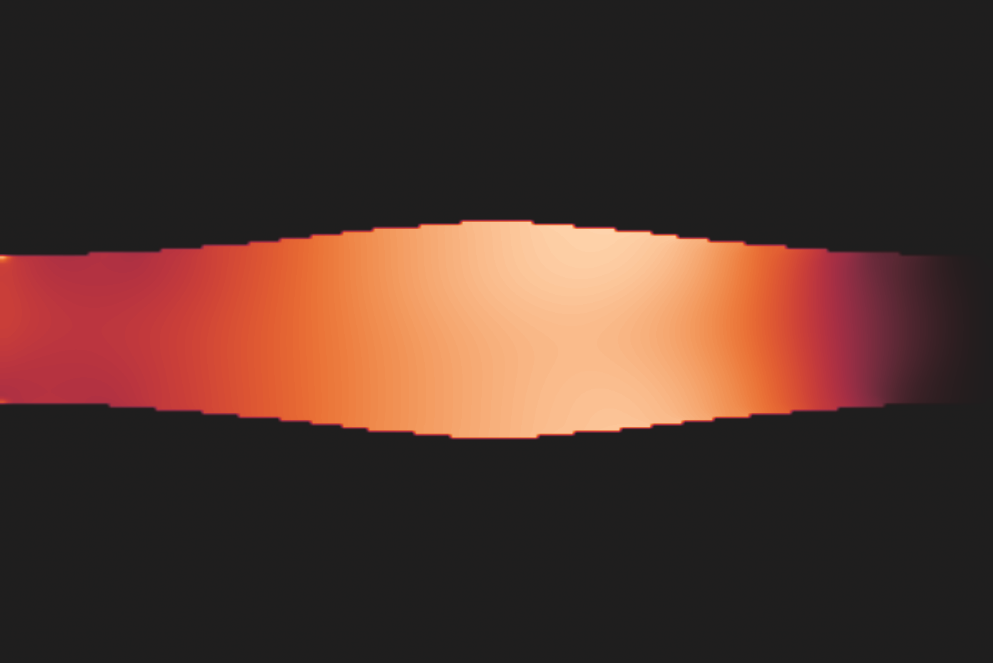}
\caption{Our reconstruction $p^\circ$}
\end{subfigure}%
\hfill
\begin{subfigure}{0.49\textwidth}
\centering
\includegraphics[height=0.525\textwidth,trim=-20 0 0 0, clip]{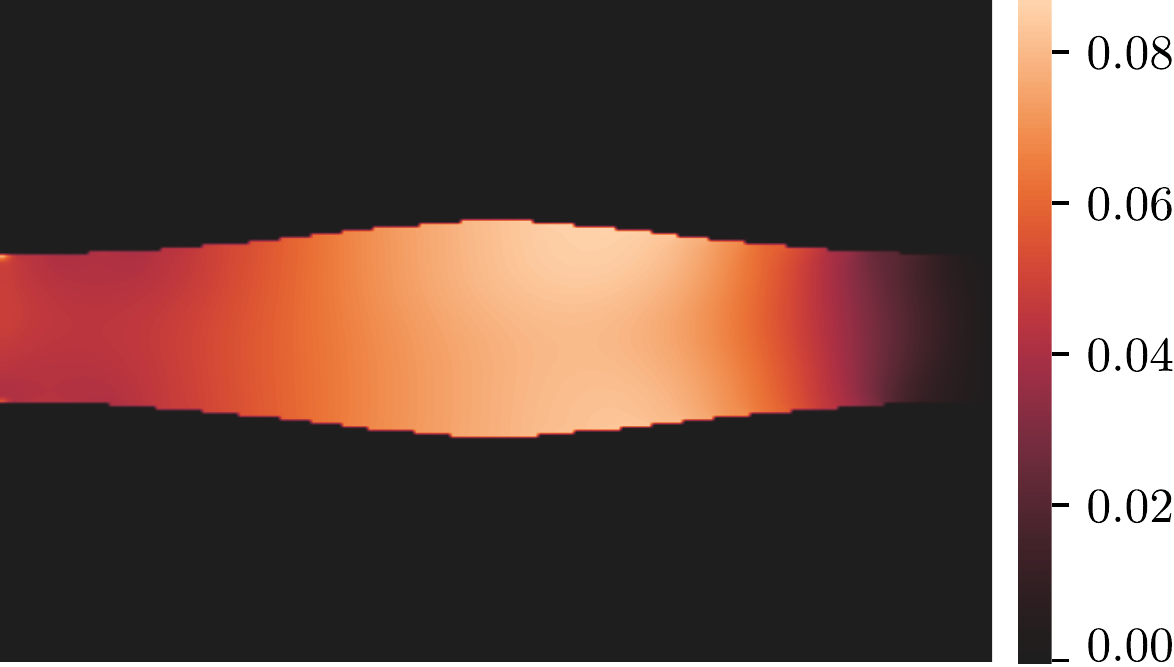}
\caption{Ground truth $p^\bullet$}
\end{subfigure}%
\caption{(a) Reconstructed and (b) ground truth reduced hydrodynamic pressure ($p$) for the flow (from left to right) in the simulated 2D model of an abdominal aortic aneurysm in figure \ref{fig:exp_rec_results2}.}
\label{fig:pres_toy_problem_2}
\end{figure}

The algorithm manages to reconstruct and segment the noisy flow images in 39 iterations, with total reconstruction error $\mathcal{E}^\bullet \simeq 2.87\%$. The results are presented in figures \ref{fig:exp_rec_results1} and \ref{fig:exp_rec_results2}. We observe that the discrepancy (figures \ref{fig:exp_ux_rec_err}, \ref{fig:exp_uy_rec_err}) consists mainly of Gaussian white noise. Again, some correlations are visible in the discrepancy of the $y$-velocity component at the upper inlet corner and the upper boundary of the simulated abdominal aortic aneurysm. The latter correlations (figure \ref{fig:exp_uy_rec_err}) can be explained by the associated uncertainty in the predicted shape $\partial\Omega^\circ$ (figure \ref{fig:exp_shape_rec}), which is well estimated for the upper boundary but slightly underestimated for the upper inlet corner. It is interesting to note that the upward skewed velocity profile at the inlet creates a region of low velocity magnitude on the lower boundary. The velocity profiles in this region produce low wall shear stresses, as seen in figures \ref{fig:exp_slices_rec}, \ref{fig:exp_slices_gt}, and \ref{fig:wss_blood_vessel_dummy}. These conditions are particularly challenging when one tries to infer the true boundary $\partial\Omega^\bullet$ because the local SNR is low ($\text{SNR} \ll 1$), meaning that there is considerable information loss there. Despite the above difficulties, algorithm \ref{algo:reconstruction} manages to approximate the posterior distribution of $\partial\Omega^\circ$ well, and successfully predicts extra uncertainty in this region (figure \ref{fig:exp_shape_rec}). \revv{Again, the reconstructed pressure $p^\circ$ is indistinguishable from the ground truth $p^\bullet$ (figure \ref{fig:pres_toy_problem_2}).}

Using the reconstructions $\bm{u}^\circ$ and $\partial\Omega^\circ$ we compute the wall shear rate and we compare it with the ground truth in figure \ref{fig:wss_blood_vessel_dummy}. We observe that the reconstructed solution approximates the ground truth well, even for very low signal-to-noise ratios ($\text{SNR}=3$). Note that the waviness of the ground truth $\gamma_w^\bullet$ is due to the relatively poor resolution of the level set function that we intentionally used to implicitly define this domain.

\begin{figure}
\centering
\begin{subfigure}{0.49\textwidth}
\includegraphics[height=0.45\textwidth]{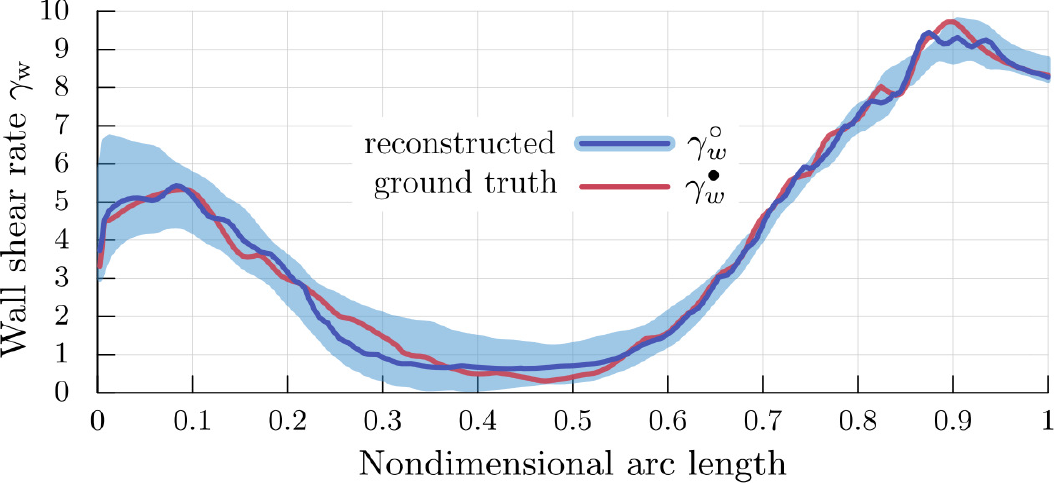}
\caption{Lower boundary}
\label{fig:wss_blood_vessel_dummy}
\end{subfigure}%
\hfill
\begin{subfigure}{0.49\textwidth}
\includegraphics[height=0.45\textwidth]{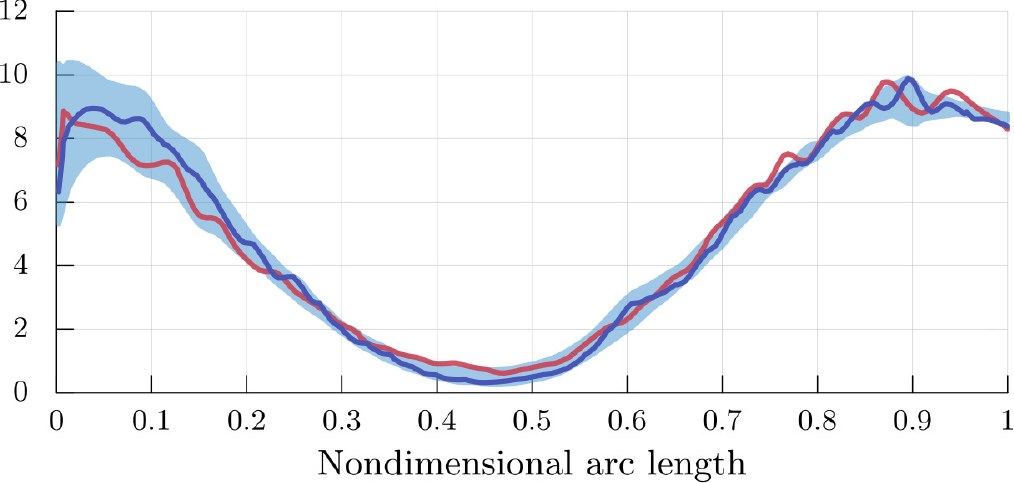}
\caption{Upper boundary}
\label{fig:wss_blood_vessel_dummy}
\end{subfigure}%
\caption{As for figure \ref{fig:wss_toy_problem_1} but for the synthetic images depicting the flow (from left to right) in the simulated 2D model of an abdominal aortic aneurysm in figure \ref{fig:exp_rec_results2}.}
\label{fig:wss_toy_problem_2}
\end{figure}

\subsection{\revv{Synthetic data for 2D flow in a simulated aortic aneurysm}}
\label{sec:aortic_aneurysm}

\begin{figure}
\centering
\begin{subfigure}{0.49\textwidth}
\centering
  \includegraphics[height=.85\linewidth]{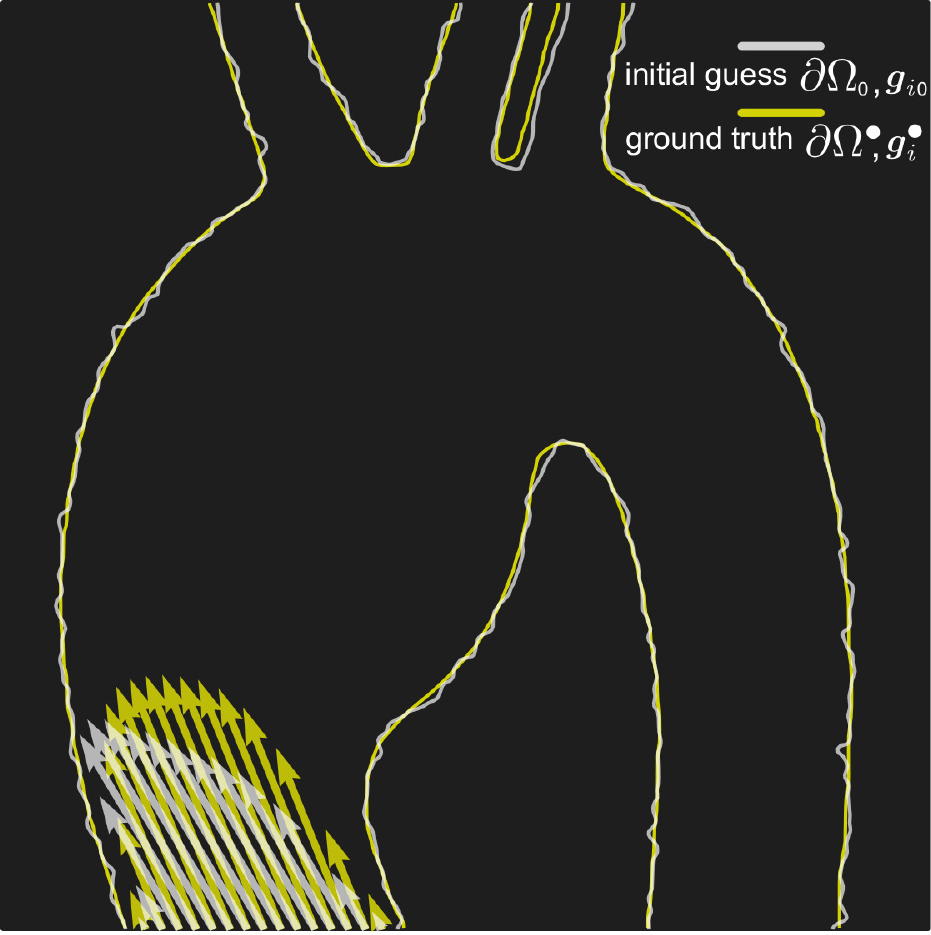}
  \caption{Initial guesses (priors) for $\partial\Omega$ and $\bm{g}_i$}
  \label{fig:aorta_init_guess}
\end{subfigure}%
\hfill
\begin{subfigure}{0.49\textwidth}
\centering
  \includegraphics[height=.85\linewidth,trim=0 0 0 0,clip]{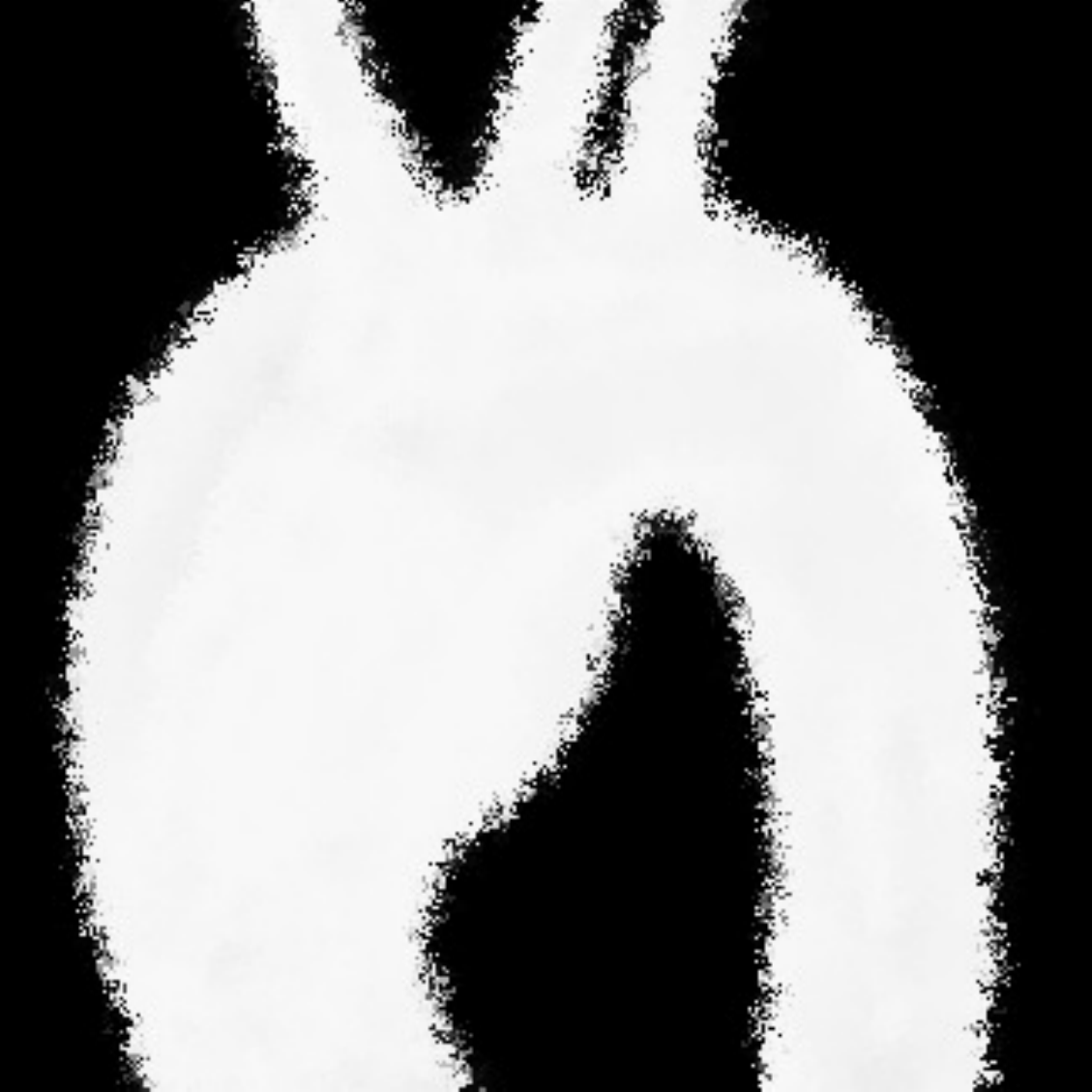}
  \caption{Noisy mask of $\Omega^\bullet$}
  \label{fig:aorta_noisy_mask}
\end{subfigure}%
\caption{Initial guesses (input for algorithm \ref{algo:reconstruction}) for the geometry ($\partial\Omega_0$) and the inlet velocity profile (${\bm{g}_i}_0$) versus their corresponding ground truth (figure \ref{fig:aorta_init_guess}), for the flow in the simulated 2D model of an aortic aneurysm. The initial guess $\partial\Omega_0$ (figure \ref{fig:aorta_init_guess}) is generated by segmenting the noisy mask (figure \ref{fig:aorta_noisy_mask}) of the ground truth domain $\Omega^\bullet$.}
\label{fig:aorta_init_guesses}
\end{figure}

\begin{figure}
\begin{subfigure}{.32\textwidth}
  \includegraphics[height=1.\linewidth,angle=90]{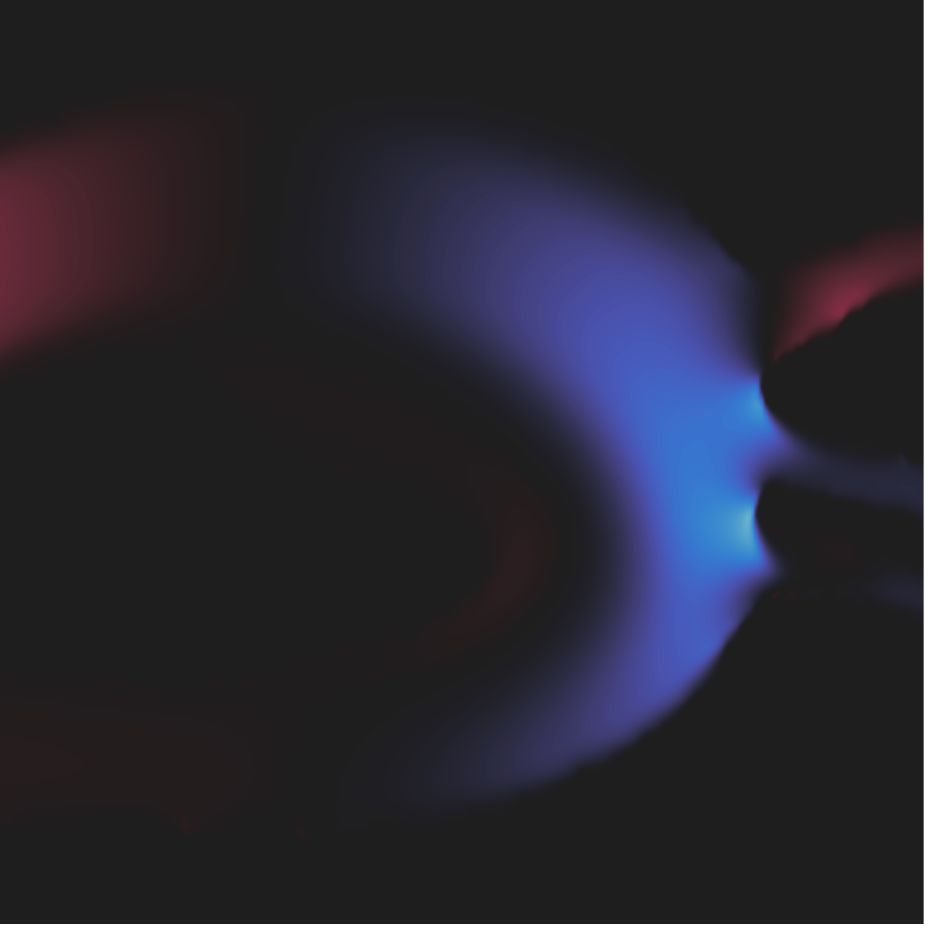}
  \caption{$(u_x)_0$}
  \label{fig:aorta_u_x_0}
\end{subfigure}%
\hfill
\begin{subfigure}{.32\textwidth}
  \includegraphics[width=1.\linewidth,angle=90]{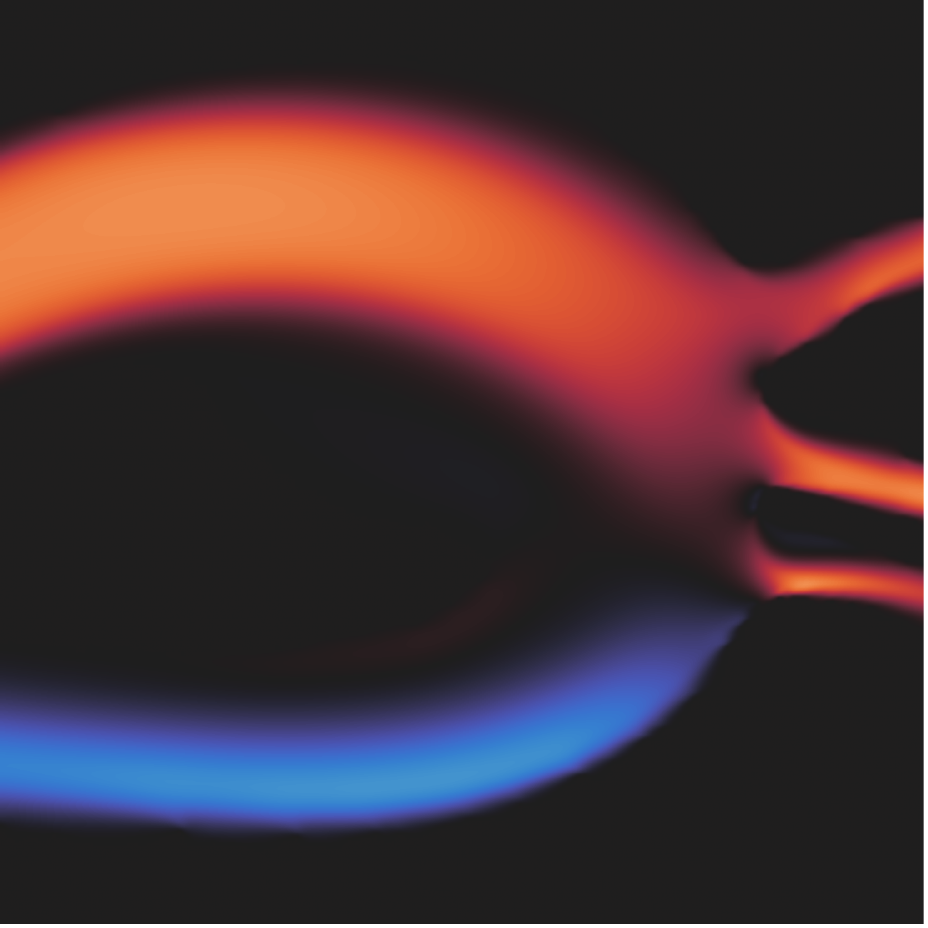}
  \caption{$(u_y)_0$}
  \label{fig:aorta_u_y_0}
\end{subfigure}%
\hfill
\begin{subfigure}{.32\textwidth}
  \includegraphics[width=1.\linewidth,angle=90]{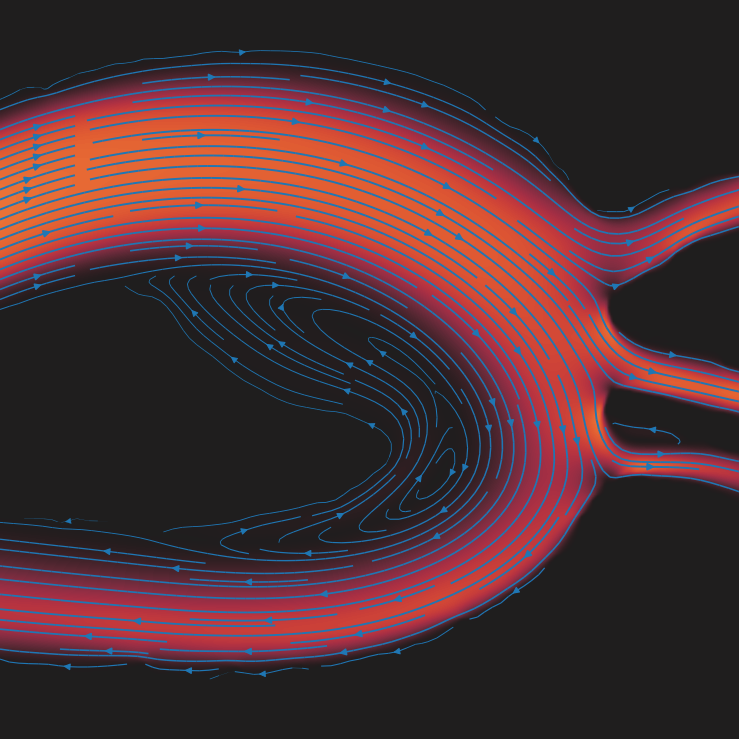}
  \caption{$\bm{u}_0$}
  \label{fig:aorta_stream_0}
\end{subfigure}
\begin{subfigure}{.32\textwidth}
  \includegraphics[width=1.\linewidth,angle=90]{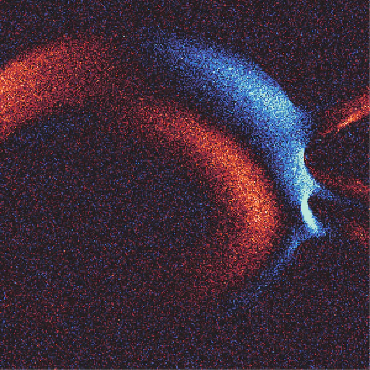}
  \caption{$\sigma_{u_x}^{-1}\big(u^\star_x - \mathcal{S}(u_x)_0\big)$}
  \label{fig:aorta_u_x_0_discr}
\end{subfigure}%
\hfill
\begin{subfigure}{.32\textwidth}
  \includegraphics[width=1.\linewidth,angle=90]{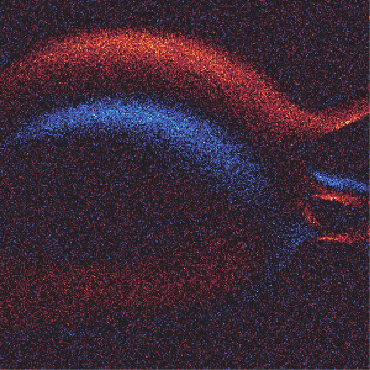}
  \caption{$\sigma_{u_y}^{-1}\big(u^\star_y - \mathcal{S}(u_y)_0\big)$}
  \label{fig:aorta_u_y_0_discr}
\end{subfigure}%
\hfill
\begin{subfigure}{.32\textwidth}
  \includegraphics[width=1.\linewidth,angle=90]{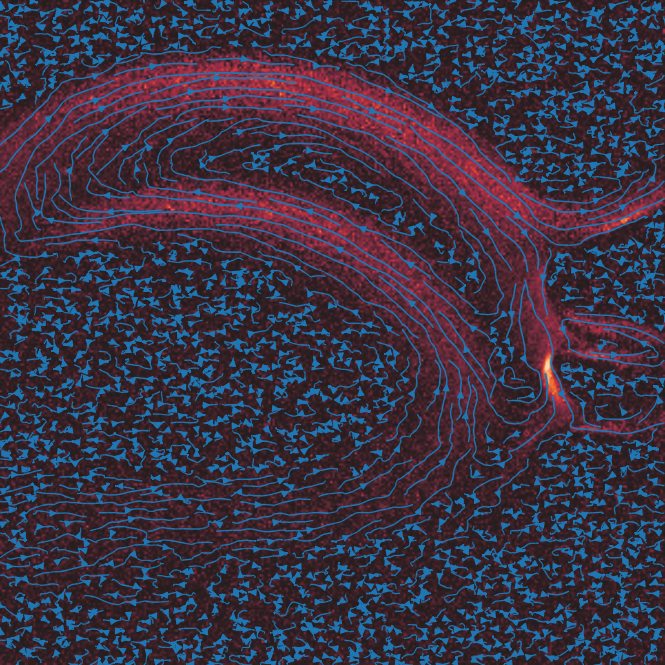}
  \caption{$\invcu(\bm{u}^\star-\mathcal{S}\bm{u}_0)$}
  \label{fig:aorta_stream_0_discr}
\end{subfigure}
\caption{Zeroth iteration (N--S solution for the initial guesses in figure \ref{fig:aorta_init_guesses}) velocity images (figures \ref{fig:aorta_u_x_0}, \ref{fig:aorta_u_y_0}), streamlines (figure \ref{fig:aorta_stream_0}), and discrepancies with the data (figures \ref{fig:aorta_u_x_0_discr}-\ref{fig:aorta_stream_0_discr}), for the flow  in the simulated 2D model of an aortic aneurysm. Streamlines are plotted on top of the velocity/discrepancy magnitude image, and streamline thickness increases as the velocity magnitude increases. Figures \ref{fig:aorta_u_x_0}-\ref{fig:aorta_stream_0} (colorbar not shown) and \ref{fig:aorta_u_x_0_discr}-\ref{fig:aorta_stream_0_discr} (colorbar shown on the right) share the same colormap.}
\label{fig:aorta_zeroth_iter_results}
\end{figure}

\revv{Next, we test algorithm \ref{algo:reconstruction} in a channel that resembles the cross-section of an aorta that has an aneurysm in its ascending part. This test case is designed to demonstrate that the algorithm is applicable to realistic geometries with multiple inlets/outlets and for abnormal flow conditions (e.g. separation and recirculation zones). We generate synthetic images for $\bm{u}^\star$ as in section \ref{sec:converging_channel}, but for ${\text{SNR}_x=\text{SNR}_y=2.5}$, and for $\Rey = 500$. \revvv{For increased Reynolds numbers ($\Rey = 1000, 1500$), we observed vortex shedding within the aneurysm and we could not find a steady flow solution to generate synthetic images of steady flow}. The inverse problem is the same as that in section \ref{sec:converging_channel} but with different input parameters (see table \ref{tab:input_params_aorta}). The initial guess for the boundary of $\Omega_0$ (figure \ref{fig:aorta_init_guess}) is generated by using the Chan--Vese segmentation method \citep{Chan2001,Getreuer2012,VanDerWalt2014} on the noisy mask of the ground truth domain $\Omega^\bullet$ (figure \ref{fig:aorta_noisy_mask}). The prior standard deviation $\sigma_\sdist$ corresponds to the length of approximately $7$ pixels of the noisy mask. The initial guess for the inlet velocity profile ${\bm{g}_i}_0$ is also shown in figure \ref{fig:aorta_init_guess}. Using the prior information of the boundary and the inlet velocity profile, algorithm \ref{algo:reconstruction} generates an initial guess for the Navier--Stokes velocity field (figures \ref{fig:aorta_u_x_0}, \ref{fig:aorta_u_y_0}) during its zeroth iteration.}

\begin{table}
  \begin{center}
\def~{\hphantom{0}}
  \begin{tabular}{llccccccc}
        & & image dim.   &   \multicolumn{2}{c}{model dim.} & \multicolumn{2}{c}{$\sigma_{u_x}/U$} & \multicolumn{2}{c}{$\sigma_{u_y}/U$}\\[3pt]
       simul. aortic aneurysm & (2D) & $300^2$ & \multicolumn{2}{c}{$325^2$} & \multicolumn{2}{c}{$1.17\times10^{-1}$} & \multicolumn{2}{c}{$2.62\times10^{-1}$} \\

        &   &  &   &  &  & & &\\
         \multicolumn{2}{c}{\textit{Regularization}} & $\sigma_\sdist/D$ & $\sigma_{{g_i}_x}/U$ & $\sigma_{{g_i}_y}/U$ & $\sigma_\nu/UD$ & $\Rey_\sdist$ & $\Rey_\zeta$ & $\ell/h$\\[3pt]
         simul. aortic aneurysm & (2D) & $0.025$ & $0.5$ & $0.4$ & . & $1$ & $1$ & $5$
  \end{tabular}
  \caption{Input parameters for the inverse 2D Navier--Stokes problem.}
  \label{tab:input_params_aorta}
  \end{center}
\end{table}

\begin{figure}
\begin{subfigure}{.32\textwidth}
  \includegraphics[width=1.\linewidth,angle=90]{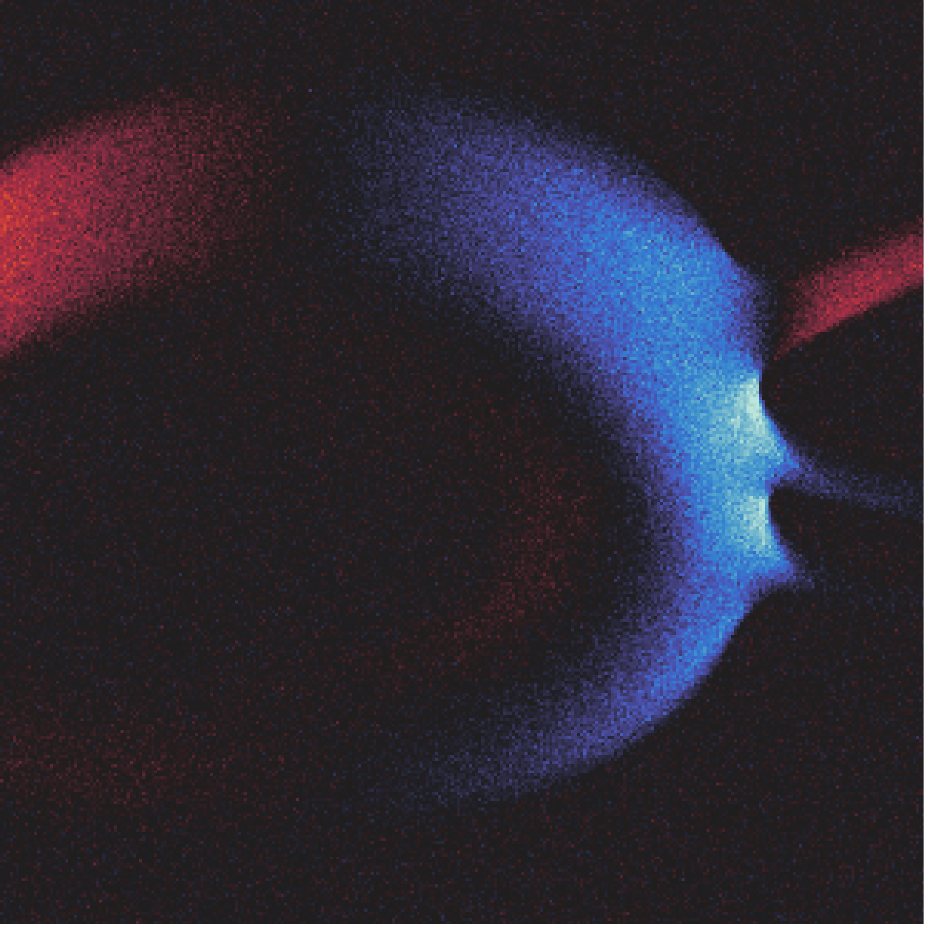}
  \caption{Synthetic image $u^\star_x$}
  \label{fig:aorta_ux_sig}
\end{subfigure}%
\hfill
\begin{subfigure}{.32\textwidth}
  \includegraphics[width=1.\linewidth,angle=90]{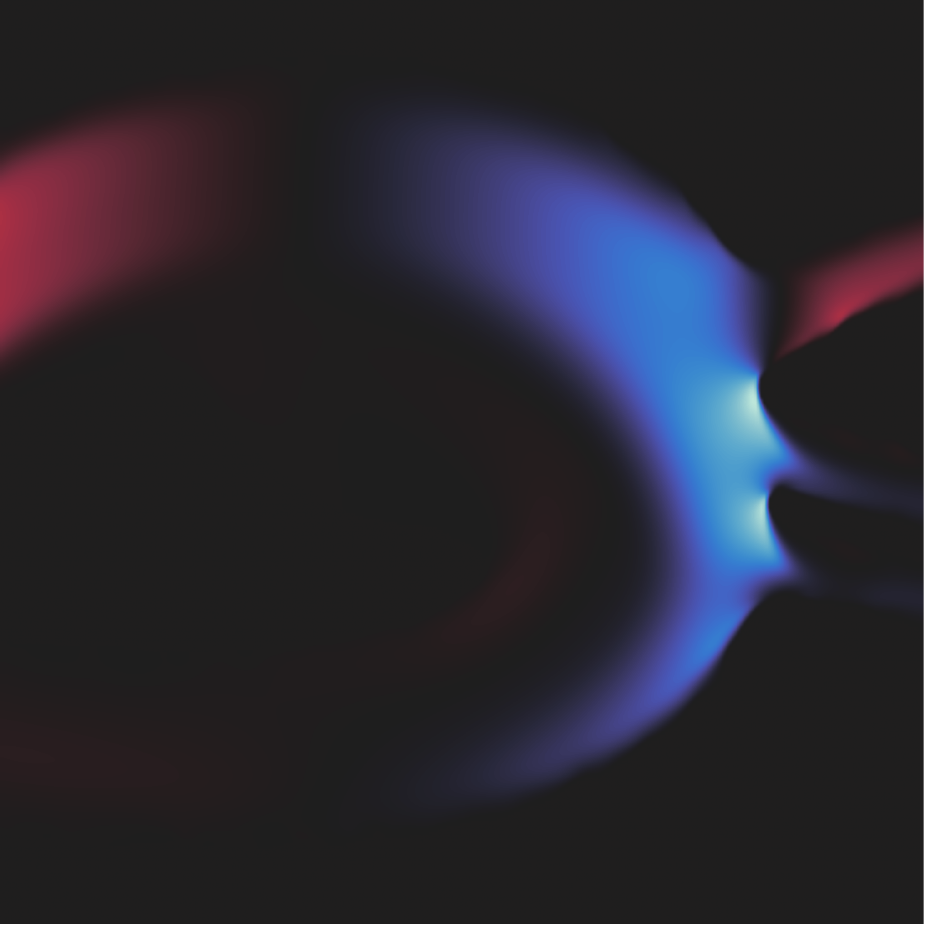}
  \caption{Our reconstruction $u_x^\circ$}
  \label{fig:aorta_ux_rec}
\end{subfigure}%
\hfill
\begin{subfigure}{.32\textwidth}
  \includegraphics[width=1.\linewidth,,angle=90]{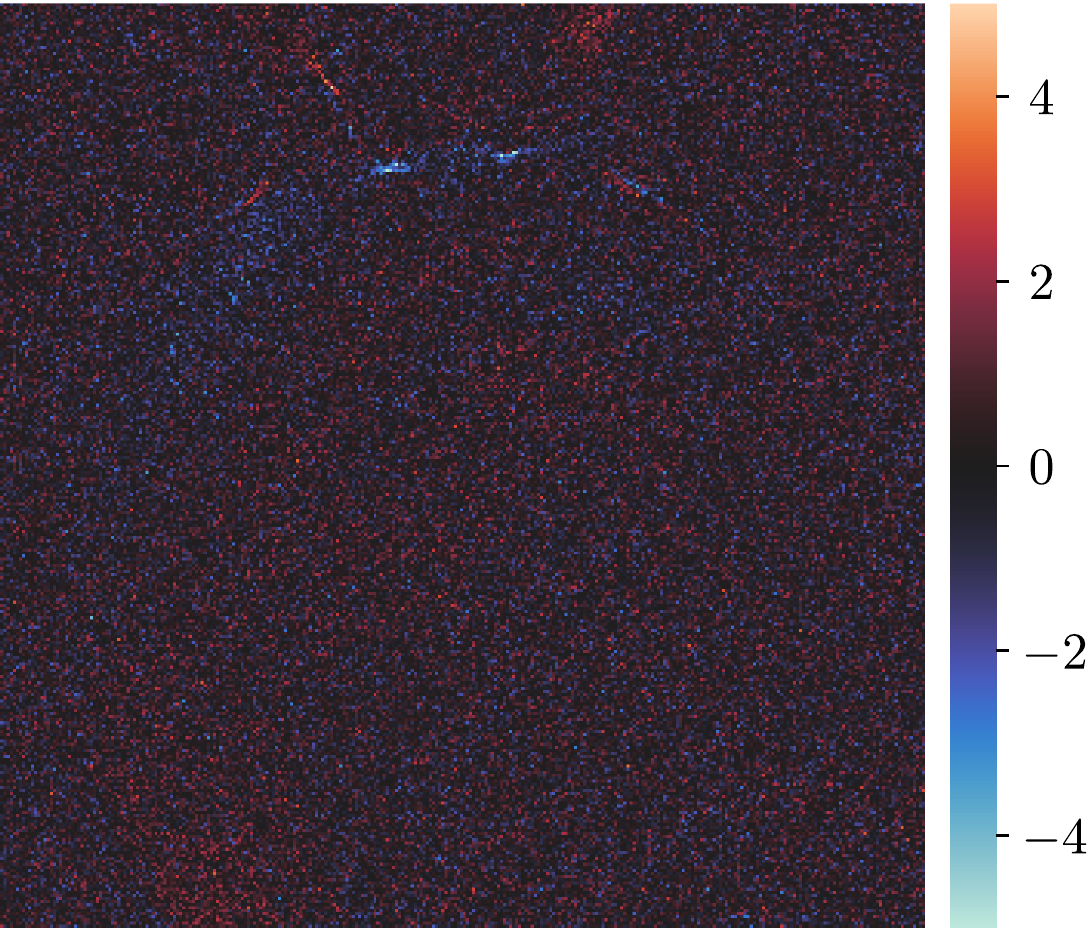}
  \caption{Discrepancy $\sigma_{u_x}^{-1}\big(u^\star_x - \mathcal{S}u_x^\circ\big)$}
  \label{fig:aorta_ux_rec_err}
\end{subfigure}
\label{fig:test}
\begin{subfigure}{.32\textwidth}
  \includegraphics[width=1.\linewidth,,angle=90]{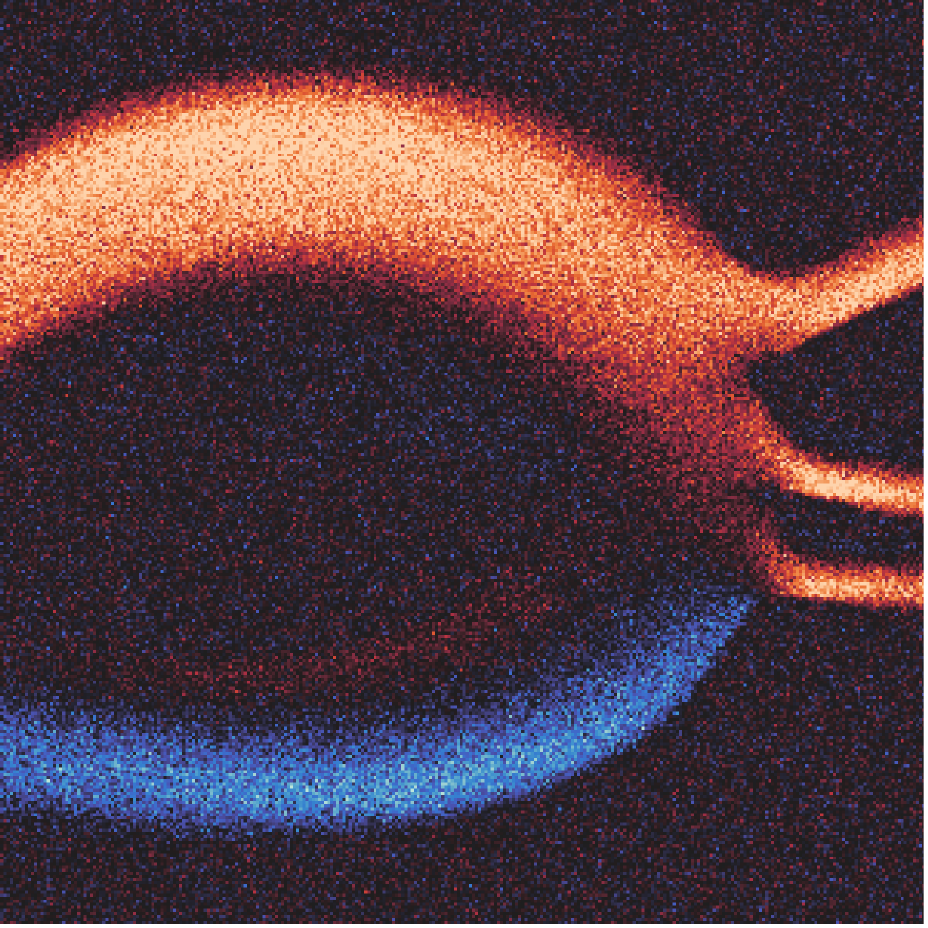}
  \caption{Synthetic image $u^\star_y$}
  \label{fig:aorta_uy_sig}
\end{subfigure}%
\hfill
\begin{subfigure}{.32\textwidth}
  \includegraphics[width=1.\linewidth,,angle=90]{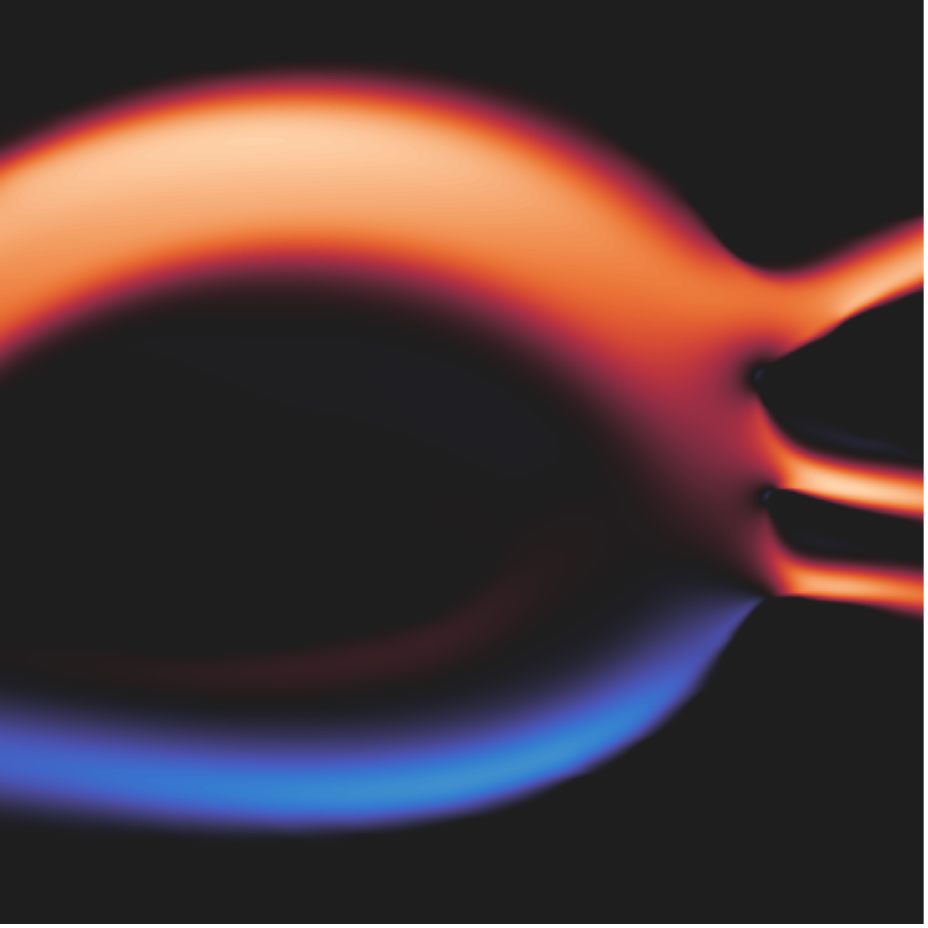}
  \caption{Our reconstruction $u_y^\circ$}
  \label{fig:aorta_uy_rec}
\end{subfigure}%
\hfill
\begin{subfigure}{.32\textwidth}
  \includegraphics[width=1.\linewidth,,angle=90]{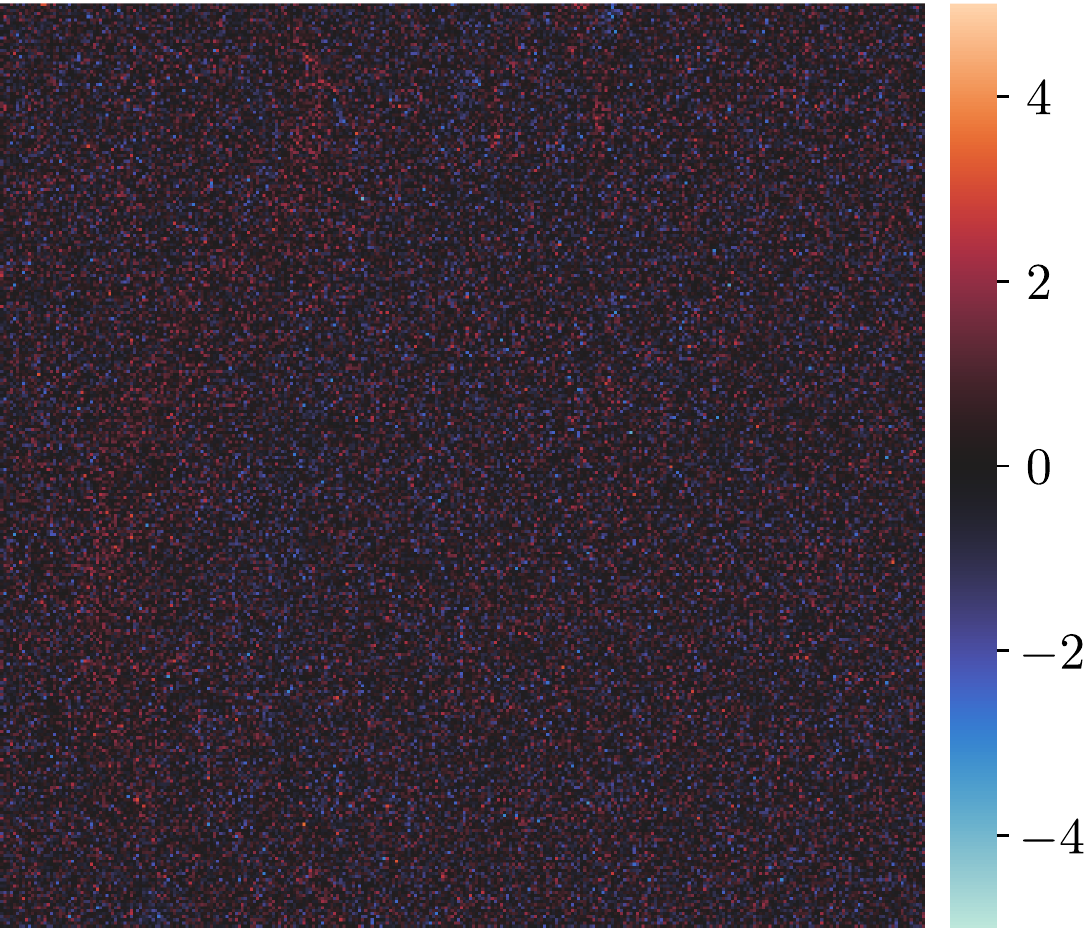}
  \caption{Discrepancy $\sigma_{u_y}^{-1}\big(u^\star_y - \mathcal{S}u_y^\circ\big)$}
  \label{fig:aorta_uy_rec_err}
\end{subfigure}%
\caption{Reconstruction (final iteration of algorithm \ref{algo:reconstruction}) of synthetic noisy velocity images depicting the flow in the simulated 2D model of an aortic aneurysm. Figures \ref{fig:aorta_ux_sig}-\ref{fig:aorta_ux_rec} and \ref{fig:aorta_uy_sig}-\ref{fig:aorta_uy_rec} show the horizontal, $u_x$, and vertical, $u_y$, velocities and share the same colormap (colorbar not shown). Figures \ref{fig:aorta_ux_rec_err} and \ref{fig:aorta_uy_rec_err} show the discrepancy between the noisy velocity images and the reconstruction (colorbars apply only to figures \ref{fig:aorta_ux_rec_err} and \ref{fig:aorta_uy_rec_err}).}
\label{fig:aorta_rec_results1}
\end{figure}

\begin{figure}
\centering
\begin{subfigure}{0.49\textwidth}
\centering
  \includegraphics[height=0.85\linewidth]{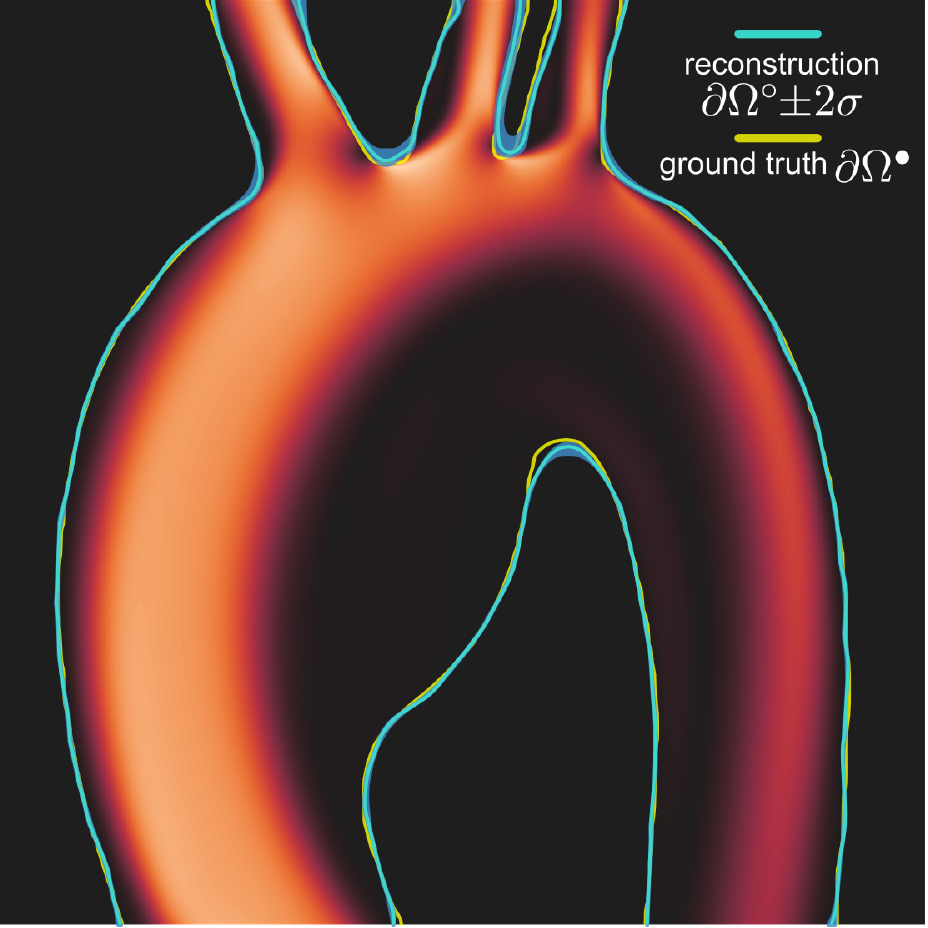}
  \caption{Velocity magnitude $\abs{\bm{u}^\circ}$ and shape $\partial\Omega^\circ$}
  \label{fig:aorta_shape_rec}
\end{subfigure}%
\hfill
\begin{subfigure}{0.49\textwidth}
\centering
  \includegraphics[height=0.85\linewidth,trim=-5 0 0 0,clip]{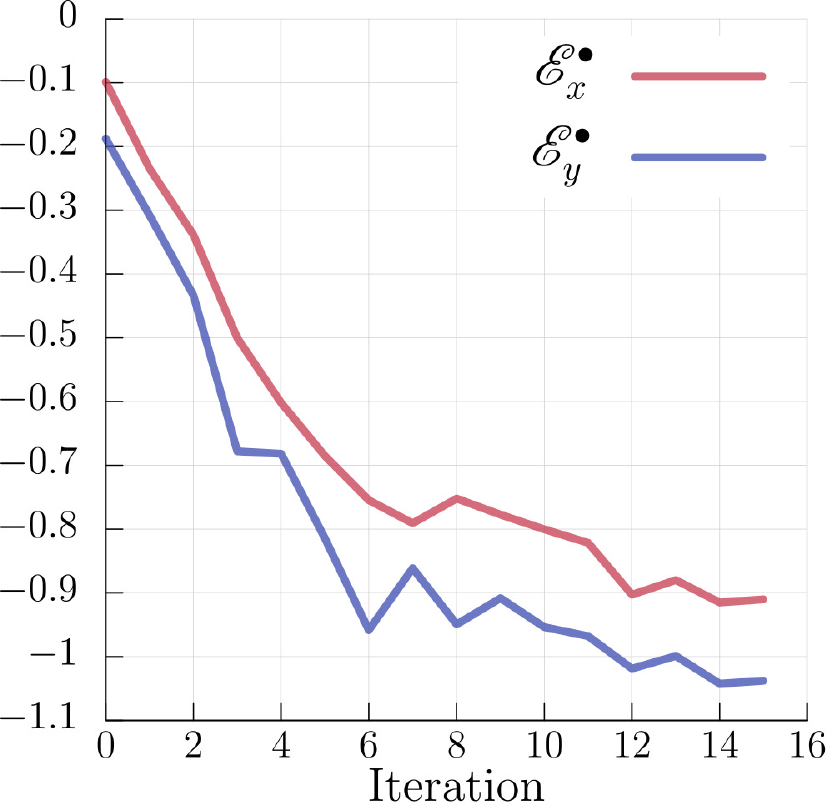}
  \caption{Reconstruction error history}
  \label{fig:aorta_conv}
\end{subfigure}%
\caption{Reconstruction (final iteration of algorithm \ref{algo:reconstruction}) of synthetic images depicting the flow in the simulated 2D model of an aortic aneurysm. Figure \ref{fig:aorta_shape_rec} depicts the reconstructed boundary $\partial\Omega^\circ$ (cyan line), the $2\sigma$ confidence region computed from the approximated posterior covariance $\widetilde{\mathcal{C}}_\zetaext \equiv \widetilde{H}_\zetaext\mathcal{C}_\sdist$ (blue region), and the ground truth boundary $\partial\Omega^\bullet$ (yellow line). Figure \ref{fig:aorta_conv} shows the reconstruction error as a function of iteration number.}
\label{fig:aorta_rec_results2}
\end{figure}

\begin{figure}
\begin{subfigure}{.32\textwidth}
\includegraphics[width=\linewidth,angle=90]{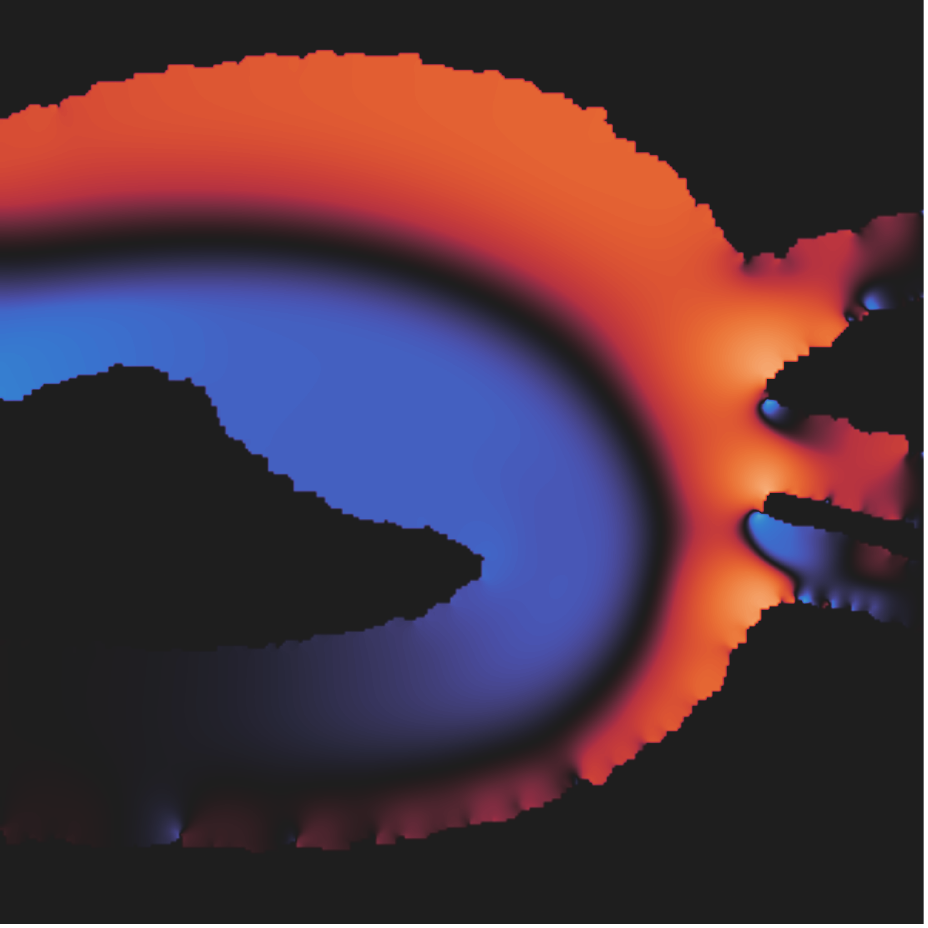}
\caption{Zeroth iteration $p_0$}
\label{fig:aorta_press_init}
\end{subfigure}
\hfill
\begin{subfigure}{.32\textwidth}
\includegraphics[width=\linewidth,angle=90]{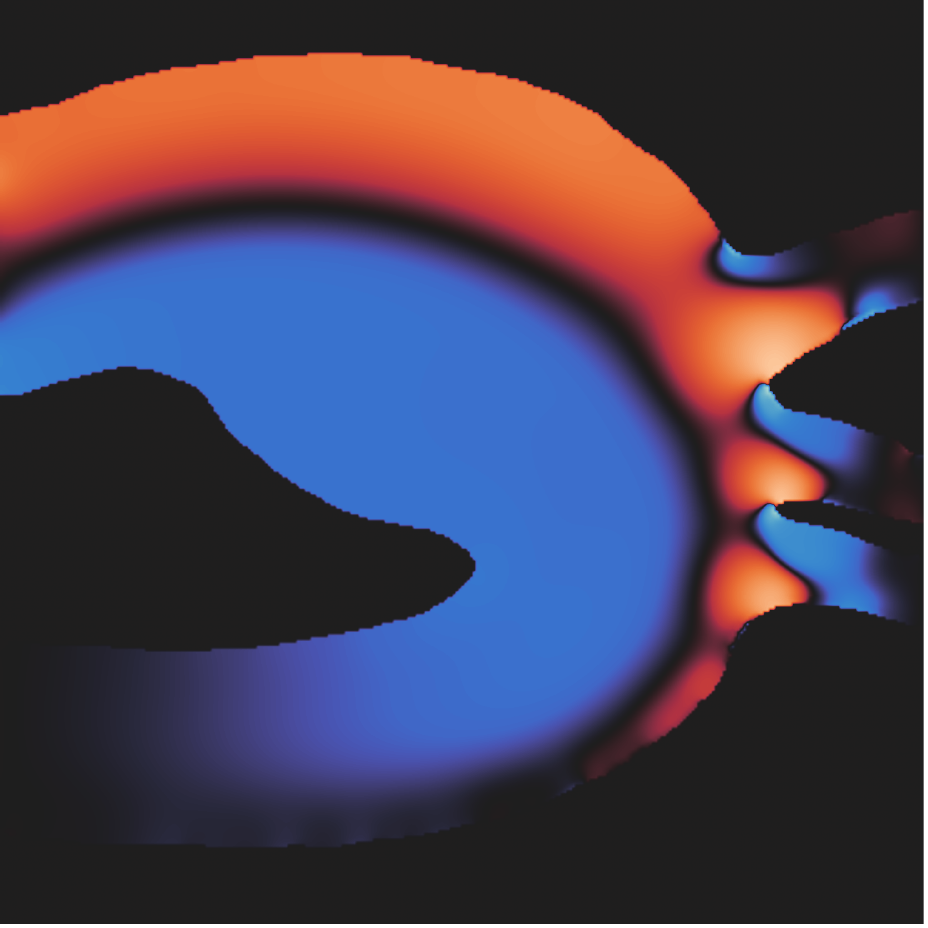}
\caption{Our reconstruction $p^\circ$}
\label{fig:aorta_press_rec}
\end{subfigure}
\hfill
\begin{subfigure}{.32\textwidth}
  \includegraphics[width=\linewidth,angle=90]{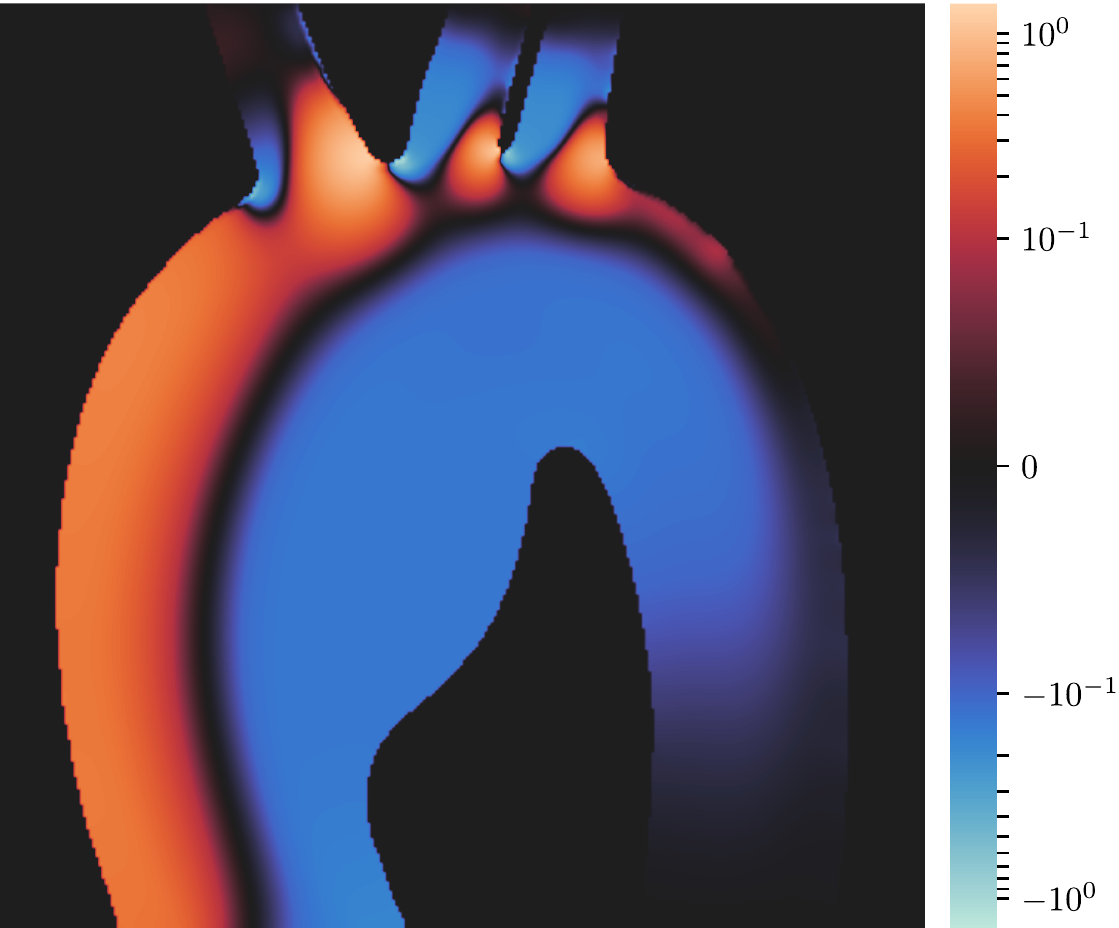}
  \caption{Ground truth $p^\bullet$}
  \label{fig:aorta_press_gt}
\end{subfigure}%
\hfill
\caption{(a) Zeroth iteration (N--S solution for the initial guesses in figure \ref{fig:aorta_init_guesses}), (b) reconstructed (final iteration of algorithm \ref{algo:reconstruction}), and (c) ground truth reduced hydrodynamic pressure for the simulated 2D model of an aortic aneurysm in figure \ref{fig:aorta_rec_results2}. All subfigures share the same colormap (symmetric logarithmic scale) and the same colorbar.}
\label{fig:aorta_press}
\end{figure}

\begin{figure}
\begin{subfigure}{.32\textwidth}
  \includegraphics[height=1.\linewidth,angle=90]{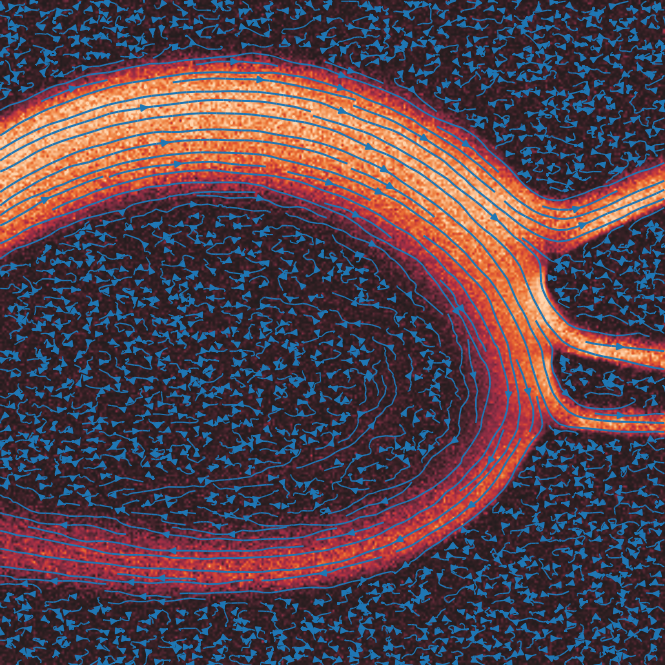}
  \caption{Synthetic data $\bm{u}^\star$}
  \label{fig:aorta_stream}
\end{subfigure}%
\hfill
\begin{subfigure}{.32\textwidth}
  \includegraphics[height=1.\linewidth,angle=90]{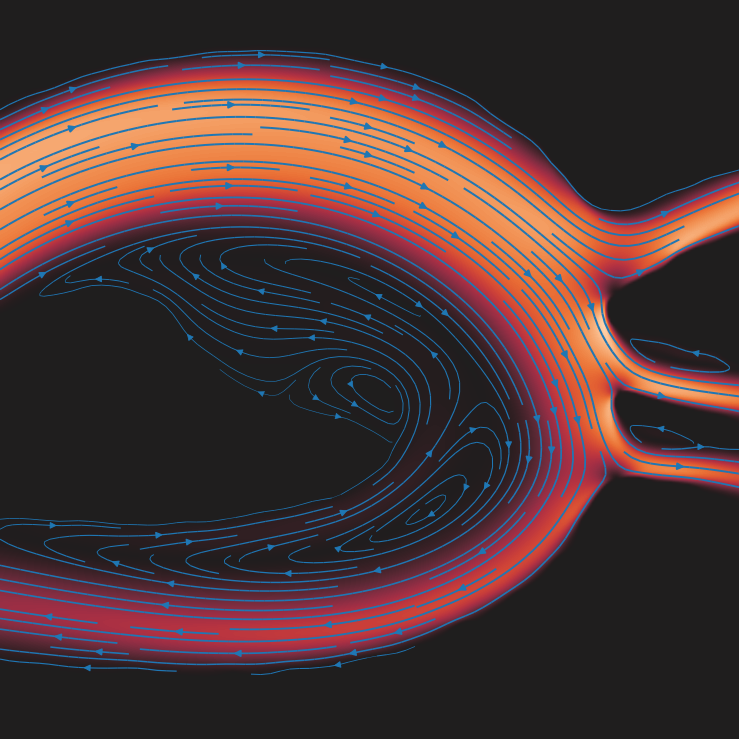}
  \caption{Our reconstruction $\bm{u}^\circ$}
  \label{fig:aorta_stream}
\end{subfigure}%
\hfill
\begin{subfigure}{.32\textwidth}
  \includegraphics[height=1.\linewidth,angle=90]{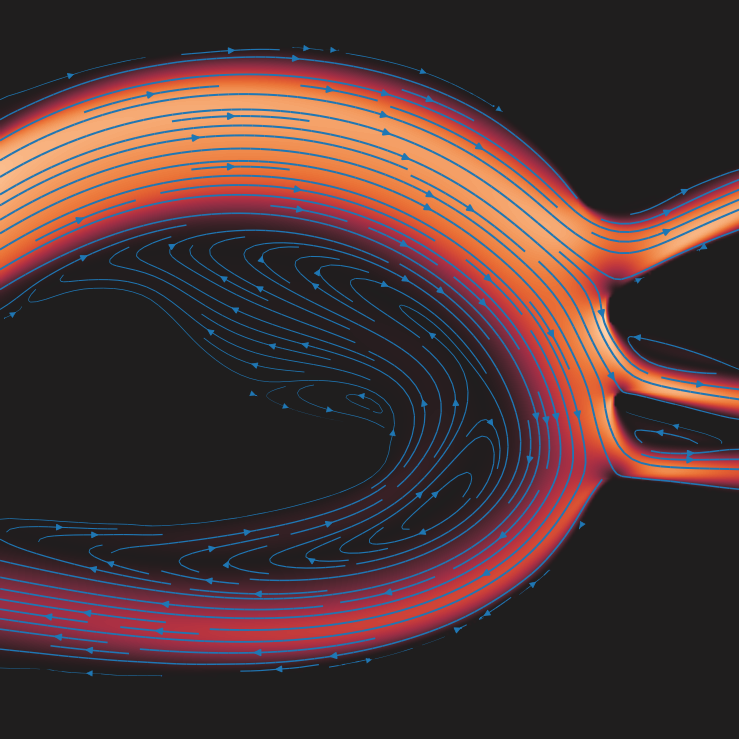}
  \caption{Ground truth $\bm{u}^\bullet$}
  \label{fig:aorta_stream}
\end{subfigure}\\
\begin{subfigure}{.32\textwidth}
  \includegraphics[height=1.\linewidth,angle=90]{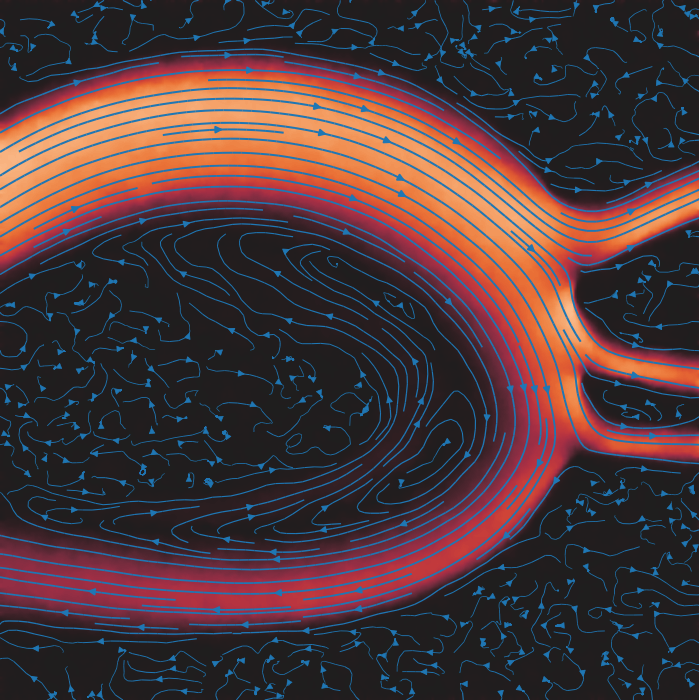}
  \caption{TV-B $\lambda/\lambda_0=0.1$}
  \label{fig:aorta_stream}
\end{subfigure}%
\hfill
\begin{subfigure}{.32\textwidth}
  \includegraphics[height=1.\linewidth,angle=90]{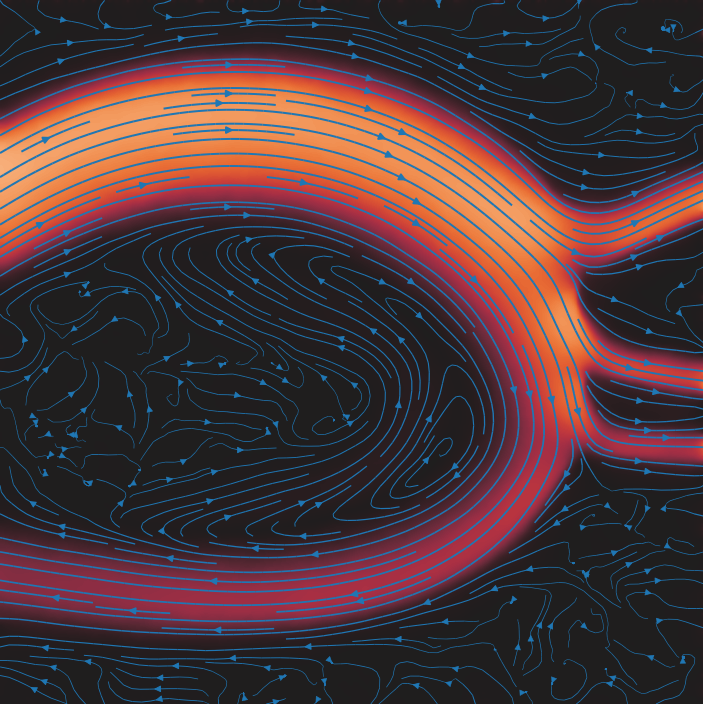}
  \caption{TV-B $\lambda/\lambda_0=0.01$}
  \label{fig:aorta_stream}
\end{subfigure}%
\hfill
\begin{subfigure}{.32\textwidth}
  \includegraphics[height=1.\linewidth,angle=90]{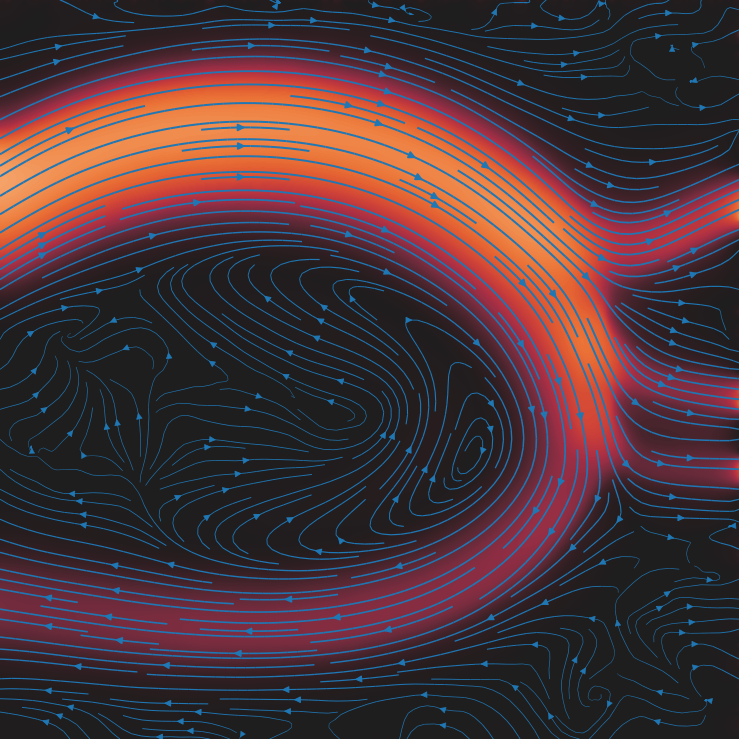}
  \caption{TV-B $\lambda/\lambda_0=0.001$}
  \label{fig:aorta_stream}
\end{subfigure}%
\caption{Streamlines for the flow in the simulated 2D model of an aortic aneurysm (figures \ref{fig:aorta_rec_results1} and \ref{fig:aorta_rec_results2}), and comparison with total variation denoising using Bregman iteration (TV-B) with different weights $\lambda$. Streamlines are plotted on top of the velocity magnitude image, and streamline thickness increases as the velocity magnitude increases.}
\label{fig:aorta_tvb_comp}
\end{figure}

\revv{The algorithm manages to reconstruct and segment the noisy flow images in 15 iterations, with total reconstruction error $\mathcal{E}^\bullet \simeq 5.73\%$. The results are presented in figures \ref{fig:aorta_rec_results1} and \ref{fig:aorta_rec_results2}. We observe that the discrepancy of the last iteration (figures \ref{fig:aorta_ux_rec_err}, \ref{fig:aorta_uy_rec_err}) consists mainly of Gaussian white noise. Some correlations are visible in the discrepancy of the $x$-velocity component near the stagnation points of the upper branches, but these correlations are explained by the extra uncertainty in the predicted shape $\partial\Omega^\circ$ (figure \ref{fig:aorta_shape_rec}). By comparing figures \ref{fig:aorta_ux_rec_err} and \ref{fig:aorta_uy_rec_err} with figures \ref{fig:aorta_u_x_0_discr} and \ref{fig:aorta_u_y_0_discr}, we confirm that the algorithm has successfully assimilated the remaining information from the noisy velocity measurements. \revv{Figure \ref{fig:aorta_press} shows the pressure of the zeroth iteration (figure \ref{fig:aorta_press_init}), and the reconstructed pressure $p^\circ$ (figure \ref{fig:aorta_press_rec}), which compares well to the ground truth pressure $p^\bullet$ (figure \ref{fig:aorta_press_gt}).}}

\revv{
We further compare the performance of algorithm \ref{algo:reconstruction} with a state-of-the-art image denoising algorithm, namely total variation denoising using Bregman iteration (\mbox{TV-B}) \citep{Pascal2012,VanDerWalt2014}, in figure \ref{fig:aorta_tvb_comp}. We first observe that algorithm \ref{algo:reconstruction} denoises the velocity field without losing contrast near the walls of the aorta, and accurately identifies the low-speed vortical structure within the aneurysm, which is obscured by noise. We then test three different values of the TV-B parameter $\lambda/\lambda_0$\footnote{The parameter $\lambda_0=\lambda_0(\sigma)$, where $\sigma$ is the noise standard deviation in the image, is given by \cite{Pascal2012} as an optimal value for $\lambda$.}, which controls the total variation regularization, and observe that, even though TV-B manages to denoise the velocity field and reveal certain large scale vortices, there is considerable loss of contrast near the walls of the aorta and a systematic error (e.g. decreasing peak velocity) that increases as $\lambda$ decreases.}

\begin{figure}
\begin{subfigure}{.32\textwidth}
\includegraphics[width=\linewidth,angle=90]{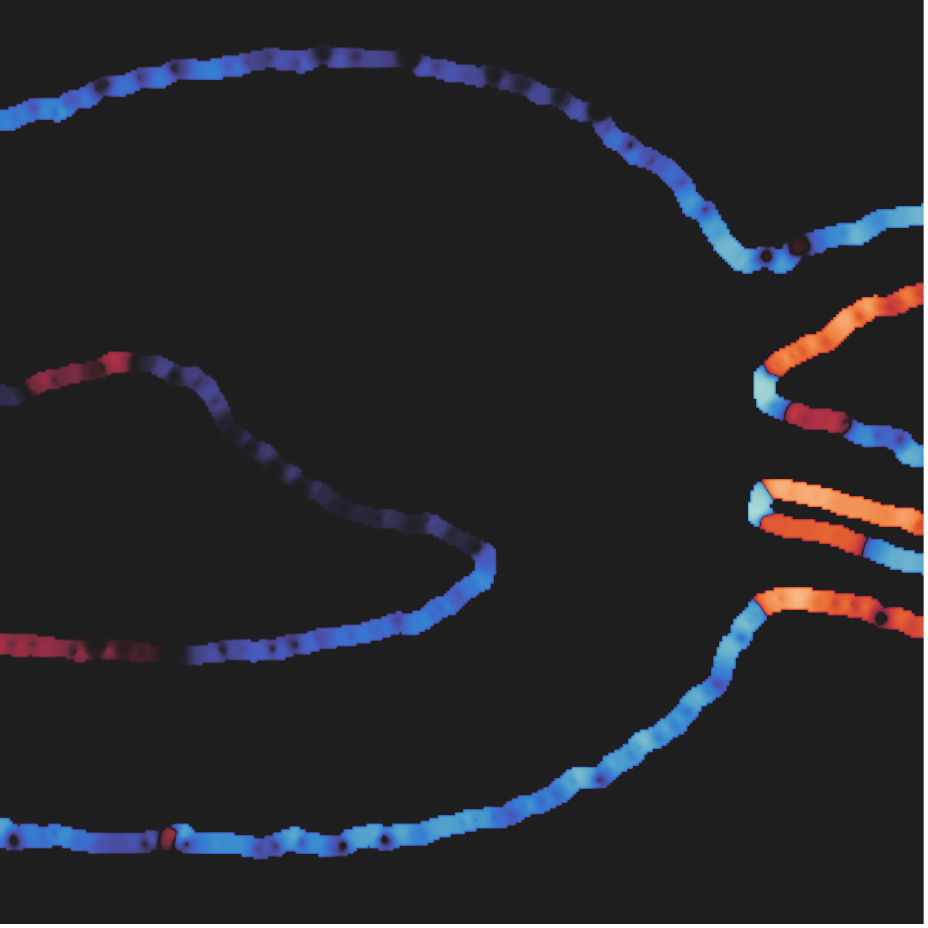}
\caption{Zeroth iteration $(\gamma_w)_0$}
\label{fig:wss_aorta_init}
\end{subfigure}
\hfill
\begin{subfigure}{.32\textwidth}
  \includegraphics[width=\linewidth,angle=90]{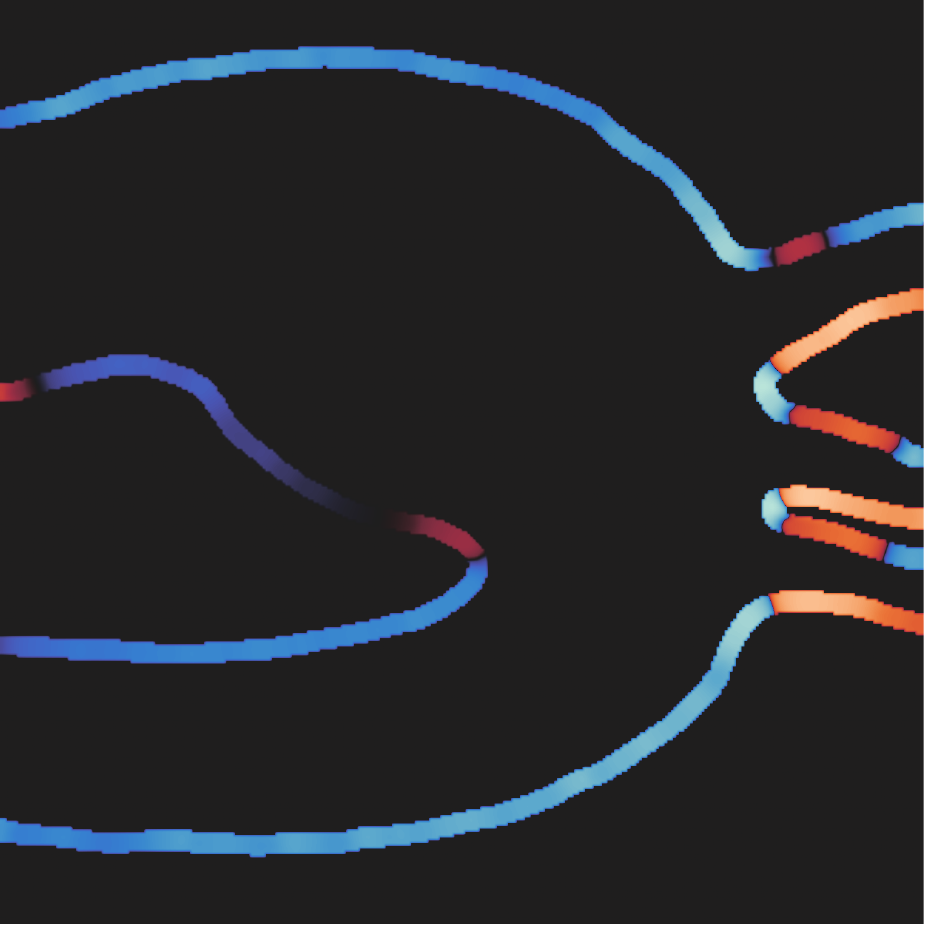}
  \caption{Our reconstruction $\gamma^\circ_w$}
  \label{fig:wss_aorta_rec}
\end{subfigure}%
\hfill
\begin{subfigure}{.32\textwidth}
  \includegraphics[width=\linewidth,angle=90]{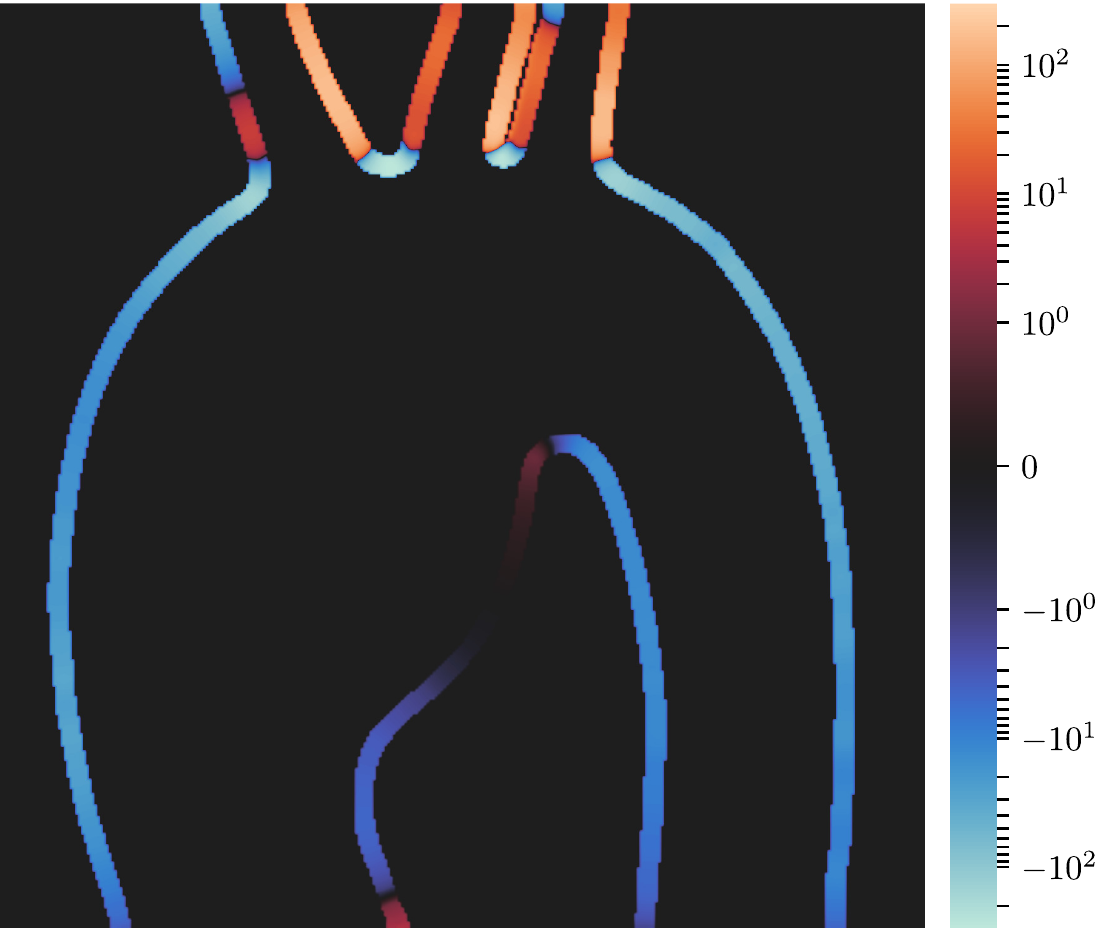}
  \caption{Ground truth $\gamma^\bullet_w$}
  \label{fig:wss_aorta_gt}
\end{subfigure}\\
\centering
\begin{subfigure}{.32\textwidth}
  \includegraphics[width=\linewidth,angle=90]{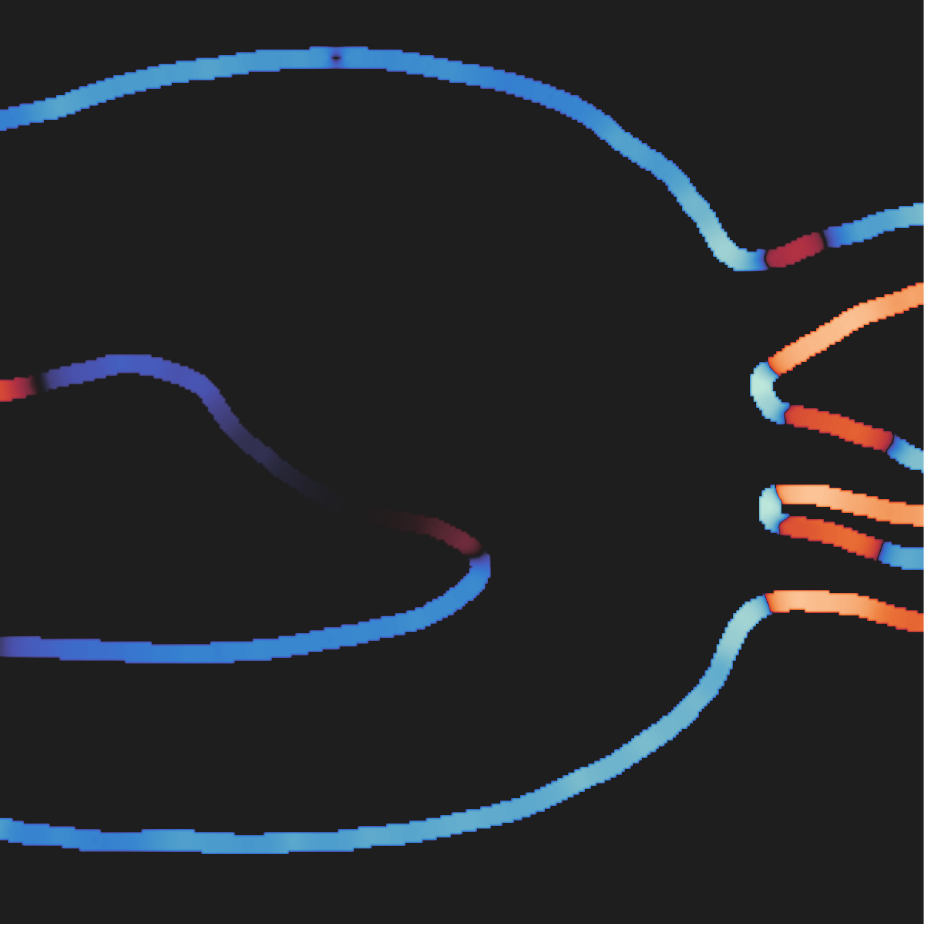}
  \caption{Lower conf. bound $\gamma^\circ_w-2\sigma$}
  \label{fig:wss_aorta_lower_conf_bound}
\end{subfigure}%
\hspace{0.2cm}
\begin{subfigure}{.32\textwidth}
  \includegraphics[width=\linewidth,angle=90]{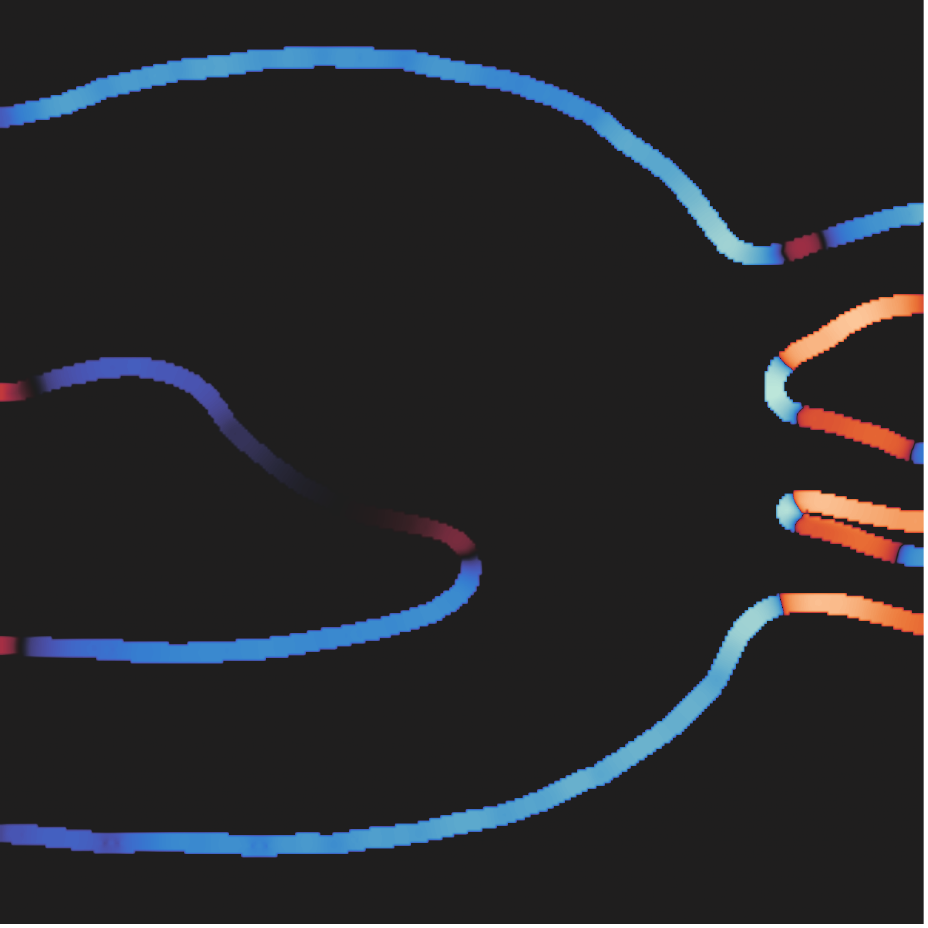}
  \caption{Upper conf. bound $\gamma^\circ_w+2\sigma$}
  \label{fig:wss_aorta_upper_conf_bound}
\end{subfigure}%
\hfill
\caption{Wall shear rate $\gamma_w \equiv \bm{\tau}\bm{\cdot}\partial_{\bm{\nu}}\bm{u}$, where $\bm{\tau}$ is the unit tangent vector of $\partial\Omega$, for the flow in the simulated 2D model of an aortic aneurysm in figure \ref{fig:aorta_rec_results2}. The wall shear stress is found by multiplying this by the viscosity. The reconstructed wall shear rate ($\gamma_w^\circ$) is calculated on $\partial\Omega^\circ$ and for $\bm{u}^\circ$, while the ground truth ($\gamma_w^\bullet$) is calculated on $\partial\Omega^\bullet$ and for $\bm{u}^\bullet$. The $\pm2\sigma$-bounds are calculated on the upper ($\partial\Omega^\circ_+$) and lower ($\partial\Omega^\circ_-$) limits of the confidence region of $\partial\Omega^\circ$. All subfigures share the same colormap (symmetric logarithmic scale) and the same colorbar.}
\label{fig:wss_aorta}
\end{figure}

\revv{Using the reconstructions $\bm{u}^\circ$ and $\partial\Omega^\circ$ we compute the reconstructed wall shear rate ($\gamma_w^\circ$) and compare it with the ground truth ($\gamma_w^\bullet$) (figure \ref{fig:wss_aorta}). We observe that $\gamma_w^\circ$ approximates $\gamma_w^\bullet$ well, and that discrepancies are well accounted for by the \mbox{$\gamma_w^\circ\pm2\sigma$-bounds}.}

\subsection{Magnetic resonance velocimetry experiment}
\label{sec:mrv_experiment}
\begin{figure}
\centering
\begin{subfigure}{0.5\textwidth}
\centering
  \includegraphics[height=\textwidth]{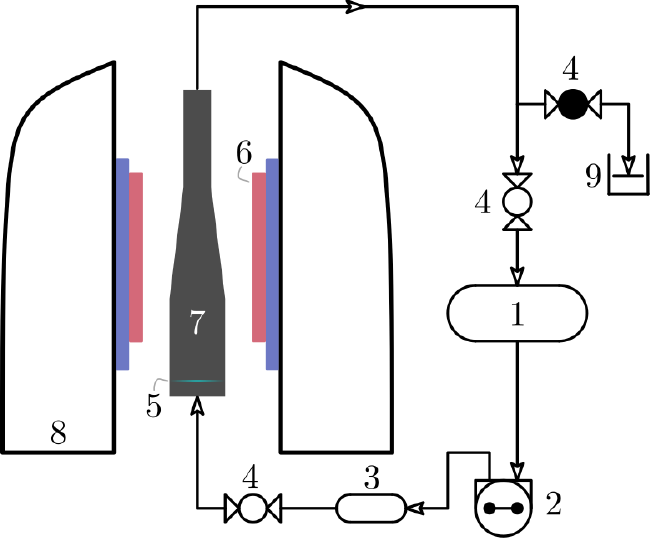}
  \caption{Magnet and flow loop}
  \label{fig:mrv_flow_loop}
\end{subfigure}%
\hfill
\begin{subfigure}{0.5\textwidth}
\centering
  \includegraphics[height=1.005\textwidth]{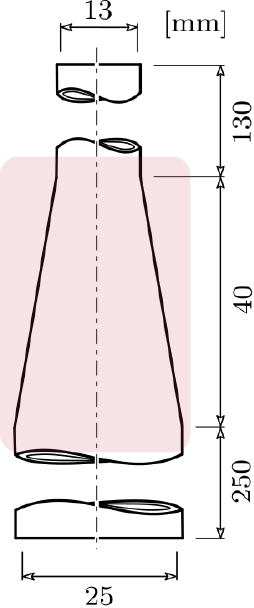}
  \caption{Converging nozzle}
  \label{fig:mrv_nozzle}
\end{subfigure}%
\hfill
\begin{subfigure}{\textwidth}
\centering
  \includegraphics[width=0.5\textwidth]{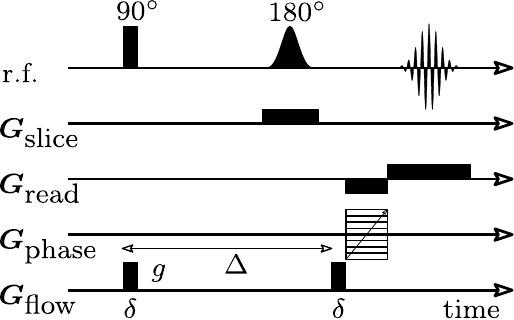}
  \caption{Spin-echo pulse sequence with slice selective refocusing and flow encoding}
  \label{fig:mrv_pulse_seq}
\end{subfigure}
\caption{Schematic of the rig that we use to conduct the MRV experiment consisting of: (1) 20L holding tank, (2) peristaltic pump, (3) 2L vessel, (4) clamp valves, (5) porous polyethylene distributor, (6) radiofrequency probe, (7) converging nozzle, (8) $4.7$T superconducting magnet, (9) volumetric cylinder for flow measurements. Figure \ref{fig:mrv_nozzle} shows a sketch of the converging nozzle with the active area of the spectrometer shown by a red box. The pulse sequence that we use for 2D velocity imaging is shown in figure \ref{fig:mrv_pulse_seq}.}
\label{fig:mrv_experiment}
\end{figure}

We measured the flow through a converging nozzle using magnetic resonance velocimetry \citep{Fukushima1999,Mantle2003,Elkins2007}. The nozzle converges from an inner diameter of 25mm to an inner diameter of 13mm, over a length of 40mm (figure \ref{fig:mrv_nozzle}). On either side of the converging section, the entrance-to-exit length equals 10 times the local diameter (figure \ref{fig:mrv_nozzle}) in order to ensure the absence of entrance/exit effects. We acquired velocity images for a Reynolds number of 162 (defined at the nozzle outlet). We used a \mbox{40 wt\%} glycerol in water solution \citep{Cheng2008,Volk2018} as the working fluid in order to increase the viscosity and minimize the effect of thermal convection in the resulting velocity field due to the temperature difference between the magnet bore and the working fluid. The nozzle is made of polyoxymethylene to minimize magnetic susceptibility differences between the nozzle wall and the working fluid \citep{Wapler2014}. Figure \ref{fig:mrv_flow_loop} depicts the schematic of the flow loop of the MRV experiment. To pump the water/glycerol solution we used a Watson Marlow 505S peristaltic pump (Watson Marlow, Falmouth UK) with a 2L dampening vessel at its outlet to dampen flow oscillations introduced by the peristaltic pump. To make the flow uniform, we installed porous polyethylene distributor plates (SPC technologies, Fakenham UK) at the entrance and the exit of the nozzle. 

We acquired the velocity images on a Bruker Spectrospin DMX200 with a $4.7$T superconducting magnet, which is equipped with a gradient set providing magnetic field gradients of a maximum strength of 13.1Gcm$^{-1}$ in three orthogonal directions, and a birdcage radiofrequency coil tuned to a ${}^{1}\mathrm{H}$ frequency of 199.7 MHz with a diameter and a length of 6.3cm. To acquire 2D velocity images we used slice-selective spin-echo imaging \citep{Edelstein1980} combined with pulsed gradient spin-echo (PGSE) \citep{Stejskal1965} for motion encoding (figure \ref{fig:mrv_pulse_seq}). We measured each of the three orthogonal velocity components in a 1mm thick transverse slice through the converging section of the nozzle, which is centered along the nozzle centerline. The flow images we acquired have a field of view of \mbox{84.2$\times$28.6mm} at \mbox{512$\times$128} pixels, giving an in-plane resolution of \mbox{165$\times$223$\mu$m}. For velocity measurements in the net flow direction, we used a gradient pulse duration, $\delta$, of 0.3 to 0.5ms and flow observation times, ${\rmDelta}$, of 9 to 12ms. For velocity measurements in the perpendicular to the net flow direction, we used an increased gradient pulse duration, $\delta$, of 1.0ms and an increased observation time, ${\rmDelta}$, of 25 to 30ms, due to the lower velocity magnitudes in this direction. We set the amplitude, $g$, of the flow encoding gradient pulses to $\pm$3Gcm$^{-1}$ for the direction parallel to the net flow and to $\pm$1.5Gcm$^{-1}$ for the direction perpendicular to the net flow, in order to maximize phase contrast whilst avoiding velocity aliasing by phase wrapping. To obtain an image for each velocity component, we took the phase difference between two images acquired with flow encoding gradients having equal magnitude $g$ but opposite signs. To remove any phase shift contributions that are not caused by the flow, we corrected the measured phase shift of each voxel by subtracting the phase shift measured under zero-flow conditions. The gradient stabilization time that we used is 1ms and we acquired the signal with a sweep width of 100kHz. We used hard 90$^\circ$ excitation pulses with a duration of 85$\mu$s, and a 512$\mu$s Gaussian-shaped soft 180$^\circ$ pulse for slice selection and spin-echo refocusing. We found the $T_1$ relaxation time of the glycerol solution to be 702ms, as measured by an inversion recovery pulse sequence. To allow for magnetization recovery between the acquisitions, we used a repetition time of 1.0s. To eliminate unwanted coherences and common signal artefacts, such as DC offset, we used a four step phase cycle. 

\begin{figure}
\centering
\begin{subfigure}{.5\textwidth}
\centering
  \includegraphics[height=0.5\linewidth]{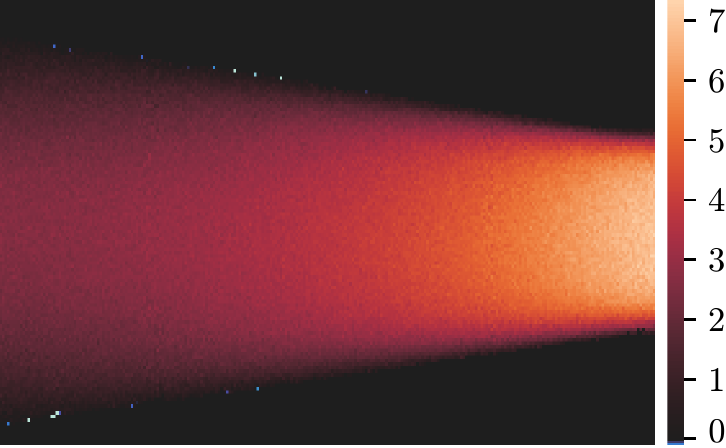}
  \caption{$u^\bullet_z$}
  \label{fig:sub2}
\end{subfigure}%
\hfill
\begin{subfigure}{.5\textwidth}
\centering
  \includegraphics[height=0.5\linewidth]{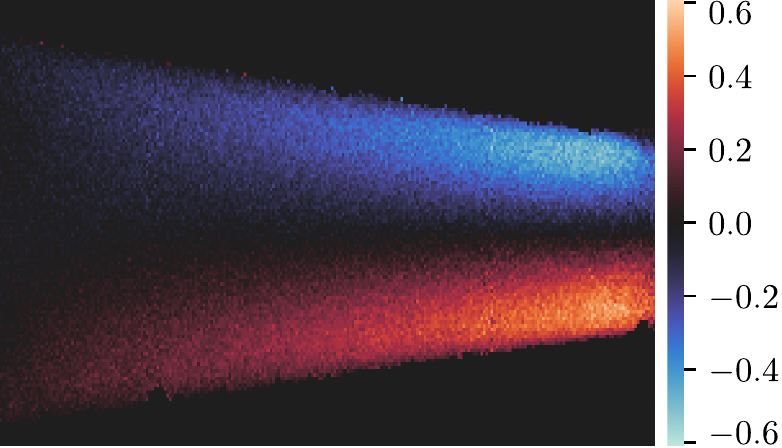}
  \caption{$u^\bullet_r$}
  \label{fig:sub2}
\end{subfigure}
\caption{High SNR images (average of 32 scans with $\text{SNR}_z\simeq44,\text{SNR}_r\simeq34$) that we acquired for the flow through the converging nozzle using MRV (units in [cm/s]).}
\label{fig:high_qual_axisymm_nozzle_re162}
\end{figure}

To be consistent with the standard definition used in MRI/MRV, we define the SNR of each MRV image using \eqref{eq:snr_def}, but with $\mu_x$ replaced by the mean signal intensity (images of the $^1\mathrm{H}$ spin density) over the nozzle domain ($\mu_I$), and $\sigma_{u_x}$ replaced by the standard deviation of the Rayleigh distributed noise in a region with no signal ($\sigma_I$) \citep{Gudbjartsson1995}. The standard deviation for the phase is therefore ${\sigma_\varphi = 1/\text{SNR}}$. The MRV images are acquired by taking the sum/difference of four phase images, and then multiplying by the constant factor $1/2\gamma g\delta\Delta$, where $\gamma$ is the gyromagnetic ratio of $^1\mathrm{H}$ (linear relation between the image phase and the velocity). The error in the MRV measured velocity is therefore $\sigma_u = \sigma_\varphi/\gamma g \delta\Delta$. To acquire high SNR images (figure \ref{fig:high_qual_axisymm_nozzle_re162}), we averaged 32 scans, resulting in a total acquisition time of 137 minutes per velocity image ($\sim4.6$ hours for both velocity components). To evaluate the denoising capability of the algorithm we acquired poor SNR images by averaging only 4 scans (the minimum requirement for a full phase cycle) and decreasing the repetition time to 300ms, resulting in a total acquisition time of 5.1 minutes per velocity image ($10.2$ minutes for both velocity components). 

To verify the quantitative nature of the MRV experiment we compared the volumetric flow rates calculated from the MRV images (using 2D slice-selective velocity imaging in planes normal to the direction of net flow) with the volumetric flow rates measured from the pump outlet. The results agree with an average error of $\pm$1.8\%.

\subsection{Magnetic resonance velocimetry data in a converging nozzle}
\label{sec:axisymm_nozzle_rec}
We now use algorithm \ref{algo:reconstruction} to reconstruct and segment the low SNR images ($\bm{u}^\star$) that we acquired during the MRV experiment (section \ref{sec:mrv_experiment}), and compare them with the high SNR images of the same flow ($\bm{u}^\bullet$ in figure \ref{fig:high_qual_axisymm_nozzle_re162}). The flow is axisymmetric with zero swirl. The subscript `$x$' is replaced by `$z$', which denotes the axial component of velocity, and the subscript `$y$' is replaced by `$r$', which denotes the radial component of velocity. The low SNR images ($\text{SNR}_z = 6.7$, $\text{SNR}_r = 5.8$) required a total scanning time of 5.1 minutes per velocity image (axial and radial components), and the high SNR images ($\text{SNR}_z = 44.2$, $\text{SNR}_r = 34.4$) required a total scanning time of 137 minutes per velocity image. Since the signal intensity of an MRV experiment corresponds to the $^1\mathrm{H}$ spin density, we segment the spin density image using a thresholding algorithm \citep{Otsu1979} in order to obtain a mask $\psi$, such that $\psi = 1$ inside $\Omega$ (the nozzle) and $\psi = 0$ outside $\Omega$. We consider $\psi$ to be the prior information for the geometry of the nozzle, which also serves as an initial guess for $\Omega$ $(\Omega_0)$. For ${\bm{g}_i}_0$ we take a parabolic velocity profile with a peak velocity of $0.6 U$, where $U \simeq 5$ cm/s is the characteristic velocity for this problem. In this case we treat the kinematic viscosity as an unknown, with a prior distribution $\mathcal{N}\big(\bar{\nu},(0.1\bar{\nu})^2\big)$, and $\bar{\nu} = 4\times10^{-6}$m$^2$/s. Note that the axis of the nozzle is not precisely known beforehand, and since we only solve an axisymmetic Navier--Stokes problem on the $z-r$ half-plane, we also introduce an unknown variable for the vertical position of the axis (see appendix \ref{app:axisymm}).

\begin{table}
  \begin{center}
\def~{\hphantom{0}}
  \begin{tabular}{llcccccc}
        & & image dimension   &   model dimension & \multicolumn{2}{c}{$\sigma_{u_z}/U$} & \multicolumn{2}{c}{$\sigma_{u_r}/U$}\\[3pt]
        nozzle & (3D)  & $255\times 128$ & $300\times130$ (half-plane)  & \multicolumn{2}{c}{$1.4168\times10^{-1}$} & \multicolumn{2}{c}{$3.0679\times10^{-2}$}\\

        &   &  &   &  &  & &\\
         \multicolumn{2}{c}{\textit{Regularization}} & $\sigma_\sdist/D$ & $\sigma_{\bm{g}_i}/U$ & $\sigma_\nu/UD$ & $\Rey_\sdist$ & $\Rey_\zeta$ & $\ell/h$\\[3pt]
        nozzle & (3D) &  0.25  &  0.5  &  $6.2\times10^{-4}$    & 0.025 & 0.025 & 3\\
  \end{tabular}
  \caption{Input parameters for the inverse 3D axisymmetric Navier--Stokes problem.}
  \label{tab:input_params_axisymm}
  \end{center}
\end{table}
\begin{figure}
\begin{subfigure}{.32\textwidth}
  \includegraphics[height=0.68\linewidth]{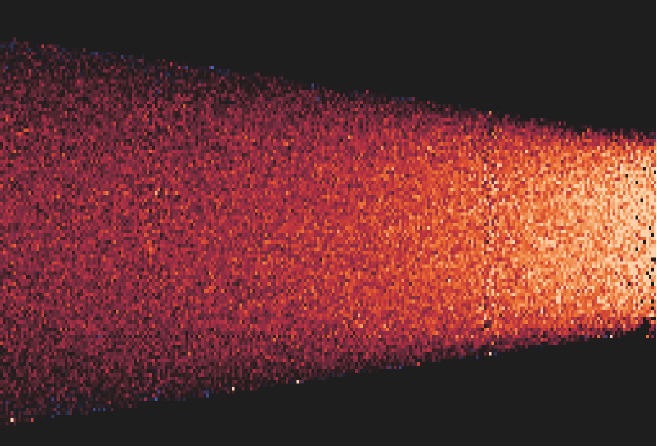}
  \caption{Low SNR MRV image $u^\star_z$}
  \label{fig:axi_ux_sig}
\end{subfigure}%
\hfill
\begin{subfigure}{.32\textwidth}
  \includegraphics[height=0.68\linewidth]{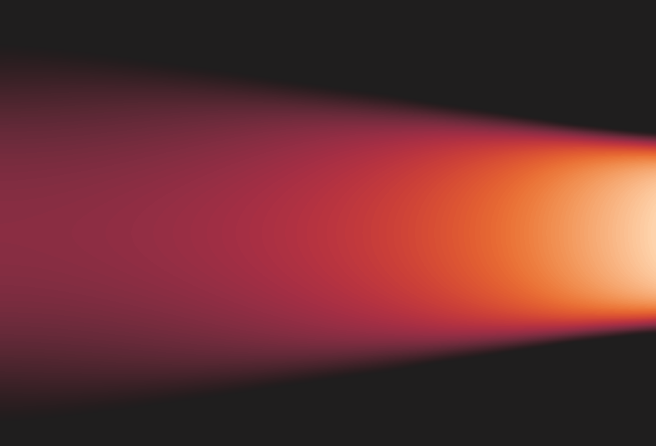}
  \caption{Our reconstruction $u_z^\circ$}
  \label{fig:axi_ux_rec}
\end{subfigure}%
\hfill
\begin{subfigure}{.32\textwidth}
  \includegraphics[height=0.68\linewidth]{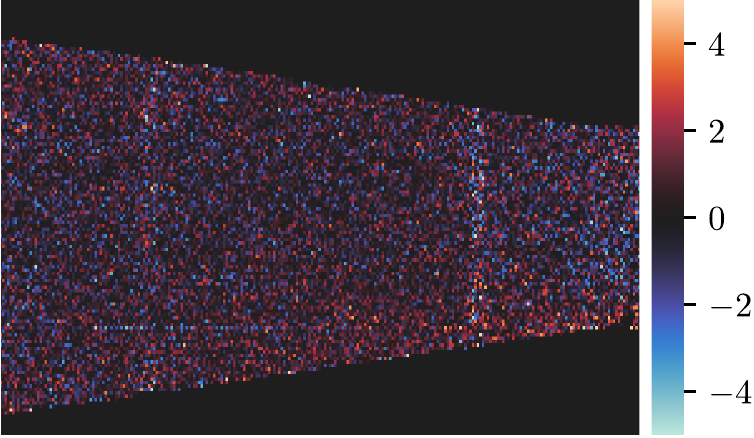}
  \caption{Discrepancy $\sigma_{u_z}^{-1}\big(u^\star_z - \mathcal{S}u_z^\circ\big)$}
  \label{fig:axi_ux_rec_err}
\end{subfigure}
\hfill
\begin{subfigure}{.32\textwidth}
  \includegraphics[height=0.68\linewidth]{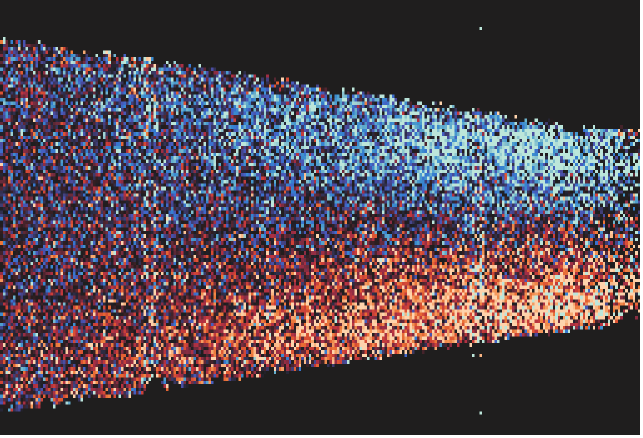}
  \caption{Low SNR MRV image $u^\star_r$}
  \label{fig:axi_uy_sig}
\end{subfigure}%
\hfill
\begin{subfigure}{.32\textwidth}
  \includegraphics[height=0.68\linewidth]{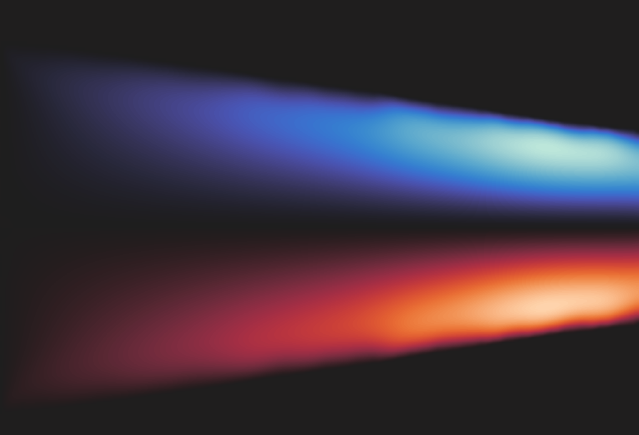}
  \caption{Our reconstruction $u_r^\circ$}
  \label{fig:axi_uy_rec}
\end{subfigure}%
\hfill
\begin{subfigure}{.32\textwidth}
  \includegraphics[height=0.68\linewidth]{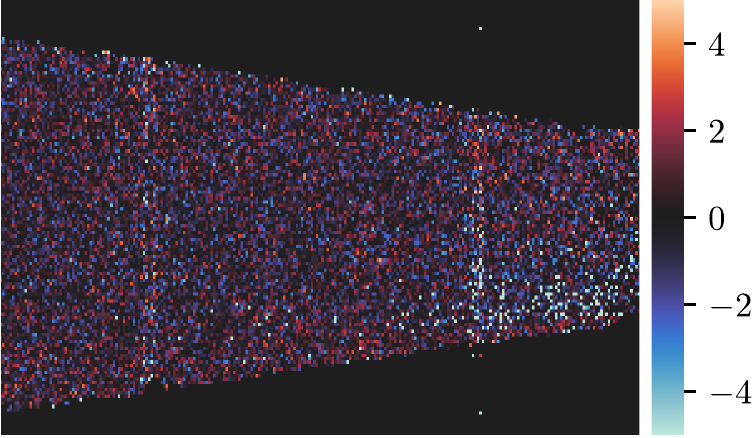}
  \caption{Discrepancy $\sigma_{u_r}^{-1}\big(u^\star_r - \mathcal{S}u_r^\circ\big)$}
  \label{fig:axi_uy_rec_err}
\end{subfigure}
\caption{Reconstruction (algorithm \ref{algo:reconstruction}) of low SNR MRV velocity images depicting the axisymmetric flow (from left to right) in the converging nozzle (figure \ref{fig:mrv_nozzle}). Figures \ref{fig:axi_ux_sig}-\ref{fig:axi_ux_rec} and \ref{fig:axi_uy_sig}-\ref{fig:axi_uy_rec} show the horizontal, $u_x$, and vertical, $u_y$, velocities and share the same colormap (colorbar not shown). Figures \ref{fig:axi_ux_rec_err} and \ref{fig:axi_uy_rec_err} show the discrepancy between the noisy velocity images and the reconstruction (colorbars apply only to figures \ref{fig:axi_ux_rec_err} and \ref{fig:axi_uy_rec_err}) . The reconstructed flow $\bm{u}^\circ$ is axisymmetric by construction, therefore $u^\circ_z$ depicts an even reflection and $u^\circ_r$ depicts an odd reflection, so that they can be compared with the MRV images (see appendix \ref{app:axisymm}).}
\label{fig:axisymm_nozzle_results_1}
\end{figure}

\begin{figure}
\begin{subfigure}{0.55\textwidth}
\centering
  \includegraphics[height=0.525\linewidth]{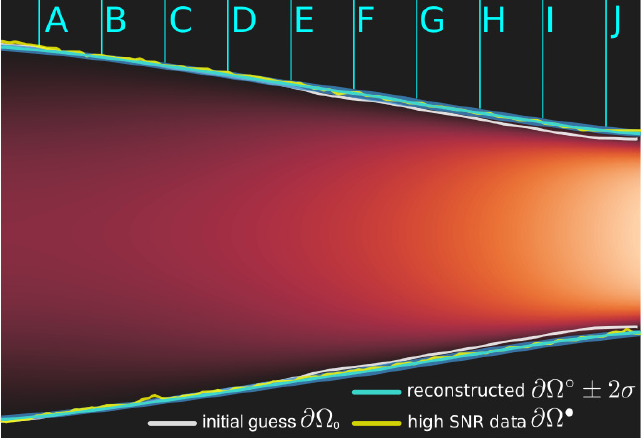}
  \caption{Velocity magnitude $\abs{\bm{u}^\circ}$ and shape $\partial\Omega^\circ$}
  \label{fig:axi_shape_rec}
\end{subfigure}%
\hfill
\begin{subfigure}{0.45\textwidth}
\centering
  \includegraphics[height=0.6416\linewidth,trim=0 0 0 -10,clip]{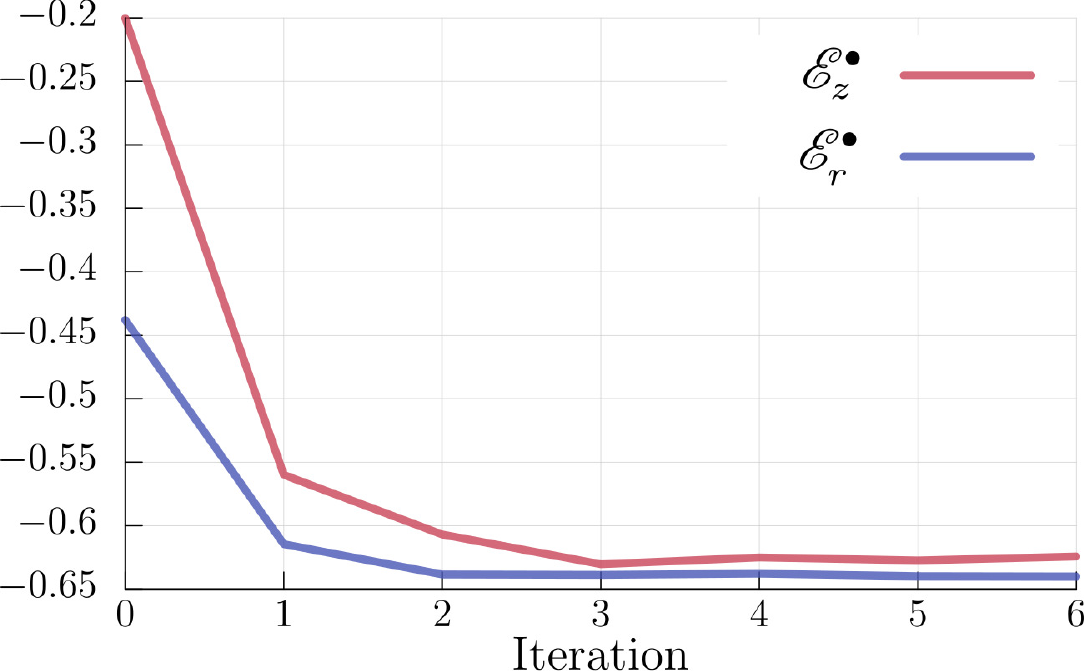}
  \caption{Reconstruction error history}
  \label{fig:axi_conv_plot}
\end{subfigure}%
\hfill
\begin{subfigure}{0.49\textwidth}
\centering
\includegraphics[height=0.82\linewidth]{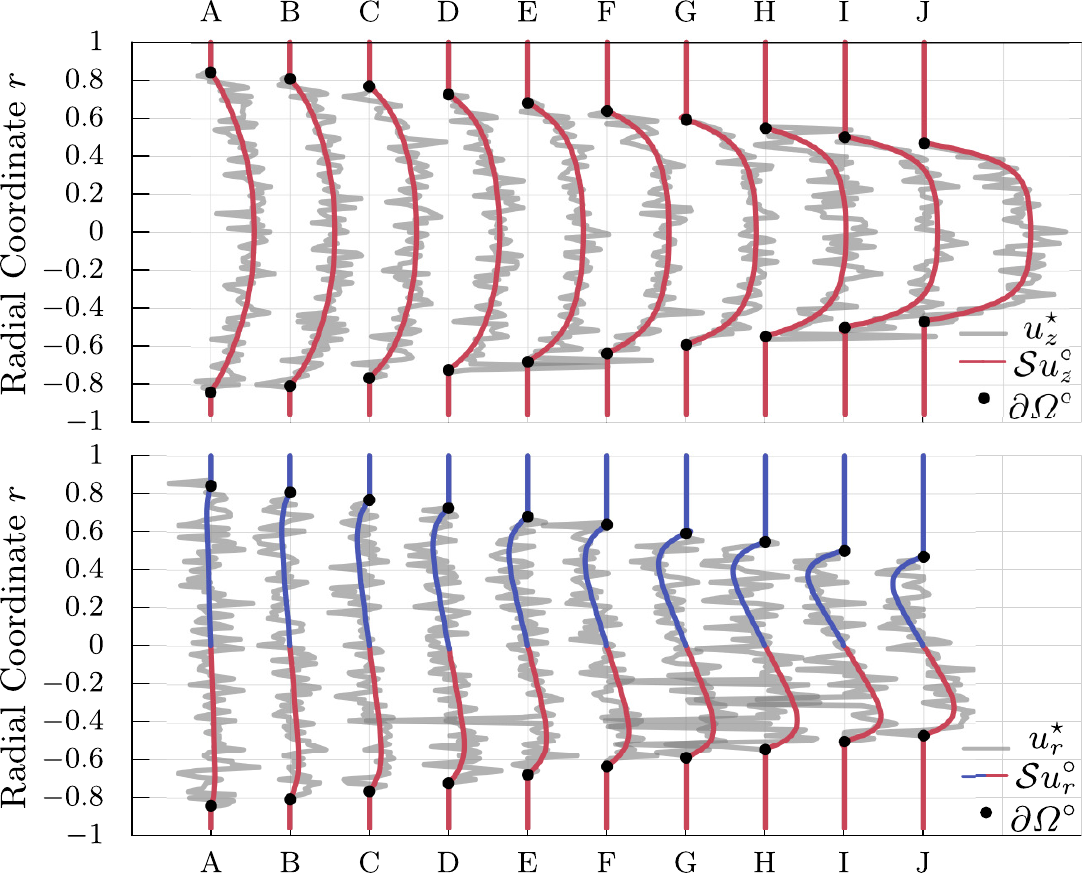}
\caption{Low SNR data (grey) and reconstruction}
\label{fig:axi_slices}
\end{subfigure}%
\hfill
\begin{subfigure}{0.49\textwidth}
\centering
\includegraphics[height=0.82\linewidth]{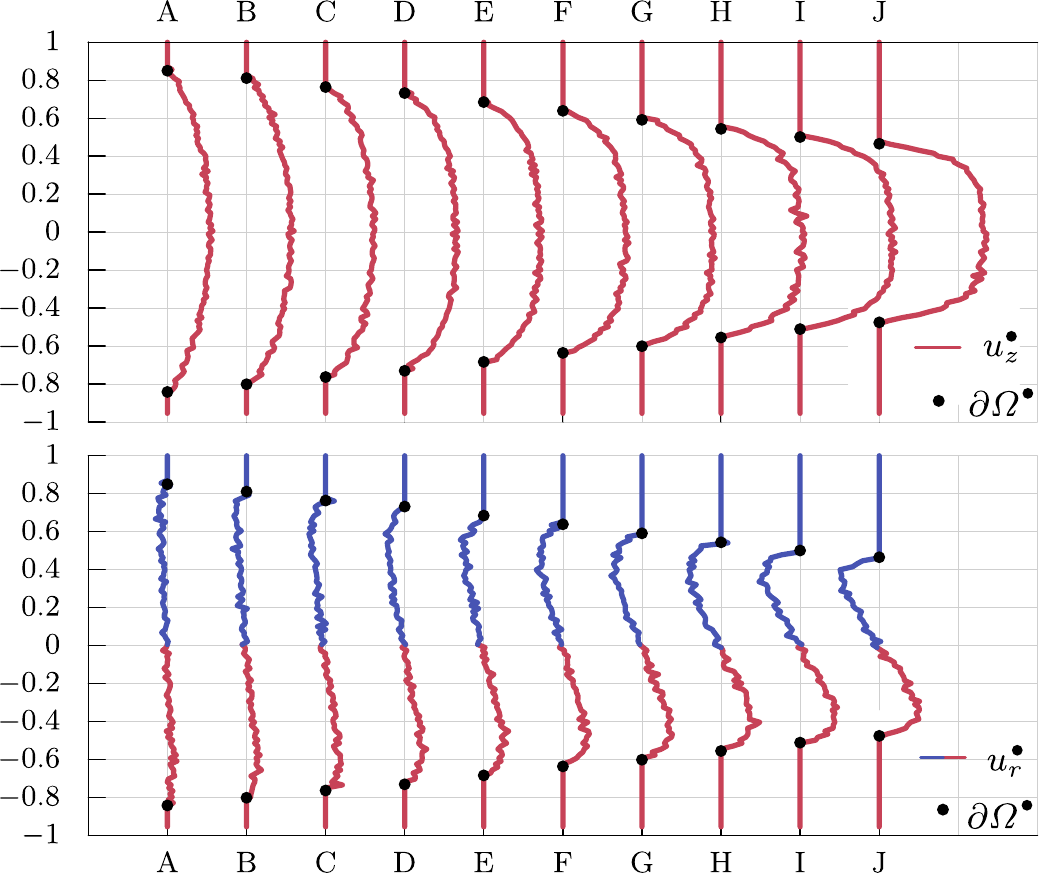}
\caption{High SNR velocity data}
\label{fig:axi_slices_gt}
\end{subfigure}%
\caption{Reconstruction (algorithm \ref{algo:reconstruction}) of synthetic images depicting the axisymmetric flow (from left to right) in the converging nozzle (figure \ref{fig:mrv_nozzle}). Figure \ref{fig:axi_shape_rec} depicts the reconstructed boundary $\partial\Omega^\circ$ (cyan line), the $2\sigma$ confidence region computed from the approximated posterior covariance $\widetilde{\mathcal{C}}_\zetaext \equiv \widetilde{H}_\zetaext\mathcal{C}_\sdist$ (blue region), the ground truth boundary $\partial\Omega^\bullet$ (yellow line), and the initial guess $\partial\Omega_0$ (white line). Figure \ref{fig:axi_conv_plot} shows the reconstruction error as a function of iteration number. Velocity slices are drawn for 10 equidistant cross-sections (labelled with the letters A to J) for both the reconstructed images (figure \ref{fig:axi_slices}) and the high SNR images (figure \ref{fig:axi_slices_gt}), colored red for positive values and blue for negative.}
\label{fig:axisymm_nozzle_results_2}
\end{figure}

\begin{figure}
\begin{subfigure}{.32\textwidth}
  \includegraphics[width=1.\linewidth]{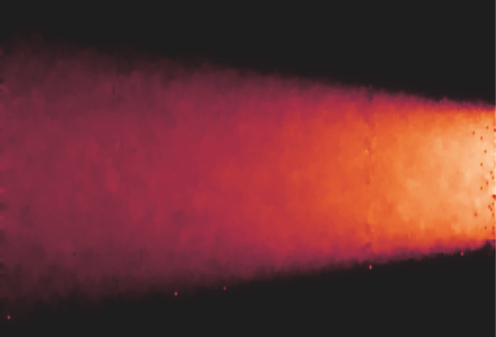}
  \caption{$u_z$, TV-B $\lambda/\lambda_0=0.1$}
  \label{fig:uz_rec_tvb_0p1}
\end{subfigure}%
\hfill
\begin{subfigure}{.32\textwidth}
  \includegraphics[width=1.\linewidth]{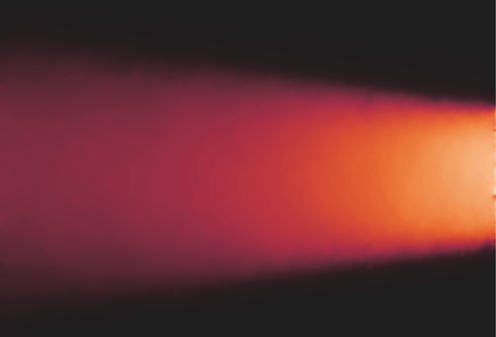}
  \caption{$u_z$, TV-B $\lambda/\lambda_0=0.01$}
  \label{fig:uz_rec_tvb_0p01}
\end{subfigure}%
\hfill
\begin{subfigure}{.32\textwidth}
  \includegraphics[width=1.\linewidth]{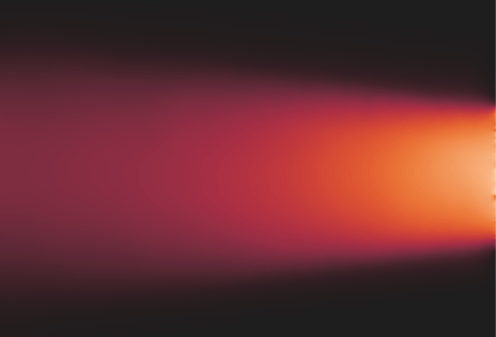}
  \caption{$u_z$, TV-B $\lambda/\lambda_0=0.001$}
  \label{fig:uz_rec_tvb_0p001}
\end{subfigure}\\
\begin{subfigure}{.32\textwidth}
  \includegraphics[width=1.\linewidth]{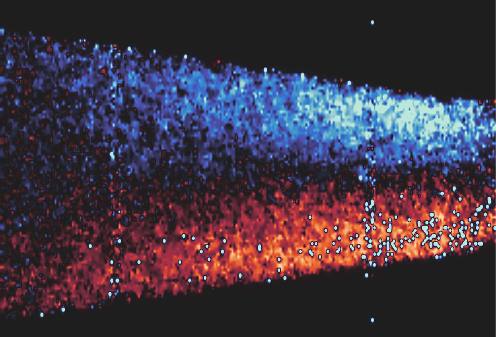}
  \caption{$u_r$, TV-B $\lambda/\lambda_0=0.1$}
  \label{fig:ur_rec_tvb_0p1}
\end{subfigure}%
\hfill
\begin{subfigure}{.32\textwidth}
  \includegraphics[width=1.\linewidth]{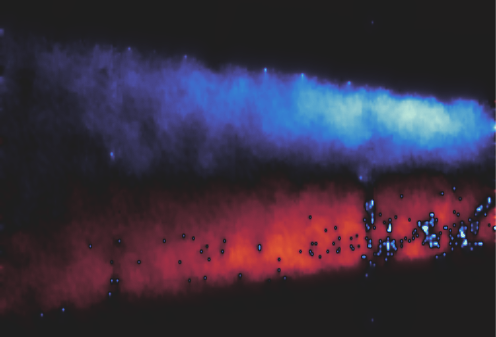}
  \caption{$u_r$, TV-B $\lambda/\lambda_0=0.01$}
  \label{fig:ur_rec_tvb_0p01}
\end{subfigure}%
\hfill
\begin{subfigure}{.32\textwidth}
  \includegraphics[width=1.\linewidth]{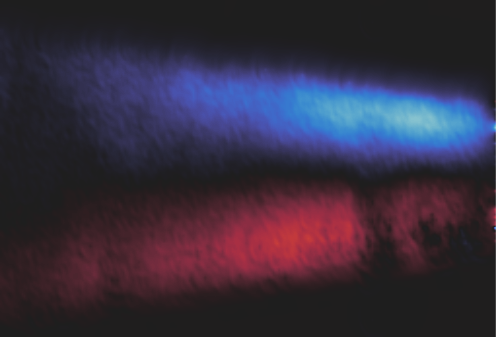}
  \caption{$u_r$, TV-B $\lambda/\lambda_0=0.001$}
  \label{fig:ur_rec_tvb_0p001}
\end{subfigure}%
\caption{Total variation denoising using Bregman iteration with different weights $\lambda$ for the low SNR MRV images (figures \ref{fig:axi_ux_sig} and \ref{fig:axi_uy_sig}) depicting the axisymmetric flow (from left to right) in the converging nozzle.}
\label{fig:axisymm_nozzle_tvb_comp}
\end{figure}

\begin{figure}
\centering
\begin{subfigure}{.6\textwidth}
\includegraphics[height=0.427\textwidth]{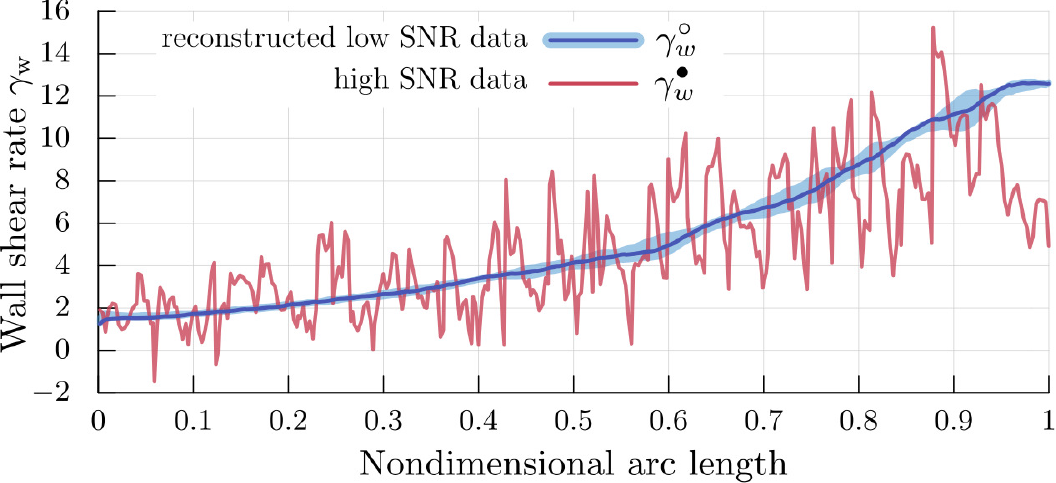}
\caption{Wall shear rates $\gamma^\circ_w$ and $\gamma^\bullet_w$}
\label{fig:axisymm_wss}
\end{subfigure}
\hfill
\begin{subfigure}{.39\textwidth}
\includegraphics[height=0.657\textwidth,trim=0 -5 -5 -5,clip]{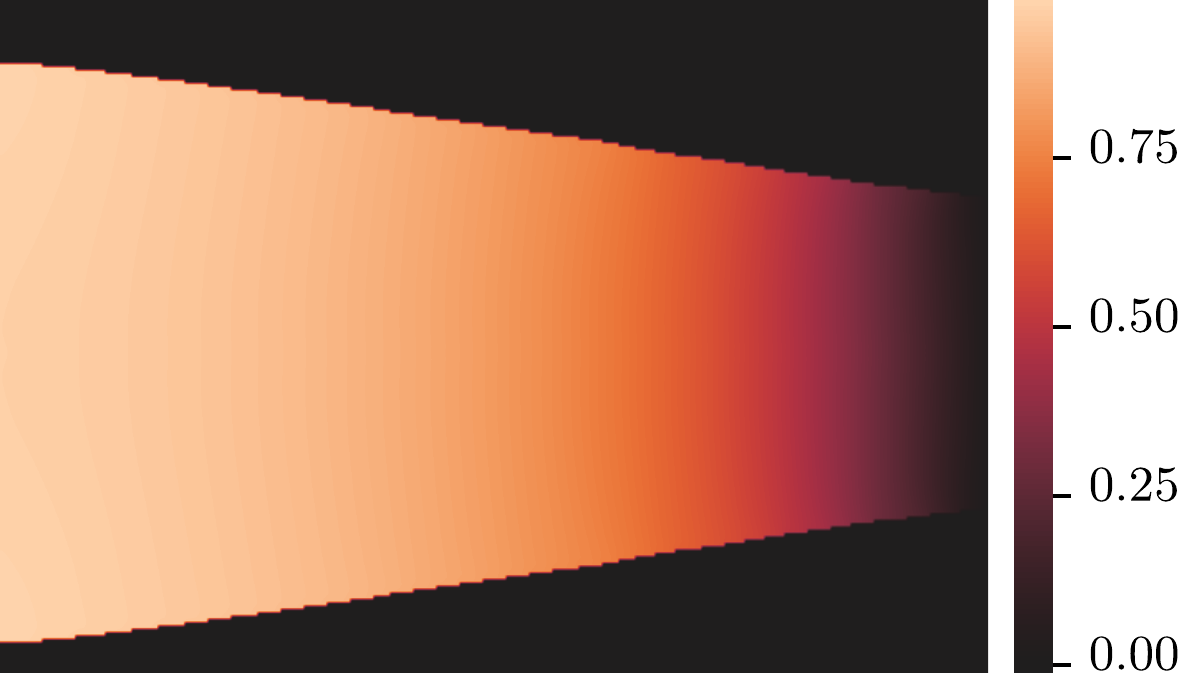}
\caption{Our pressure reconstruction $p^\circ$}
\label{fig:axisymm_pres}
\end{subfigure}
\caption{(a) Wall shear rates (as for figure \ref{fig:wss_toy_problem_1}) and (b) reduced hydrodynamic pressure inferred from the MRV images depicting the axisymmetric flow in the converging nozzle.}
\end{figure}

Using the input parameters of table \ref{tab:input_params_axisymm}, the algorithm manages to reconstruct the noisy velocity image and reduce segmentation errors in just 6 iterations, with total reconstruction error $\mathcal{E}^\bullet \simeq 5.94\%$. The results are presented in figures \ref{fig:axisymm_nozzle_results_1} and \ref{fig:axisymm_nozzle_results_2}. We observe that algorithm \ref{algo:reconstruction} manages to filter out the noise, the outliers, and the acquisition artefacts of the low SNR MRV images depicting the axial $u^\star_z$ (figure \ref{fig:axi_ux_sig}) and the radial  $u^\star_r$ (figure \ref{fig:axi_uy_sig}) component of velocity. A notable difference between these real MRV images and the synthetic MRV images in sections \ref{sec:converging_channel} and \ref{sec:blood_vessel_dummy}, is that the real MRV images display artefacts and contain outliers. We have not pre-processed the MRV images for example by removing outliers. The estimated posterior uncertainty of $\partial\Omega^\circ$ is depicted in figure \ref{fig:axi_shape_rec}, in which we observe that regions with gaps in the data coincide with regions of higher uncertainty. Although we treat the kinematic viscosity $\nu$ as an unknown parameter, the posterior distribution of $\nu$ remains effectively unchanged. More precisely, we infer a kinematic viscosity of ${\nu^\circ = 3.995\times10^{-6}\simeq \mean{\nu}}$, with a posterior variance of $(0.1005\mean{\nu})^2$. This is because we use a Bayesian approach to this inverse problem, where the prior information for $\nu$ is already rich enough. Technically, the reconstruction functional $\mathscr{E}$ is insensitive to small changes of $\nu$ (or $1/\Rey$), and, as a result, the prior term in the gradient of $\nu$ (equation \eqref{eq:grad_nu}) dominates; i.e. the model $\mathscr{M}$ is not informative. Physically, it is not possible to infer $\nu$ (with reasonable certainty) for this particular flow without additional information on pressure.

As in section \ref{sec:aortic_aneurysm}, we compare the denoising performance of algorithm \ref{algo:reconstruction} (figure \ref{fig:axisymm_nozzle_results_1}) with TV-B \citep{Pascal2012,VanDerWalt2014} (figure \ref{fig:axisymm_nozzle_tvb_comp}). We again observe that algorithm \ref{algo:reconstruction} has managed to filter out both the noise and the artefacts, while the TV-B-denoised images present artefacts, loss of contrast, and a systematic error that depends on the parameter $\lambda$.

Figure \ref{fig:axisymm_wss} shows the reconstructed wall shear rate $\gamma^\circ_w$, computed for the reconstructed velocity field $\bm{u}^\circ$ on the segmented shape $\partial\Omega^\circ$, and compares it with the ground truth wall shear rate $\gamma^\bullet_w$, computed for the high SNR velocity field $\bm{u}^\bullet$ (figure \ref{fig:high_qual_axisymm_nozzle_re162}) on the high SNR shape $\partial\Omega^\bullet$ ($^1$H spin density). We observe that the ground truth wall shear rate is particularly noisy, as MRV suffers from low resolution and partial volume effects \citep{Bouillot2018,Saito2020} near the boundaries $\partial\Omega$. Certainly, it is possible to smooth the boundary $\partial\Omega^\bullet$ (which we obtained using the method of \cite{Otsu1979} for the $^1\mathrm{H}$ spin density) using conventional image processing algorithms. However, the velocity field $\bm{u}^\bullet$ will not be consistent with the new smoothed boundary (the no-slip boundary condition will not be satisfied). The method that we propose here for the reconstruction and segmentation of MRV images tackles exactly this problem: it infers the most likely shape of the boundary ($\partial\Omega^\circ$) from the velocity field itself, without requiring an additional experiment (e.g. CT, MRA) or manual segmentation using another software. Furthermore, in this Bayesian setting we can use the $^1\mathrm{H}$ spin density to introduce $\textit{a priori}$ knowledge of $\partial\Omega$ in the form of a prior, which would prove useful in areas of low velocity magnitudes where the velocity field itself does not provide enough information in order to segment the boundaries, \revv{e.g. flow within an aneurysm or a heart ventricle 
\citep{Demirkiran2021}}. As a result, algorithm \ref{algo:reconstruction} performs very well in estimating the posterior distribution of wall shear rate, a quantity which depends both on the velocity field and the boundary shape, and which is hard to measure otherwise.

\subsection{{Choosing the regularization parameters}}
\rev{
Regularization is crucial in order to successfully reconstruct the velocity field and segment the geometry of the nozzle in the presence of noise, artefacts, and outliers. Regularization comes from the Navier--Stokes problems (primal and adjoint) ($\mathscr{M}$), and the regularization of the model parameters ($\mathscr{R}$).}

\subsubsection{Notes on prior information for the Navier--Stokes unknowns}
\rev{
By adopting a Bayesian inference framework, we assume that the prior information of an unknown $x$ is a Gaussian random field with mean $\mean{x}$ and covariance $\mathcal{C}_x$, i.e. ${x \sim \mathcal{N}(\mean{x},\mathcal{C}_x)}$ (see section \ref{sec:regularization} and appendix \ref{app:gaussian_meas}). We, therefore, need to provide algorithm \ref{algo:reconstruction} with a prior mean and a prior covariance for every N--S unknown. For the inlet velocity boundary condition, $\mean{\bm{g}}_i$ can be a smooth approximation to the noisy velocity data at the inlet, and then $\sigma_{\bm{g}_i}$ is the prior standard deviation around this mean. For the outlet natural boundary condition, $\mean{\bm{g}}_o$ can be $0$, and then $\sigma_{\bm{g}_o}$ determines the confidence of the user regarding whether or not the outlet is a pseudotraction-free boundary. For both the inlet and the outlet boundary conditions, the parameter $\ell$, which can be different for each boundary condition, controls the regularity of the functions ${\bm{g}_i}$ and ${\bm{g}_o}$, i.e. length scales smaller than $\ell$ are suppressed. For the shape, $\mean{\phi}_{\pm}$ can be a rough segmentation of the original geometry, and then $\sigma_\sdist$ is the prior standard deviation around this mean. For example, in section \ref{sec:aortic_aneurysm}, we set $\sigma_\sdist$ approximately equal to a length of 7 pixels by visually inspecting the noisy mask \mbox{(figure \ref{fig:aorta_noisy_mask})}. The same methodology applies to the determination of prior information regarding the kinematic viscosity $\nu$.}

\rev{
The advantage of this probabilistic framework is that when prior information is available it can be readily exploited in order to regularize the inverse problem and facilitate its numerical solution. On the other hand, if there is no prior information available regarding an unknown, we can assume that this unknown is distributed according to a zero-mean Gaussian distribution with a sufficiently large standard deviation $\sigma$.}

\subsubsection{Notes on shape regularization and the choice of $\Rey_{\sdist},\Rey_\zeta$}
\rev{
For the axisymmetric nozzle (see section \ref{sec:axisymm_nozzle_rec}), we avoid overfitting the shape $\partial\Omega$ by choosing the Reynolds numbers for the geometric flow to be ${\Rey_{\sdist} = \Rey_\zeta = 0.025}$. Increasing these Reynolds numbers to around $1.0$, we start noticing that the assimilated boundary becomes more susceptible to noise in the image. However, for the simulated aortic aneurysm (see section \ref{sec:aortic_aneurysm}) we chose ${\Rey_{\sdist} = \Rey_\zeta = 1.0}$ in order to preserve high curvature regions. From numerical experiments we have observed that typical successful values for the Reynolds numbers $\Rey_{\sdist},\Rey_\zeta$ lie in the interval $(0.01,0.1)$ for low SNR images ($\text{SNR} < 10$) with relatively flat boundaries, in $(0.1,1.0)$ for higher SNR images ($\text{SNR} \geq 10$) with relatively flat boundaries, and ${\geq 1}$ for geometries with regions of high curvature. Physical intuition that justifies the use of $\Rey_{\sdist},\Rey_\zeta$ as the preferred shape regularization parameters is provided in section \ref{sec:prop_boundary_geomflow}.
}

\section{Conclusions}
We have formulated a generalized inverse Navier--Stokes problem for the joint reconstruction and segmentation of noisy velocity images of \revv{steady incompressible flow}. To regularize the inverse problem, we adopt a Bayesian framework by assuming Gaussian prior distributions for the unknown model parameters. Although the inverse problem is formulated using variational methods, every iteration of the nonlinear problem is actually equivalent to a Gaussian process in Hilbert spaces. We implicitly define the boundaries of the flow domain in terms of signed distance functions and use Nitsche's method to weakly enforce the Dirichlet boundary condition on the moving front. The moving of the boundaries is expressed by a convection-diffusion equation for the signed distance function, which allows us to control the regularity of the boundary by tuning an artificial diffusion coefficient. We use the steepest ascent directions of the model parameters in conjunction with a quasi-Newton method (BFGS), and we show how the posterior Gaussian distribution of a model parameter can be estimated from the reconstructed inverse Hessian. 

We devise an algorithm that solves this inverse Navier--Stokes problem and test it for noisy ($\text{SNR}=2.5, 3$) 2D synthetic images of Navier--Stokes flows. The algorithm successfully reconstructs the velocity images, infers the most likely boundaries of the flow and estimates their posterior uncertainty. We then design a magnetic resonance velocimetry (MRV) experiment to obtain images of a 3D axisymmetric Navier--Stokes flow in a converging nozzle. We acquire MRV images of poor quality ($\text{SNR}\simeq6$), intended for reconstruction/segmentation, and images of higher quality ($\text{SNR}>30$) that serve as the ground truth. We show that the algorithm performs very well in reconstructing and segmenting the poor MRV images, which were obtained in just $10.2$ minutes, and that the reconstruction compares well to the high SNR images, which required a total acquisition time of $\sim 4.6$ hours. Lastly, we use the reconstructed images and the segmented (smoothed) domain to estimate the posterior distribution of the wall shear rate and compare it with the ground truth. Since the wall shear rate depends on both the shape and the velocity field, we note that our algorithm provides a consistent treatment to this problem by jointly reconstructing and segmenting the flow images, avoiding the design of an additional experiment (e.g. CT, MRA) for the measurement of the geometry, or the use of external (non physics-informed) segmentation software.  

The present method has several advantages over general image reconstruction and segmentation algorithms, which do not respect the underlying physics and the boundary conditions, and, at the same time, provides additional knowledge of the flow physics (e.g. pressure field and wall shear stress), which is otherwise difficult to measure. \revvv{It can be used to substantially decrease signal acquisition times and provides additional knowledge of the physical system being imaged. Although our current implementation is restricted to 2D planar and axisymmetric flows, the method naturally extends to periodic and unsteady Navier--Stokes problems in complicated 3D geometries.}

\vspace{1cm}
\textbf{Declaration of Interests}. The authors report no conflict of interest.

\appendix
\section{Gaussian measures in Hilbert spaces}\label{app:gaussian_meas}
The mean of a Gaussian measure $\gamma \sim \mathcal{N}(m,\mathcal{C})$ in $L^2$ is given by
\begin{gather}
m \equiv \mathbb{E} h := \int_{L^2} h\ \gamma(dh)\quad.
\end{gather}
The covariance operator $\mathcal{C}:L^2\to L^2$ and the covariance $C : L^2\times L^2 \to \R$ are defined by
\begin{gather}
\cov x := \int_{L^2} h \binner{x,h}\ \gamma(dh)\quad,\quad C(x,x') := \int_{L^2}\binner{x,h}\binner{x',h}\ \gamma(dh)\quad,
\end{gather}
noting that $\binner{\cov x,x'} = C(x,x')$. The above (Bochner) integrals define integration over the function space $L^2$, and under the measure $\gamma$, and are well defined due to Fernique's theorem \citep{Hairer2009}. These integrals can be directly computed by sampling the Gaussian measure $\gamma$ with Karhunen--Lo\`eve expansion (as in section \ref{sec:uncertainty}).

\section{Euler--Lagrange system}\label{app:eul_lag_sys}
The integration by parts formulae for the nonlinear term (equation \eqref{eq:dMdu}) are 
\begin{align}
\int_\Omega \big(\bm{u}'\bm{\cdot}\nabla\bm{u}\big)\bm{\cdot}\bm{v} &= \int_\Omega u_j'\partial_ju_i\ v_i = -\int_\Omega u_j'u_i\ \partial_jv_i + \partial_ju_j'u_i\ v_i + \int_{\partial\Omega} u'_j \nu_j\ u_iv_i \nonumber \\ &= -\int_\Omega \big(\bm{u}\bm{\cdot}(\nabla\bm{v})^\dagger)\bm{\cdot}\bm{u}' + {\color{black}\int_{\partial\Omega} (\bm{u}\bm{\cdot}\bm{v})(\bm{\nu}\bm{\cdot}\bm{u}')}\quad,\\
\int_\Omega \big(\bm{u}\bm{\cdot}\nabla\bm{u}'\big)\bm{\cdot}\bm{v} &= \int_\Omega u_j\partial_j u'_i\ v_i = -\int_\Omega \partial_ju_j u'_i\ v_i + u_j u'_i\ \partial_jv_i  + \int_{\partial\Omega} u_j u'_i\ \nu_j v_i\nonumber\\
&=-\int_\Omega \big(\bm{u}\bm{\cdot}\nabla\bm{v}\big)\bm{\cdot}\bm{u}' + {\color{black}\int_{\partial\Omega} (\bm{u}\bm{\cdot}\bm{\nu})(\bm{v}\bm{\cdot}\bm{u}')}\quad .
\end{align}

\section{Axisymmetric inverse Navier--Stokes problem}\label{app:axisymm}
The axisymmetric Navier--Stokes problem is
\begin{gather}
\bm{u}\bm{\cdot}\nabla\bm{u} - \nu\rmDelta \bm{u} + \nabla p + \bm{f} = \bm{0}, \quad \nabla\bm{\cdot}\bm{u} = 0\quad,
\end{gather}
where
\begin{gather*}
\bm{u} = u_z \hat{\bm{z}} + u_r \hat{\bm{r}}\quad, \quad \nabla \bm{u}= (\partial_z \bm{u},~\partial_r \bm{u})\quad, \quad \rmDelta \bm{u} = \partial_z^2 u_z + \partial_r^2 u_r + \frac{1}{r}~\partial_r u_r\quad, \\
\nabla\bm{\cdot}\bm{u} = \partial_z u_z + \partial_r u_r + \frac{u_r}{r}\quad,\quad \bm{f} = \Big(0,\frac{\nu u_r}{r^2}\Big)\quad,
\end{gather*}
and the nonlinear term $\bm{u}\bm{\cdot}\nabla\bm{u}$ retains the same form as in the Cartesian frame.

In order to compare the axisymmetric modeled velocity field with the MRV images, we introduce two new operators: i) the reflection operator ${\mathcal{R} : \mathbb{R}^+\times\mathbb{R} \to \mathbb{R}\times\mathbb{R}}$, and ii) a rigid transformation ${\mathcal{T} : \mathbb{R}^2\to\mathbb{R}^2}$. The reconstruction error is then expressed by
\begin{equation}
\mathscr{E}(\bm{u}) \equiv \frac{1}{2}\norm{\bm{u}^\star-\mathcal{S}{\color{black}\mathcal{T}\mathcal{R}}\bm{u}}^2_{\cu} := \frac{1}{2}\int_I \big(\bm{u}^\star-\mathcal{S}{\color{black}\mathcal{T}\mathcal{R}}\bm{u}\big)\invcu\big(\bm{u}^\star-\mathcal{S}{\color{black}\mathcal{T}\mathcal{R}}\bm{u}\big)~\mathrm{d}x\mathrm{d}y\quad.
\label{eq:rec_error}
\end{equation}
We introduce an unknown variable for the vertical position of the axisymmetry axis by letting
$\mathcal{T}u = u(x,y+y_0)$, for $y_0 =\text{const}$. Then, the generalized gradient for $y_0$ is
\begin{gather}
\Big\langle D_{y_0}\mathscr{J}, y'_0 \Big\rangle_\R = \Big\langle -\int_{I}\invcu\big(\bm{u}^\star-\mathcal{S}\mathcal{T}\mathcal{R}\bm{u}\big)\big(\mathcal{S}\mathcal{T}\mathcal{R}~\partial_y\bm{u}\big),~ y'_0\Big\rangle_\R\quad,
\end{gather}
and $y_0$ is treated in the same way as the inverse Navier--Stokes problem unknowns $\bm{x}$.
\bibliographystyle{jfm}
\bibliography{main.bib}

\end{document}